\documentclass[twocolumn]{aastex631}

\usepackage{longtable}
\usepackage{upgreek}
\usepackage{lineno}
\shorttitle{Probing the Star Formation Main Sequence down to $10^{7}$~M$_\odot$ at $1.0<z<3.0$}
\shortauthors{M\'erida et al.}
\graphicspath{{./}{figures/}}

\begin{document}

\title{Probing the Star Formation Main Sequence down to $10^{8}$~M$_\odot$ at $1.0<z<3.0$}

\author[0000-0001-8115-5845]{Rosa M. M\'erida}
\affiliation{Centro de Astrobiolog\'ia (CAB), CSIC-INTA, Ctra. de Ajalvir km 4, Torrej\'on de Ardoz, E-28850, Madrid, Spain}
\affiliation{Departamento de F\'isica Te\'orica, Universidad Aut\'onoma de Madrid, E-28049, Cantoblanco (Madrid), Spain}

\author[0000-0003-4528-5639]{Pablo G. P\'erez-Gonz\'alez}
\affiliation{Centro de Astrobiolog\'ia (CAB), CSIC-INTA, Ctra. de Ajalvir km 4, Torrej\'on de Ardoz, E-28850, Madrid, Spain}
\affiliation{Departamento de Física de la Tierra y Astrof\'isica, Fac. CC. F\'isicas, Universidad Complutense de Madrid, Plaza de las Ciencias 1, Madrid, E-28040, Spain}

\author[0000-0003-0651-0098]{Patricia S\'anchez-Bl\'azquez}
\affiliation{Departamento de Física de la Tierra y Astrof\'isica, Fac. CC. F\'isicas, Universidad Complutense de Madrid, Plaza de las Ciencias 1, Madrid, E-28040, Spain}
\affiliation{IPARCOS (Instituto de Física de Partículas y del Cosmos)
Facultad de Ciencias F\'{\i}sicas, Ciudad Universitaria, Plaza de las Ciencias, 1, Madrid,E-28040, Spain}

\author[0000-0002-8365-5525]{\'Angela Garc\'ia-Argum\'anez}
\affiliation{Departamento de Física de la Tierra y Astrof\'isica, Fac. CC. F\'isicas, Universidad Complutense de Madrid, Plaza de las Ciencias 1, Madrid, E-28040, Spain}

\author[0000-0002-8053-8040]{Marianna Annunziatella}
\affiliation{Centro de Astrobiolog\'ia (CAB), CSIC-INTA, Ctra. de Ajalvir km 4, Torrej\'on de Ardoz, E-28850, Madrid, Spain}

\author[0000-0001-6820-0015]{Luca Costantin}
\affiliation{Centro de Astrobiolog\'ia (CAB), CSIC-INTA, Ctra. de Ajalvir km 4, Torrej\'on de Ardoz, E-28850, Madrid, Spain}

\author[0000-0002-6696-7834]{Alejandro Lumbreras-Calle}
\affiliation{Centro de Estudios de F\'isica del Cosmos de Arag\'on, Plaza San Juan 1, E-44001 Teruel, Spain}

\author[0000-0002-4140-0428]{Bel\'en Alcalde-Pampliega}
\affiliation{European Southern Observatory (ESO), Alonso de C\'ordova 3107, Vitacura, Casilla 19001, Santiago de Chile, Chile}

\author[0000-0001-6813-875X]{Guillermo Barro}
\affiliation{Department of Physics, University of the Pacific, 3601 Pacific Ave., Stockton, CA 95211, USA}

\author[0000-0001-6426-3844]{N\'estor Espino-Briones}
\affiliation{Instituto Nacional de Astrof\'isica, \'Optica y Electr\'onica, Luis E. Erro No. 1, Tonantzintla, C.P. 72840, Puebla, M\'exico}

\author[0000-0002-6610-2048]{Anton M. Koekemoer}
\affiliation{Space Telescope Science Institute, 3700 San Martin Dr., Baltimore, MD 21218, USA}



\begin{abstract}

We investigate the star formation main sequence (MS) (SFR-M$_{\star}$) down to 10$^{8-9}\mathrm{M}_\odot$ using a sample of 34,061 newly-discovered ultra-faint ($27\lesssim i \lesssim 30$~mag) galaxies at 1$<z<$3 detected in the GOODS-N field. Virtually these galaxies are not contained in previous public catalogs, effectively doubling the number of known sources in the field. The sample was constructed by stacking the optical broad-band observations taken by the HST/GOODS-CANDELS surveys as well as the 25 ultra-deep medium-band images gathered by the GTC/SHARDS project. Our sources are faint (average observed magnitudes $<i>\sim28.2$~mag, $<H>\sim27.9$~mag), blue (UV-slope $<\upbeta>\sim-1.9$), star-forming (rest-frame colors $<U-V>\sim0.10$~mag, $<V-J>\sim0.17$~mag) galaxies. These observational characteristics are identified with young (mass-weighted age $<\mathrm{t_{M-w}}>\sim0.014$~Gyr) stellar populations subject to low attenuations  ($<\mathrm{A(V)}>\sim0.30$~mag). Our sample allows us to probe the MS down to $10^{8.0}\,\mathrm{M}_\odot$ at $z=1$ and $10^{8.5}\,\mathrm{M}_\odot$ at $z=3$, around 0.6 dex deeper than previous analysis. In the low-mass galaxy regime, we find an average value for the slope of 0.97 at $1<z<2$ and 1.12 at $2<z<3$. Nearly $\sim$60\% of our sample presents stellar masses in the range $10^{6-8}$ M$_\odot$ between $1<z<3$. If the slope of the MS remained constant in this regime, the sources populating the low-mass tail of our sample would qualify as starburst galaxies. 

\end{abstract}



\section{Introduction} 

The progressive increase in the amount of public astronomical data obtained by many observatories at multiple wavelengths is nowadays ruling the astrophysical paradigm: science has entered a data-intensive mode. Large photometric (and, to a lesser extent, spectroscopic) surveys, both ground- and space-based, have led to the gathering of vast multi-wavelength galaxy catalogs, which have enabled to obtain many statistically robust results, shedding light on how galaxies formed and evolved from very early epochs in the history of the Universe.

Among those surveys, we benefit in this paper from two of the deepest ever carried out: CANDELS and SHARDS. The Cosmic Assembly Near-infrared Deep Extragalactic Legacy Survey (CANDELS; \citealt{CANDELS}, \citealt{WFC3}) is a 902-orbit Hubble Space Telescope (HST) Multi-Cycle Treasury (MCT) program that aimed to study the evolution of galaxies during the first third of the Universe evolution combining previously obtained  Advanced Camera for Surveys (ACS) optical data with newly observed Wide Field Camera 3 (WFC3) near-infrared (NIR) images. CANDELS gathered multi-wavelength deep images in five sky fields: the Great Observatories Origins Deep Survey fields (GOODS-N and GOODS-S, \citealt{ACS}, \citealt{GOODS}), the UKIDSS Ultradeep Survey field  (UDS, \citealt{UDS1} \& \citealt{2007MNRAS.380..585C}), the Extended Groth Strip (EGS, \citealt{EGS}) and the Cosmological Evolution Survey field (COSMOS, \citealt{COSMOS}).

Multi-wavelength photometric catalogs have been built for all the CANDELS fields: \citet{Guo} for GOODS-S; \citet{uds_cat} for UDS; \citet{cosmos_cat} for COSMOS; \citet{egs_cat} for EGS, and \citet{Barro} for GOODS-N (hereafter B19). The sources in each catalog were selected in the NIR filter $F160W$. Point-like source $5\sigma$ depths range from 26.0 to 27.6~mag in all 5 CANDELS fields, except for CANDELS GOODS-S, where the depths achieved in the CANDELS/Deep region and the Hubble Ultra Deep Field (HUDF) rise to 28.2 and 29.8~mag, respectively. These observations allowed for the detection of over 250,000 galaxies in the 5 fields.

A selection based on a single band and biased to the NIR is necessarily missing some objects that, not being bright enough at the observed wavelength, could be detected in other bands. In particular, faint sources with a blue spectral energy distribution (SED) peaking in the near-UV could not have enough flux in the rest-frame optical to be detected by NIR-based searches \citep{Santos}. Among these blue sources, we could expect to find low-mass star-forming galaxies, whose study becomes more and more relevant as we move to higher redshifts and reach the epoch of the first star formation episodes in the Universe. For example, the study of faint low-mass sources is crucial because they are considered to be candidates for the reionization \citep{Anderson}, and also because a very large fraction of the unobscured UV luminosity density ($\geq$90\%) could be due to their emission \citep{Reddy}.

In one of the CANDELS fields, GOODS-N, another very relevant survey was carried out, the Survey for High-z Absorption Red and Dead Sources, SHARDS \citep{SHARDS}, which also covers 2 Frontier Fields (\citealt{SHARDSFF1}, \citealt{SHARDSFF2}). SHARDS obtained data in 25 medium-band filters in the 500 to 950~nm spectral range, reaching at least magnitude 27 at the 3$\sigma$ level at subarcsec seeing in each one of them, and providing spectral resolution $R=50$ in the optical. The final depth of the SHARDS observations nicely matches the CANDELS/GOODS ACS observations, providing better sampled SEDs for most CANDELS-selected sources in the GOODS-N region \citep{Barro}. The spectral resolution of SHARDS expedites the detection of emission-line galaxies (\citealt{Cava}, \citealt{SHARDSFF1}, \citealt{Arrabal}, \citealt{Lumbreras}, \citealt{Rodriguez-Munoz}) as well as the detailed analysis of the SEDs to obtain singularly accurate photometric redshifts \citep{Barro}, absorption band measurements (\citealt{Helena}, \citealt{Hernan_absorption}, \citealt{Hernan_absorption_2}) or properties of the stellar population of bulges and disks in massive galaxies (\citealt{constantin_1}, \citealt{constantin_2}).

We will concentrate on the study of low-mass galaxies at $z>1$. Previous studies managed to recover low-luminosity/mass galaxies at high redshift, missed by major catalogs, using gravitational lensing (e.g., \citealt{Stark}, \citealt{Alavi} at $z\sim2$; \citealt{Karman} at $z>3$; \citealt{Santini} at $z\leq$4; \citealt{Caputi} at $z=4.3$; \citealt{Sun} at $z\sim$4; \citealt{Rinaldi} at $z\approx$3.0-6.5), or searching for the Lyman-$\alpha$ line combining broad and narrow band filters (\citealt{Matthee}, \citealt{Nilsson}). However, the presence of interlopers at intermediate and low redshifts can lead to the contamination of the high-z samples. \citet{MUSE_2} used an optimal extraction scheme that took the published HST source locations as prior and then performed a blind search of emission line galaxies based on advanced test statistics and filter matching. Another option for gravitational lensing or searching for emission lines is to look for low-mass galaxies in the proximity of absorption systems (\citealt{Arrigoni}, \citealt{Diaz}).

In this paper, we present a catalog of galaxies undetected by CANDELS/GOODS, described in \citet{Barro} (and also by the 3D-HST public catalog, \citealt{3DHST}). Our sources do not show prominent emission in the NIR, e.g., in the $H$-band, but are bright and can be easily selected through their UV/optical emission, e.g., in the $bviz$ ACS bands or the SHARDS data. Stacking UV/optical images allows the enhancement of faint features when present in different bands and improves the signal-to-noise ratio (SNR). We will detect these galaxies using stacked images centered at $\sim$700~nm. Given its spectral resolution, by using the SHARDS medium-band observations, we can also recover emission line objects.

In principle, the faint emission from optically-bright NIR-faint objects that we expect to find can be due to the presence of blue, star-forming, young, low-mass, and unobscured galaxies \citep[e.g.,][]{Stark, Mcgaugh, MUSE}, or massive star-forming systems suffering from severe dust attenuation \citep[e.g.,][]{Alcalde, Yamaguchi, Sun}. 

Apart from presenting a catalog of faint sources in the GOODS-N field, we concentrate on the analysis of the relationship between the star formation rate (SFR) and the stellar mass (M$_{\star}$), what is known as the main sequence (MS; see \citealt{Noeske}, \citealt{Elbaz}, \citealt{Whitaker_12}, \citealt{Whitaker_15}, \citealt{Speagle}, \citealt{Lee}, \citealt{Schreiber}, \citealt{Santini}, \citealt{Tomczak}, \citealt{Popesso}, \citealt{Barro}). In particular, we probe low stellar masses, trying to reach the dwarf regime at the highest redshift possible. This kind of MS study is linked to understanding the type of star formation histories (SFHs) followed by low-mass galaxies, whether they are smooth or bursty \citep[see][and also simulations in \citealt{Weisz} and \citealt{Teyssier}]{Guo_low_mass,Karman,Caputi,Rinaldi}.

Indeed,  precedent studies of the MS available in the literature up to $z\sim8$ (see references in the previous paragraph) are biased towards the behavior of the SFR-M$_{\star}$ relationships at stellar masses above $10^{9-10}\,\mathrm{M}_\odot$ at $z>1$. In this mass regime, there is a tight relation between SFR and stellar mass, with an intrinsic scatter of 0.2-0.3 dex \citep{Speagle}. The tightness of the relation is indicative of an SFH that traces stellar mass growth more smoothly, rather than an SFH with many discrete bursts (\citealt{Salmon} and references therein). As an example of this type of discussion, \citet{bursty} claimed that the SFHs of massive galaxies can be described by slowly varying functions of time, on timescales larger than 100~Myr, but stochastic processes rule the dwarf regime. The smooth/bursty behavior of the SFHs should be directly linked to feedback processes, where AGN and supernovae play a major role and probably distinct at different masses (see \citealt{DeGraf}, \citealt{Torrey}, \citealt{Koudmani}, \citealt{Dekel}, \citealt{Chaikin}). 

Many of the previously mentioned works affirm that the MS cannot be well described by a single slope and that it exhibits a turnover point at high masses, which is redshift dependent, more specifically at $\sim$10$^{10.2-10.5}$M$_\odot$ (e.g., \citealt{Whitaker_15}, \citealt{Schreiber}, \citealt{Tomczak}, \citealt{Lee}). Below the turn-over point, the slope lies between 0.6-1.0 (\citealt{Speagle} and references therein), and above it turns shallower. In this paper, we aim at probing the low-mass regime, where very little information is available to date.

At the low-mass end of the MS, reaching down to $10^8$ M$_\odot$, there are some researches (e.g., \citealt{Reddy}, \citealt{Sawicki}, \citealt{Santini}, \citealt{MUSE}) that obtained an MS slope compatible with unity for $0<z<3$. Nevertheless, they count with a small number of objects.

In this work, we will look into the properties of low-mass galaxies between $1<z<3$ in an unprecedented way, given the completeness, size, and robustness of our sample compared to previous studies at that mass range. We aim at constraining the MS below the typical mass completeness limits used to date, shedding light on the smooth/bursty nature of the SFHs of low-mass galaxies and the feedback processes that rule them. 

The structure of the paper is as follows. In section \ref{sect:data} we present the data and sample construction methodology. We then describe the physical properties of the sample in sections \ref{sect:properties} and \ref{sec:synthesizer}. In sections \ref{sec:SFR} and \ref{sec:discussion}, we present and discuss our results regarding the SFR-M$_{\star}$ MS relation. The conclusions are summarized in section \ref{sect:conclusions}. 

Throughout the paper we assume a flat cosmology with $\mathrm{\Omega_M\, =\, 0.3,\, \Omega_{\Lambda}\, =\, 0.7}$, and a Hubble constant $\mathrm{H_0\, =\, 70\, km\,s^{-1} Mpc^{-1}}$, and use AB magnitudes \citep{okegunn1987}. All stellar mass and SFR estimations refer to a \citet{Chabrier} IMF.

\section{Data and sample construction}
\label{sect:data}
\subsection{Data}

We base this work on the analysis of the CANDELS HST images combined with the 10.4-m Gran Telescopio de Canarias (GTC) ultra-deep imaging data from SHARDS.

The primary CANDELS data consist of imaging obtained in the ACS optical bands and the WFC3 infrared bands (WFC3/IR), with a total of 8 broad-band filters from 0.4 to 1.6~$\upmu$m.
We use version v3.0 of the mosaics provided by the GOODS HST/ACS Treasury Program \citep{ACS, CANDELS, WFC3}, and the v1.0 data release for the WFC3/IR bands \citep{CANDELS, WFC3}. We exclude the $F140W$ band because it is significantly noisier than the other WFC3 bands. The HST imaging reaches limiting magnitudes that range from 27.4 to 28.1~mag with a point spread function (PSF) full-width half maximum (FWHM) that ranges from 0.1\arcsec\, to 0.2\arcsec.

SHARDS \citep{SHARDS} and SHARDS Frontier Fields (SHARDS-FF; \citealt{SHARDSFF1}, \citealt{SHARDSFF2}) are 2 large programs carried out using the OSIRIS \citep{Osiris} instrument to image the GOODS-N field, and the MACS1149 and A370 cluster fields respectively, with 25 contiguous, medium-band filters ($\sim$15-16~nm wide except for two of the reddest ones). These data cover the spectral range between 0.5 and 0.95~$\upmu$m. SHARDS surveyed the GOODS-N field with two pointings (p1, p2), summing up an area of $\sim$120 arcmin$^2$. The data reach AB magnitudes of $\sim$26.8-27.8~mag at the 5$\sigma$ level with subarcsec seeing in all bands.

\begin{table*}
\addtolength{\tabcolsep}{-3pt}
\caption{Filters used in this work. The 5$\sigma$ depths are calculated as the value below which we find 75\% of the objects with a magnitude uncertainty $<0.2$ mag. The r$_{\textrm{ap}}$ column shows the median radius of the apertures used to measure the photometry within each image. In order to take into account the elliptical apertures provided by \texttt{SExtractor}, this radius is calculated as $\sqrt{\textrm{SMA*sma}}$, where SMA and sma stand for semi-major and semi-minor axes, respectively. For each filter, we have two sets of values, one corresponding to the SHARDS-selected sample (left) and another one corresponding to the HST-selected sample (right). In the case of  the MIPS, PACS, and SPIRE bands, measurements are made by PSF fitting (see more details in \citealt{PZETA}).}   
\label{tab:filters}     
\centering                                     
\begin{tabular}{c c c c c c}         
\hline\hline                        
Band & $\lambda_{\textrm{central}}$(µm)  & FWHM(arcsec) & 5$\sigma$ depth (mag) &r$_{\textrm{ap}}$ (arsec) & Ref. \\  
\hline                                  
U                        & 0.36                    & 1.25     & 27.6& 0.84/0.55&\citet{Capak}   \\
\smallskip
B                        &  0.44                          & 0.83        & 27.6&0.84/0.55 &\citet{Capak}  \\
\smallskip
ACS HST bands                & 0.43-0.90            &        0.10-0.11     & 27.4-28.1& 0.84/0.20 &\citet{ACS}, \citet{CANDELS} \\
 & & & & &\citet{WFC3} \\
 \smallskip
WFC3/IR HST bands                & 1.05-1.54            &        0.18-0.19    & 27.8-28.0& 0.84/0.20&  \citet{CANDELS}, \citet{WFC3}  \\
\smallskip
SHARDS bands             & 0.49-0.92          &  0.75-1.14  & 26.8-27.8&0.84/0.55  &\citet{SHARDS} \\
\smallskip
SHARDS ALL-bands stack         & 0.73                    & 0.90       & 28.1&0.84/0.55 &\textcolor{blue}{This work}   \\
\smallskip
SHARDS SDSS $r$ band stack       & 0.59                   & 0.95        &27.7 &0.84/0.55  &\textcolor{blue}{This work} \\
\smallskip
SHARDS SDSS $i$ band stack       & 0.76                    & 0.91      & 27.5 &0.84/0.55 &\textcolor{blue}{This work}    \\
\smallskip
SHARDS SDSS $z$ band stack       & 0.88                    & 0.92        &27.4 &0.84/0.55 &\textcolor{blue}{This work} \\
\smallskip
ACS HST stack            & 0.72                    &  0.16           & 28.7& 0.84/0.20&\textcolor{blue}{This work}  \\
\smallskip
ACS+WFC3/IR HST stack       & 1.07                    &  0.18        &28.9 & 0.84/0.20&\textcolor{blue}{This work}  \\
\smallskip
WFC3/IR HST stack           & 1.28                    &   0.20         &28.6 & 0.84/0.20&\textcolor{blue}{This work}  \\
\smallskip
WIRCAM K                 & 2.13                     & 0.88        & 25.4&0.84/0.55 &\citet{WIRCAM} \\
\smallskip
IRAC bands               & 3.60-8.00            &      1.70-2.00  &24.2-25.7 & 1.50/1.50 &\citet{IRAC1}, \citet{IRAC2}, \\
 & & & & &\citet{Barro} \\
MIPS bands& 24-70& 6-18&30$^*$-1.2$^{**}$ & &\citet{SYNTHESIZER}\\
PACS bands&100-160 &7-11 &1.7-3.6$^{**}$ & &\citet{Elbaz}, \citet{Magnelli}\\
SPIRE bands& 250-500&14-17 &9-13$^{**}$ &  &\citet{Elbaz}, \citet{Magnelli}\\
\hline     
\end{tabular}
\tablecomments{$^*$\label{note1} $\mu$Jy}
\tablecomments{$^{**}$\label{note2} mJy}
\end{table*}

Additionally, we complement the HST and GTC datasets with NIR observations from the Wide-field InfraRed Camera at the Canada-France-Hawaii Telescope (CFHT/WIRCAM) (K-band; \citealt{WIRCAM}), Spitzer/IRAC (3.6~$\upmu$m, 4.5~$\upmu$m, 5.8~$\upmu$m, and 8.0~$\upmu$m bands; \citealt{IRAC1,IRAC2}), and    $U$, $B$ images from SUBARU \citep{Capak}. 

For the calculation of SFRs, we also take into account measurements from the Spitzer/MIPS 24 and 70~$\upmu$m mosaics presented in \citet{SYNTHESIZER}, the Herschel PACS 100 and 160~$\upmu$m, and the SPIRE 250, 350, and 500~$\upmu$m catalogs described in \citet{Elbaz} and \citet{Magnelli}. See B19 for more details on this dataset in the GOODS-N region.

We obtain the photometry from masked versions of all these images except for IRAC, for which we use the residual images from the deconvolution method presented in \citet{Barro}. The masked images are obtained by removing the objects in B19 from the original images. The masked regions are then filled with sky emission to avoid possible contamination on the selection and photometry of our sample of faint sources. The IRAC images were obtained using {\tt TFIT} \citep{TFIT}. {\tt TFIT} is a template-fitting code to measure galaxy photometry using prior information about the position of the sources from high-resolution observations (in this case, the positions measured from $F160W$). 
The code creates mock images and accounts for the contamination of nearby objects by comparing them with the real ones. Then, {\tt TFIT} measures the magnitudes, providing residual images after subtracting the emission of all the sources included in the input catalog (constructed with the high-resolution image). Any flux coming from sources not present in that input catalog would be measurable in the {\tt TFIT} residual image.

To optimize the detection of intrinsically faint sources, we combine the images of all the 25 filters of SHARDS in a stacked frame, which we will call the SHARDS ALL-bands stack hereafter. We also build stacks with central wavelengths and widths similar to the SDSS $r$ (10 SHARDS filters from $F500W17$ to $F670W17$, excluding $F568W17$ -affected by a sodium skyline-), $i$ (10 SHARDS filters from $F687W17$ to $F840W17$), and $z$ (7 SHARDS filters from $F806W17$ to $F941W17$) bands to allow for the detection of emission line galaxies with a low stellar continuum whose signal could be diluted when summing up the 25 bands. These stacks will also be used to  build SEDs for each selected galaxy. For HST, we construct 3 images stacking only the ACS filters, the WFC3/IR filters, and using all HST bands (ACS+WFC3/IR).
Table \ref{tab:filters} lists the filters used in this work, together with some of their main features.

\subsection{Sample selection}

\begin{deluxetable}{ccc}
\label{tab:sextractor}
\caption{\texttt{SExtractor} parameters used for  the detection of objects in the HST and SHARDS stacked images.}
\tablehead{\colhead{} & \colhead{SHARDS ALL-bands} & \colhead{ACS HST}} 
\startdata
  FILTER & Mexhat & none \\
  FILTER FWHM (pixels)  & 3   &   \\
 \texttt{DETEC\_THRESH} & 1.5 & 2.0\\
 \texttt{ANALYSIS\_THRES}& 0.5 & 0.5\\
 \texttt{DEBLEND\_NTHRESH}& 64 & 64\\
 \texttt{DETECT\_MINIAREA} (pixels) &  4 & 4 \\
 \enddata
\end{deluxetable}

Our main goal is to study the properties of low-mass galaxies ($10^{8-9}\mathrm{M_\odot}$) at $z>1$ that were missed by the selection criteria of other works in the same field (e.g., \citealt{3DHST}, \citealt{Bouwens}, \citealt{Finkelstein}, \citealt{Maseda}, \citealt{Barro}), but that can be detected in HST and SHARDS' stacked images.

\subsection{Detection of faint sources in SHARDS and HST data}

We use \texttt{SExtractor} \citep{SExtractor} to detect sources in the SHARDS ALL-bands (the SHARDS-faint catalog) and ACS HST (the HST-faint catalog) stacks. Only sources in the area in common between SHARDS and the B19 HST-based catalog are considered. 

The thresholds and filters applied in both cases are listed in Table~\ref{tab:sextractor}. Both stacks cover a similar wavelength range, providing a coherent combined sample of faint sources detected in the optical spectral range. Some of the sources that belong to the SHARDS-faint catalog have a counterpart in the HST-faint catalog, but some of them do not. These SHARDS-faint sources are more likely emission line galaxies, that are more easily seen in SHARDS thanks to its higher spectral resolution. In addition, many sources are only detectable in the ACS HST stack and lie beyond the SHARDS magnitude limit.

There are cases where an object is detected in both stacks. This is identified when an HST aperture fits inside a SHARDS aperture. The HST emission is significantly more concentrated and the spatial resolution allows the definition of a clearer center for the sources. In the cases where both SHARDS and HST coincide in a detection, we keep the SHARDS aperture (the source thus belongs to the SHARDS-faint catalog), but re-centered to the HST aperture.

 We note that the objects detected within the SHARDS ALL-bands stack are usually more extended than those detected within the ACS HST stack: the typical aperture radius of the former is 0.84\arcsec, compared to 0.2\arcsec\, of the latter (see Table \ref{tab:filters}). This is expected as the signal dilution due to the SHARDS PSF makes some faint point-like sources found in HST stacked data to be missed in SHARDS.

 In the following sections, we only analyze SExtracted sources whose SNR$>4$ in the corresponding selection stack according to the \texttt{photutils} package from \texttt{Python}, as measured in a $r=0.7$\arcsec\, circular aperture for the SHARDS-selected sample, and a $r=0.2$~\arcsec\, circular aperture for the HST-selected sample. These radii ensure that more than 90\% of the total flux is enclosed in the aperture for point-like sources.

The sources from the SHARDS-faint catalog require additional purges. First, all objects whose ellipticities are higher than 0.8 are rejected, since we do not expect such elliptical objects in this sample. Second, in the same line, we remove all the sources with semi-major axis SMA $>2$~\arcsec. A cut in exposure time is also imposed to remove the sources located in the outermost parts of the stack, where we only count with the contribution from a few filters.

\begin{figure}
    \centering
    \includegraphics[width=8cm, height=6.9cm]{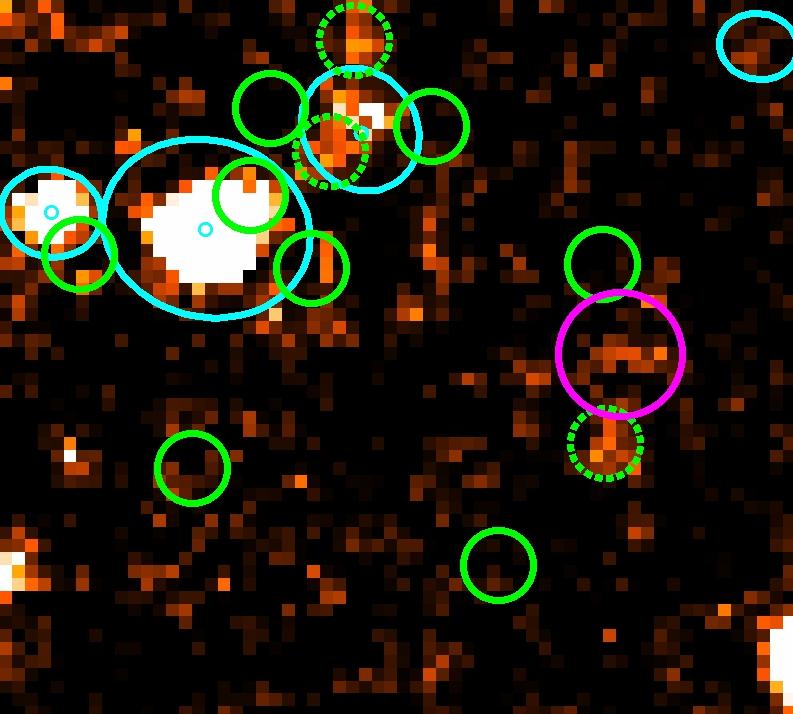}  \\     \vspace{0.3cm}
    \includegraphics[width=8cm, height=6.9cm]{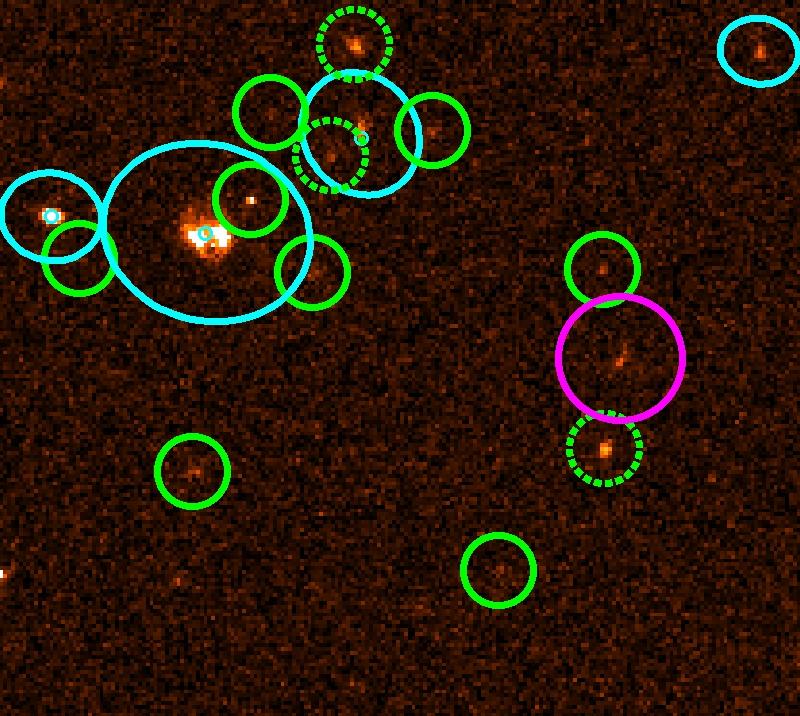}
    \caption{Cutouts (size $12"\times11"$) of the SHARDS ALL-bands stack (top) and the ACS HST stack (bottom) around coordinates $\alpha=$12:37:40.30 $\delta=$+62:14:38.66. In both images, the sources detected by B19 are marked with cyan elliptical apertures (based on \citealt{Kron1980} radii). Sources included in the 3D-HST \citep{3DHST} catalog are marked with small cyan circles. None of the \citet{Bouwens}, \citet{Finkelstein} nor \citet{Maseda} objects lie in this cutout. The sources detected in the ACS HST stack are marked in green, those with a SHARDS counterpart show a dashed contour. The objects detected in the SHARDS ALL-bands stack are encircled in magenta.}
    \label{fig:cutouts}
\end{figure}

In Fig.~\ref{fig:cutouts}, we show a cutout of the stacks for the  SHARDS ALL- and the ACS HST bands to exemplify our detection methodology and to offer a comparison of our sample with sources contained in other public catalogs. In this small part of the whole SHARDS/CANDELS surveyed area (just 0.04\% of the total area), B19 detects 4 sources, 3 of them also detected by the 3D-HST team \citep{3DHST}. None of the objects from \citet{Bouwens}, \citet{Finkelstein} nor \citet{Maseda} lie in this section of the sky. We recover 1 more source directly detected in the SHARDS ALL-bands stack, also showing an HST counterpart. The detection based on the ACS-HST stack provides 11 more objects, 3 of them also presenting a SHARDS counterpart. In total, in this region covering 0.04\% of the SHARDS+CANDELS common surveyed area (122~arcmin$^2$, 63~arcmin$^2$ in the region covered by the SHARDS first pointing and 59 arcmin$^2$ in the region covered by the second one), we detect 3 times more sources than what has ever been cataloged before.

\subsection{Multi-wavelength photometry}
\label{sect:Multi-wavelength photometry}

\begin{figure*}
    \centering
    \includegraphics[width=18cm, height=16cm]{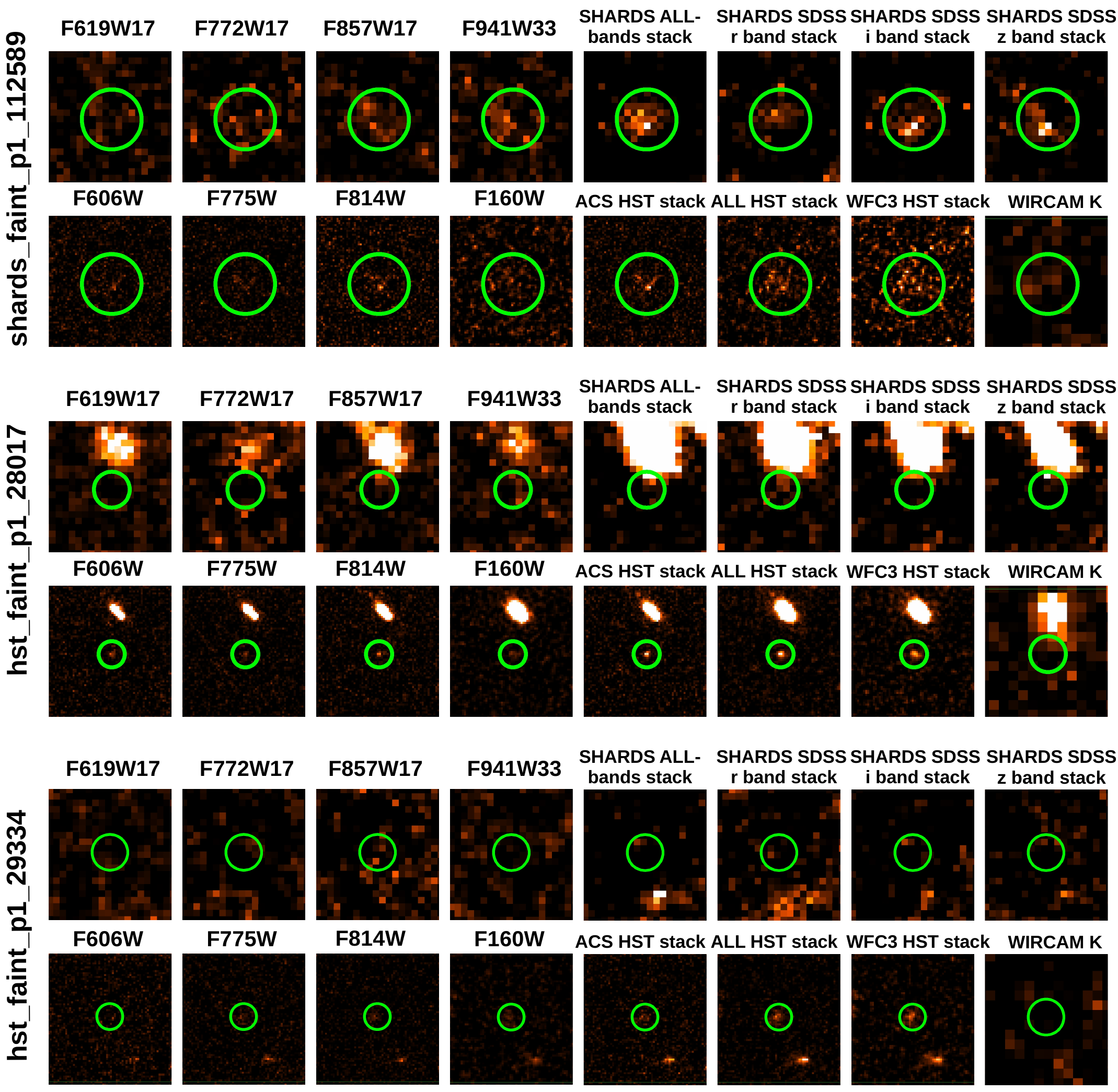}
    \caption{Cutouts ($4"\times4"$) of 3 different objects of our sample in different SHARDS and HST bands and stacks, and also the WIRCAM K band. The green circles depict the photometric apertures used in each band for each object. The object at the top is an example of a source extracted from the SHARDS stacked images ($r=0.92$\arcsec), whereas the other two are obtained from the ACS HST stack. They are measured in apertures of $r=0.4$\arcsec\, when looking at the HST images, and in apertures of $r=0.55$\arcsec\, when measuring within the ground-based images. The properties of these galaxies are listed in Tables \ref{tab:subsample} and \ref{tab:subsample_physical}.}
    \label{fig:stamps_transmission}
\end{figure*}

We measure photometry in all available individual-band images and stacks (using masked versions, see Section \ref{sect:data}) for the galaxies from the SHARDS- and HST-faint catalogs mentioned above with the \texttt{Rainbow} pipeline (\citealt{PZETA}, \citealt{SYNTHESIZER}, \citealt{Rainbow}). \texttt{Rainbow} is a software that cross-correlates multi-band catalogs and provides aperture-matched photometry on the different bands, covering the range that goes from the UV to the far infrared (FIR). It also re-centers the apertures within the images using local World Coordinate Systems solutions, calculating the offset relative to a reference catalog. This correction is always smaller than 0.01\arcsec.

Different aperture radii are used to measure within the ground- and space-based images. The details about the choice of the optimal aperture radii, as well as several tests to show the robustness of the photometry to this choice, can be found in Appendix \ref{sec:app1}.
Fig.~\ref{fig:stamps_transmission} shows representative examples of the objects of our sample  in different photometric bands with the corresponding  apertures.  The photometric properties of these galaxies can be found in Table \ref{tab:subsample}.

In order to build a robust sample, characterized by reliable SED-derived properties, and to avoid spurious measurements, we only keep those sources that are detected in two or more of the  stacked images described in Section \ref{sect:data} and two or more SHARDS and/or HST individual photometric bands. A detection in a particular band is defined as a flux measurement with an SNR$>3$ according to \texttt{Rainbow}. Measurements with a smaller SNR are rejected and the value of the sky noise is taken as an upper limit for that band, using a $5\sigma$ level. We keep this value and not the $3\sigma$ one because upper limits are crucial in the redshift and stellar population synthesis (SPS) fittings since they guarantee that models do not scale to very high stellar masses or strong breaks. Detections are, however, defined using an SNR$>3$ condition since the upper limits are already strongly constraining the fitting, and going deeper in terms of SNR for the detection makes us lose too many potential objects which turn up to be real galaxies based on detections in other bands. Additionally, if SNR$>3$, but the flux in a band is smaller than the 3$\sigma$ level of the sky, the measurement is substituted by the sky noise, also taken as an upper limit. 

\begin{deluxetable*}{cccccc}
\tabletypesize{\small}
\label{tab:subsample}
\caption{Photometric properties in the selection bands of the subsample shown in Fig.\ref{fig:stamps_transmission}. The absence of uncertainty denotes an upper limit.}
\tablehead{\colhead{Id.} & \colhead{RA}& \colhead{DEC}&\colhead{ $i$ (AB mag)} & \colhead{$H$ (AB mag)} & \colhead{\# Bands$>$3$\mathrm{\sigma}$/upper limits}}
\startdata 
shards$\_$faint$\_$p1$\_$112589&12:37:04.63&+62:20:19.75&27.0$_{26.8}^{27.2}$&26.5$_{26.7}^{26.3}$&24/23\\
hst$\_$faint$\_$p1$\_$28017&12:36:56.64&+62:15:49.66&27.0&27.9$_{27.6}^{28.1}$&8/39\\
hst$\_$faint$\_$p1$\_$29334&12:36:43.22&+62:16:07.20&27.8$_{27.7}^{28.0}$&27.7$_{27.6}^{27.9}$&8/39\\
\enddata
\end{deluxetable*}

Our final sample of {\it bona fide} sources contains (1) 5,947 SHARD-faint objects (3,377 and 2,570 in the p1 and p2 pointings, respectively); and (2) 28,114 HST-faint sources (15,787 in the region covered by the SHARDS first pointing and 12,327 in the region covered by the second one).
2,037 HST sources have a SHARDS counterpart, whereas 3,791 are only detected in SHARDS.
The sum of the SHARDS- and HST-faint catalogs, hereafter the SHARDS/CANDELS faint catalog, contains a total of 34,061 sources, a number of objects very similar to what we find in the public CANDELS and 3D-HST catalogs in GOODS-N (35,445 and 38,279 sources, respectively). Our catalog has 902, 665, and 3,646 sources in common with the \citet{Bouwens},  \citet{Finkelstein}, and \citet{3DHST} catalogs, respectively, which represent 3\%, 2\%, and 11\% of the sample. 2 of our sources coincide with galaxies from the subsample of \citet{Maseda}, but they are located at $z<1$, which is out of the redshift interval considered in this work. The rest of the targets are new detections.

The final  photometry is obtained by applying the corresponding aperture corrections, according to the PSF of each band and stack. PSF-aperture corrections can be safely applied to our targets, as they are point-like at HST and SHARDS resolutions (see Appendix \ref{sec:app2}). We check the stability of the seeing conditions across the SHARDS stacks and find a maximum variation in the PSF of only 0.04~\arcsec across them.

\begin{figure*}
    \centering
    \includegraphics[width=18.2cm, height=9.6cm]{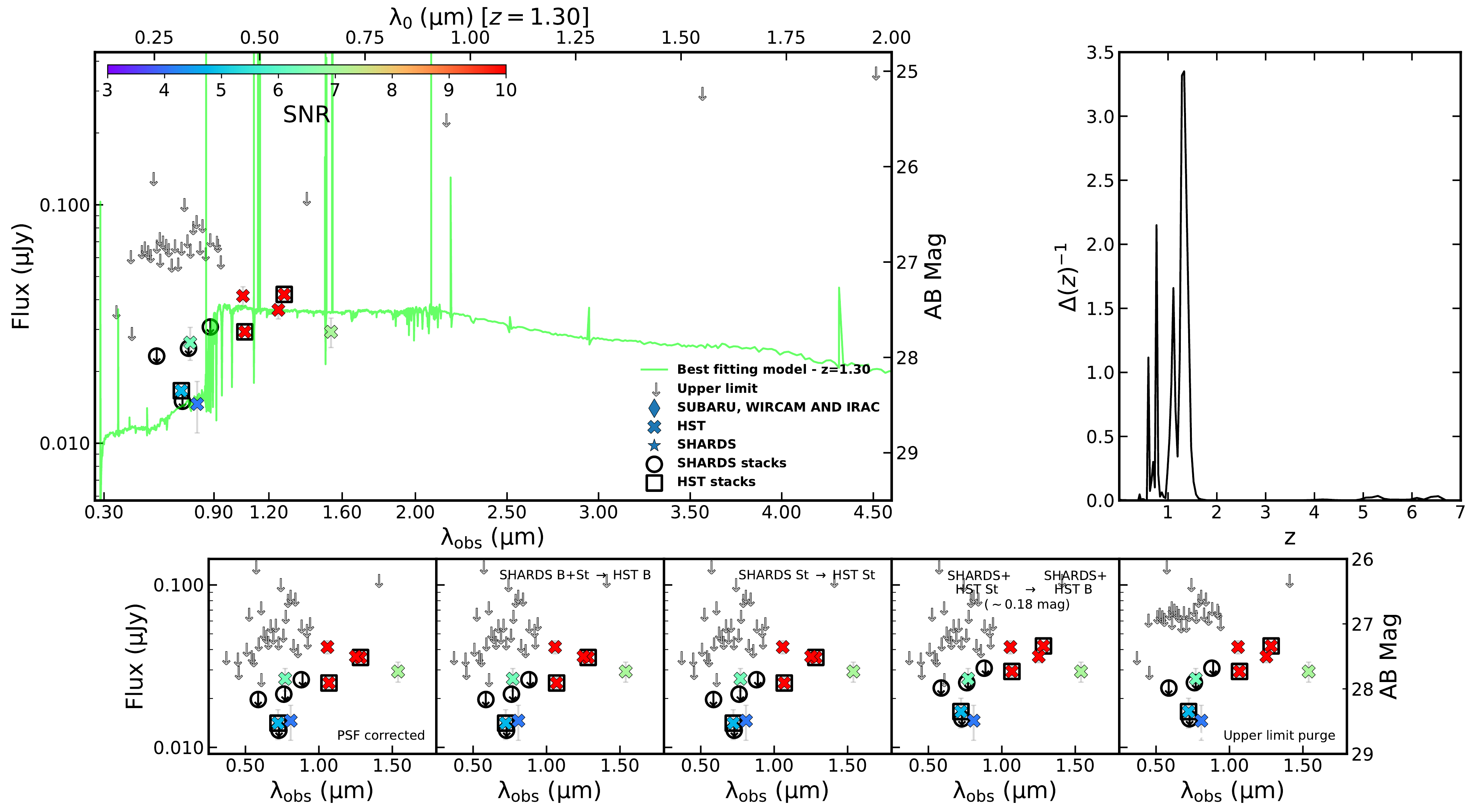}
    \caption{SED and fitting results for the galaxy hst$\_$faint$\_$p1$\_$29334. The vertical arrows on the SED plot (upper-left panel) stand upper limits. Diamonds: Subaru $U$, $B$, WIRCAM $K$, and IRAC 3.5$\upmu$m and 4.5$\upmu$m (the remaining IRAC spectral range is omitted for clarity, although it is used in the analysis); stars: SHARDS fluxes (encircled when referring to SHARDS stacked images); crosses: HST fluxes (squared when referring to HST stacked images). The markers are color-coded according to the SNR. The SED is fitted to stellar population models. We show the best-fitting one, corresponding to the following stellar population properties: age $t_0=0.72$~Gyr, timescale $\tau=130$~Myr, attenuation A(V)=0.0~mag, stellar mass log(M/M$_\odot$)=7.8, and a mass-weighted age of 0.48~Gyr (see Section \ref{sec:synthesizer}). At the top right, we show the photometric redshift probability distribution function (zPDF). In the bottom panels, we show the stages of the offset corrections applied to the photometric points obtained with \texttt{Rainbow}. The first subplot shows the photometric points after the PSF correction. The next subplots show the change in the points according to the different corrections described in Section \ref{sect:Multi-wavelength photometry}. Cutouts of this object in different bands are shown in Figure \ref{fig:stamps_transmission}. The properties of this galaxy are listed in Tables \ref{tab:subsample} and \ref{tab:subsample_physical}.}
    \label{fig:SED}
\end{figure*} 

We applied different magnitude corrections to ensure the consistency of ground- and space-based photometry, given that, in some cases, different aperture radii are used for both datasets (see Appendix \ref{sec:app1}), and that the HST image quality is more stable than that of the ground-based images.
These corrections are applied to both, individual bands and stacked images, with the following sequence: (1)  
we correct the fluxes in all the SHARDS filters, forcing the continuum level to be the same as in the HST individual bands; (2) the fluxes in the SHARDS stacks are corrected to make them comparable to those in the HST stacks; (3) the fluxes in both, HST and SHARDS stacks are corrected to match the level of the individual HST and SHARDS photometric points, and (4) HST and SHARDS upper limits are compared with the  upper limits of adjacent filters, to ensure that all of them show a similar and consistent trend. Typically, all these corrections are smaller than 0.2~mag, but applying them improves the results of the SED analysis.

Let us note that the majority of our galaxies are detected through the ACS HST stack (83\% of the galaxies) and not through the SHARDS ALL-bands stack. This means that the SEDs of most of our sources count with actual flux measurements only in HST data, and upper limits in most other bands (namely, Subaru $U$ and $B$ bands, SHARDS, $K$-band, and IRAC). 40\% of the total sample shows no detections in the SHARDS individual bands and only 33\% of it is detected in 1 or 2 bands. 55\% is not detected in any of the SHARDS stacks and 33\% is detected in at least two of them. The corrections which affect the SHARDS photometry have only an effect on $<20\%$ of our sample.
The third correction, which compares the measurements from the stacked images with the measurements taken within the individual images, and the upper limit purge are thus the noticeable corrections.

Even though the first and second corrections, which result from having different image resolutions, affect a minor part of our sample,
we test if the choice of an optimal aperture radius and additional corrections on the fluxes are appropriate by convolving the HST images with a PSF that matches the SHARDS resolution to effectively “simulate” what the HST images would look like if they had the same resolution as the SHARDS images. We then measure the fluxes of the objects in the convolved HST image using the same aperture size as in the SHARDS images and compare them with the aperture-corrected fluxes measured within the original resolution HST images. We find that, on average, the aperture-corrected fluxes are 0.1~mag higher than those measured in the convolved HST images. This difference translates into very small systematic offsets in the derived stellar mass compared to the typical error in this parameter (see Section \ref{sec:synthesizer}). Furthermore, we do not find a significant offset in the colors ($\sim$ 0.03~mag), with a scatter of 0.30~mag.

Figure~\ref{fig:SED} shows an example of a SED for a typical target of the HST-faint catalog, and the effect of sequentially applying the different corrections on the SHARDS and HST photometric points. Only the third and fourth bottom panels show a difference with respect to the previous steps since this object is not detected in any of the SHARDS bands or stacks. The third bottom panel shows that this offset correction is only of the order of 0.18~mag.

\subsection{Photometric properties of the sample and comparison with public catalogs}
\label{sect:Photometric_properties}
In this section, we briefly describe the statistical properties of our sample and compare it with the sources from the publicly available catalogs mentioned in previous sections.

\begin{figure*}
    \centering
    \includegraphics[width=8.95cm, height=7.3cm]{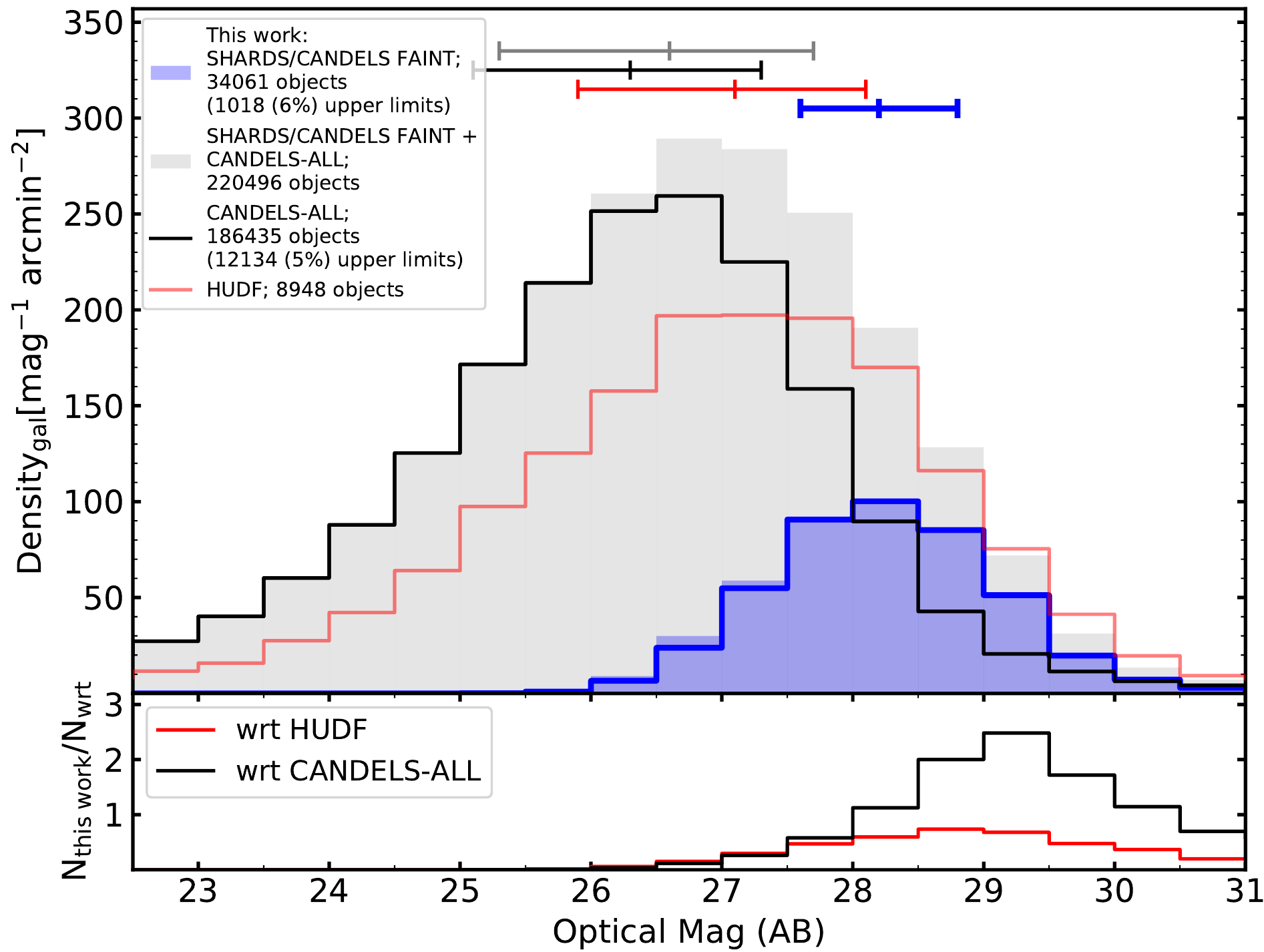}
    \includegraphics[width=8.95cm, height=7.3cm]{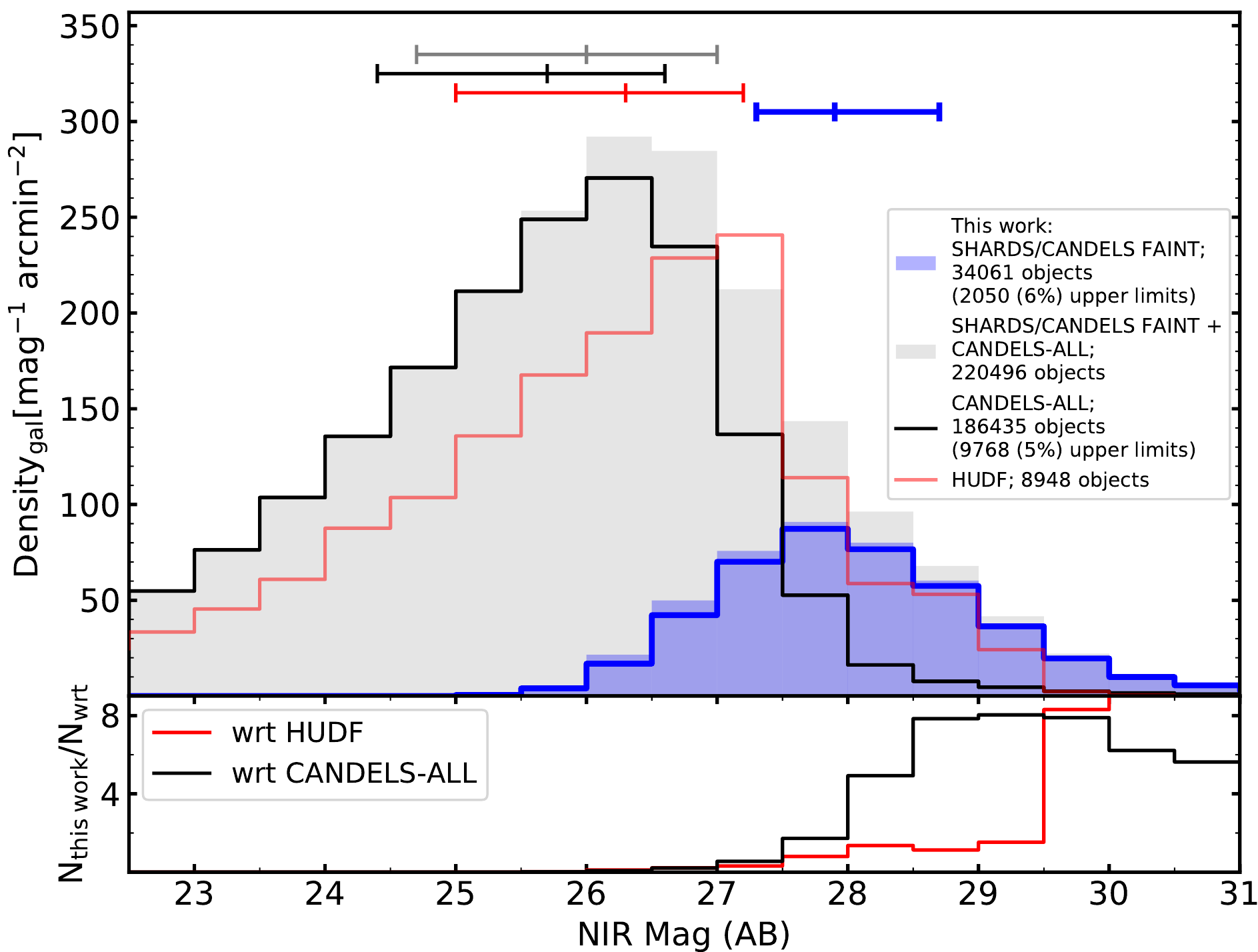}
    \caption{Top panels: histograms showing the magnitude distribution of our objects in the optical ($F775W$ HST image band; left) and in the NIR ($F160W$ HST image band; right). The black line shows the distribution that results from considering the galaxies from all the CANDELS fields, whereas in red we show only the objects that lie in the HUDF. Our sample is represented in blue: the shaded region corresponds to the whole sample, including upper limits, whereas the blue contour corresponds only to the detected galaxies. The sum of the galaxies of all the CANDELS fields, together with our sample, is shown with a grey shade. Quartiles are shown as horizontal segments. Bottom panels: ratio of our objects in each magnitude bin with respect to each comparison sample. In this case, in black, we show the ratio of our objects with respect to all CANDELS fields, and in red just with respect to the HUDF.}
    \label{fig:HISTO_MAG}
\end{figure*}

Fig.~\ref{fig:HISTO_MAG} shows the distribution of optical (HST $F775W$) and NIR (HST $F160W$) magnitudes for our sample, and the targets in the 5 CANDELS fields (hereafter, CANDELS-ALL sample). CANDELS data include CANDELS/Deep and CANDELS/Wide, with different magnitude limits. GOODS-S has, in addition, a region where measurements are even deeper than those from CANDELS/Deep, which is the HUDF; we also include in our comparison the histogram with the objects in this region. Comparing our results with the objects in all these surveys allows us to better understand how well our sample completes the faint-end of the current magnitude distribution.

In the optical, the magnitude distribution of our sample peaks at 28.3~mag, with values ranging from 26-31~mag, with a median and quartiles of 28.2$_{27.6}^{28.8}$~mag\footnote{Note that some  magnitudes are higher than the depths listed in Table~\ref{tab:filters} due to our definition of depth, based on the magnitudes below which we find 75\% of the objects with an uncertainty $<0.2$ mag}. The magnitude distribution for previously known sources according to the CANDELS-ALL sample extends to such faint values, but peaks at 26.8~mag and, statistically, they are 2~mag brighter than our sources (median and quartiles 26.3$_{25.1}^{27.3}$~mag). The combined distribution that results from the sum of the galaxies from CANDELS-ALL together with our sample extends towards fainter magnitudes thanks to our sample, with the maximum still coinciding with the peak of the CANDELS-ALL sample, and median and quartiles 26.6$_{25.3}^{27.7}$~mag. In red in Fig.~\ref{fig:HISTO_MAG}, we show the distribution of objects detected in the HUDF, that peaks at 27.3~mag, fainter than the CANDELS-ALL sample, but still 1~mag brighter than our sample, with median and quartiles 27.1$_{25.9}^{28.1}$~mag. In terms of the ratio per magnitude bin of our sample and CANDELS-ALL, our sample counts with the same number of galaxies as CANDELS-ALL at 28.3~mag, doubling CANDELS-ALL at 29.3~mag, and still overcoming it at 30.3~mag (same number of galaxies as CANDELS-ALL). When comparing with the HUDF, our sample shows around half the number of galaxies of the HUDF at 28.3~mag, with the maximum at 28.8~mag, representing three-quarters of the HUDF, and decreasing towards fainter galaxies (nearly a quarter of the HUDF at 30.3~mag).

In the NIR, our catalog peaks at 27.8~mag, 0.5~mag fainter than the HUDF sample, and covers the magnitude interval between 25.5 and 31~mag, with median and quartiles 27.9$_{27.3}^{28.7}$~mag. The CANDELS-ALL sample distribution peaks at 26.3~mag and also extends to fainter magnitudes but quickly drops beyond magnitude 28 (25.7$_{24.4}^{26.6}$~mag). The sum of CANDELS-ALL with our sample produces a histogram whose median and quartiles are 26.0$_{24.7}^{27.0}$~mag. It makes the distribution drop more slowly beyond the peak towards the limit at mag$\sim$31. The HUDF sample dramatically drops beyond the peak too, decreasing up to 29.5~mag, with median and quartiles 26.3$_{25.0}^{27.2}$~mag. Whereas in the optical, CANDELS in the HUDF always samples fainter galaxies per magnitude bin and arcmin$^2$ than our catalog, in the NIR the ratio of objects with respect to the HUDF exceeds unity for magnitudes fainter than 28.3. The ratio of sources with respect to the CANDELS-ALL sample exceeds the 1:1 ratio from 27.8~mag on, and at 29.3~mag our sample has 8 times more sources than CANDELS-ALL. Our sample is thus actually completing the current faint-end of the magnitude distribution, especially in the NIR. 

The $H$-band 50\% completeness magnitude limits for the CANDELS/Wide, Deep, and the HUDF catalogs are 25.9, 26.6, and 28.1~mag respectively (\citealt{Barro}; \citealt{Guo}). To compute the completeness in previous studies, they linearly fit the differential number density in the magnitude range where the catalogs are considered to be complete (20-24 mag for the Deep and Wide regions and 21-26 mag for the HUDF). When the number density starts deviating from the fit, the catalog begins to lose sources. According to Fig.~\ref{fig:HISTO_MAG}, and as explained in the previous paragraphs, our sample is comparable to what is measured in the HUDF, in the optical, and in the NIR, peaking later than the sum of all the galaxies of CANDELS-ALL, and extending to 29-31 AB magnitudes. It is thus reasonable to consider our catalog to be complete beyond magnitude 24, similarly to the HUDF. The combination B19 + SHARDS/CANDELS faint extends the 50\% completeness limit, computed as in the previous studies that we mention, to 27.3~mag in the CANDELS/Wide region (fitting the number density between 21.0-26.5~mag), and to 27.7~mag in the CANDELS/Deep region (fitting the number density between 23.5-25.5~mag).

\begin{figure}
    \centering
    \includegraphics[width=8.5cm, height=7cm]{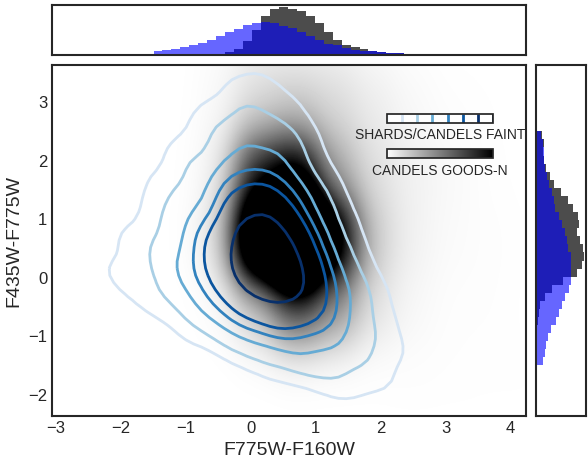}
    \caption{Color-color diagram, $i-H$ vs. $b-i$, of the CANDELS-ALL sample, in black, and the SHARDS/CANDELS faint sample, in blue contours. The density contours represent the 20th, 50th, 60th, 70th, 80th, and 90th percentiles. Histograms of both colors are shown on the right and top, following the same color code as in the main plot.}
    \label{fig:COLOR_COLOR}
\end{figure}

In Fig.~\ref{fig:COLOR_COLOR} we show a color-color diagram based on the apparent magnitudes measured in the $F435W$, $F775W$, and $F160W$ filters. In this case, we limit the comparison to the CANDELS GOODS-N field. According to this diagram, our sample is bluer than the sources from B19 in both colors. The median $i$-$H$ color and quartiles of our sample are 0.21$_{-0.32}^{0.72}$, whereas B19 median color and quartiles are 0.67$_{0.35}^{1.03}$. For $b$-$i$, these numbers are 0.32$_{-0.33}^{1.06}$ and 0.80$_{0.32}^{1.35}$ respectively. The bluer colors of our galaxies are expected, given that our selection is based on optical imaging, while the comparison catalogs (B19, 3D-HST, \citealt{Bouwens}, \citealt{Finkelstein}, \citealt{Maseda}) are biased towards the selection in the WFC3 data.

Summarizing the information in Figs.~\ref{fig:HISTO_MAG} and ~\ref{fig:COLOR_COLOR}, our sample of galaxies reaches 1-2 magnitudes deeper (in the optical and NIR) than previously published catalogs (B19, 3DHST, \citealt{Bouwens}, \citealt{Finkelstein}, \citealt{Maseda}) and the selection is biased towards bluer objects.

\section{Photometric redshifts of the sample of SHARDS/CANDELS faint sources}
\label{sect:properties}

We derive the photometric redshifts using the \texttt{PZETA} code (\citealt{PZETA}, \citealt{SYNTHESIZER}). \texttt{PZETA} is based on empirically-built SED templates spanning from UV to mid-IR rest-frame wavelengths, constructed with galaxies with known and reliable spectroscopic redshifts, which are used as a training set. The method used by this code is similar to a neural network technique in the sense that it uses the photometric data of these galaxies with known spectroscopic redshifts to train the photometric redshift algorithm. 

\begin{figure*}
    \centering
    \includegraphics[width=18cm, height=11.4cm]{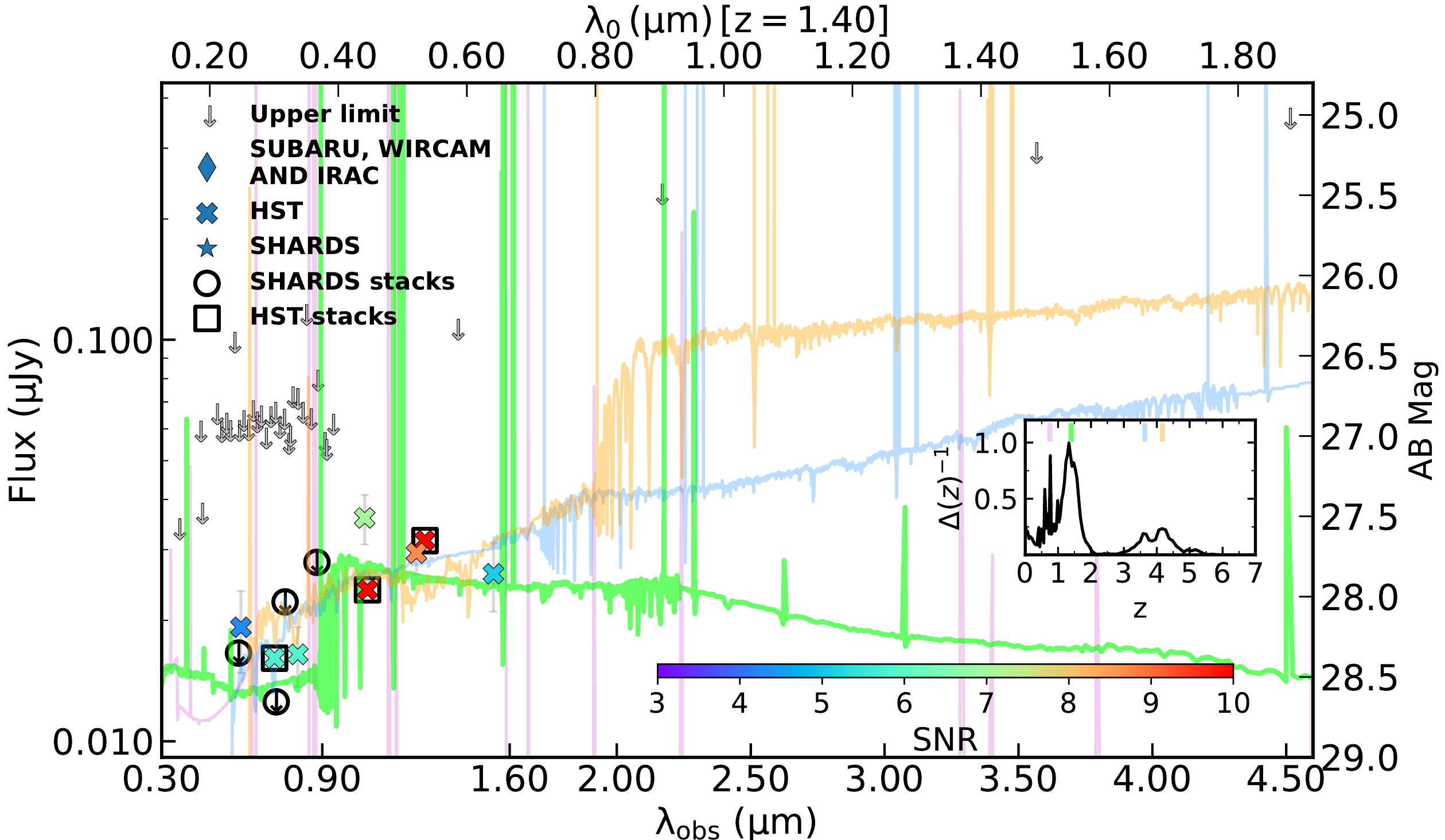}
    \caption{SED of the source hst$\_$faint$\_$p1$\_$28017 between 0.30~$\upmu$m and 4.60~$\upmu$m (the remaining IRAC spectral range is omitted for clarity, although it is used in the analysis). The coding of the different markers can be found in Fig.~\ref{fig:SED}. The SED is fitted to stellar population synthesis models in order to estimate a photometric redshift as well as stellar population properties. The photometric redshift probability distribution function, zPDF, is included as an inset. We mark several different redshift probability peaks on it and show the different models for each of those peaks. The most-probable photo-z for this galaxy is $z=1.40$, whose model is shown in green. Postage stamps for this source are shown in Fig.~\ref{fig:stamps_transmission}. The properties of this galaxy are listed in Tables \ref{tab:subsample} and \ref{tab:subsample_physical}.}
    \label{fig:zpdf}
\end{figure*}

\begin{figure*}
    \centering
    \includegraphics[width=18cm, height=13.6cm]{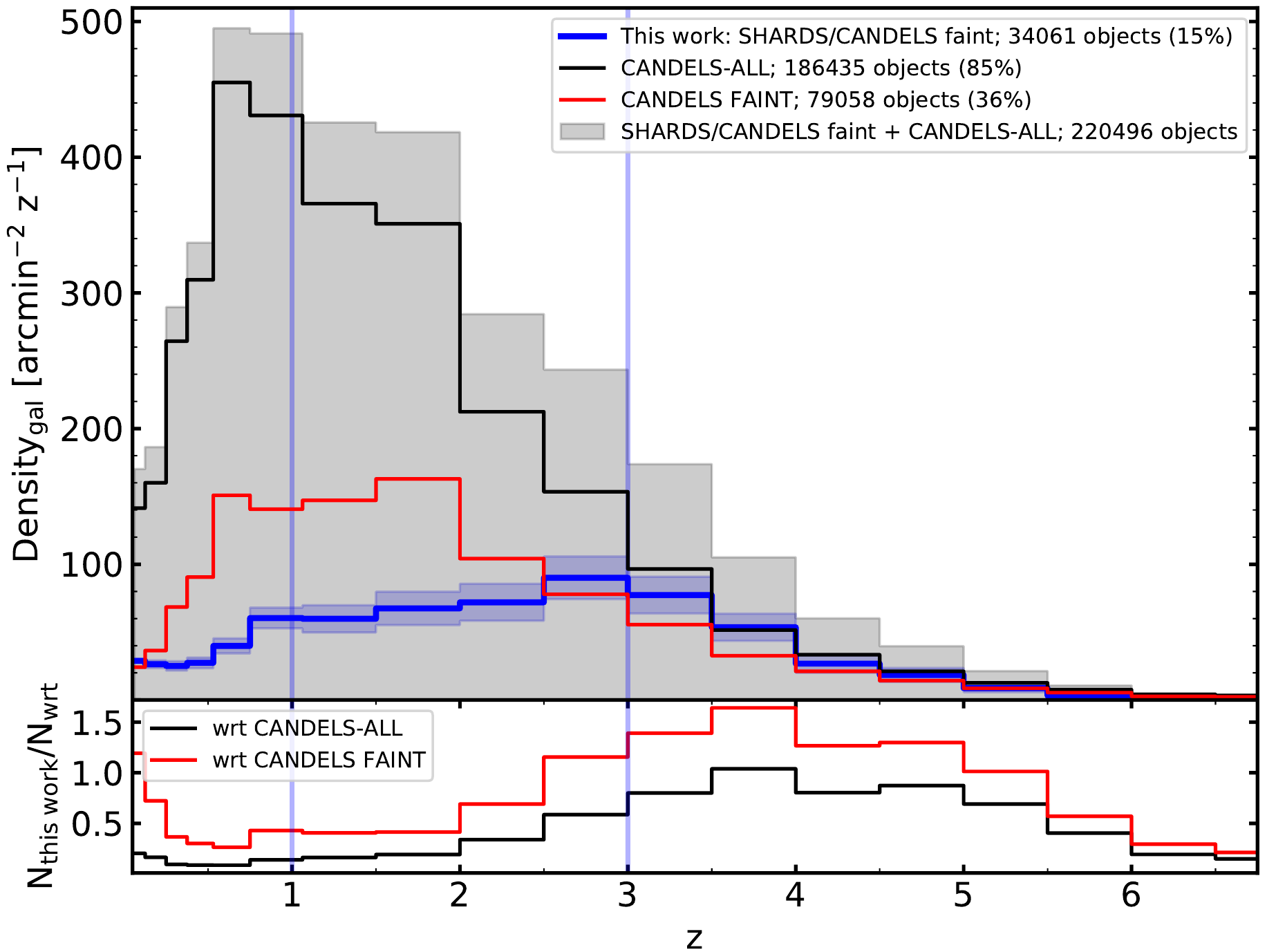}
    \caption{Redshift distribution of our sample, shown as a blue contour, compared with CANDELS-ALL (black contour) and a subsample extracted from the latter only including the faintest objects (CANDELS-FAINT, in red, defined as those galaxies with an AB magnitude in the $F160W$ filter fainter than 26~mag). Jointly with the redshift distribution of our sample, we also include the corresponding uncertainty for each redshift bin, represented by the blue shaded area, and calculated by considering the whole zPDF for each galaxy. The sum of the CANDELS-ALL sample and our galaxies is plotted in solid grey. In the lower panel, we show the ratio of our objects at different redshift bins with respect to CANDELS-ALL and CANDELS-faint samples, using the same color code as above. The vertical lines highlight the values between which the photometric redshifts of our sample are well-constrained.}
    \label{fig:z}
\end{figure*}

The code takes into account upper limits, which makes it especially suitable for very faint sources. The sky noise for non-detections can be used as hard upper limits or adding information to the $\chi^2$ calculations if a given template shows emission above the noise measurements. The restrictive role of the upper limits is a parameter that can be tweaked to improve the results and depends on the relative depth of the datasets. The upper limits for all UV-to-IRAC spectral ranges (including the K-band, remarkably) are quite important for constraining the photo-z solutions of our sample. The magnitudes of our targets are well above the magnitude limit of the deepest spectroscopic surveys ($i=25$~mag), and none of them have spectroscopic redshifts. 

For each source, a most-probable photometric redshift value is assigned based on the integration of the probability distribution function (zPDF). The zPDF also allows us to take into account uncertainties in the photo-z estimation and how those affect the other physical properties discussed in this paper, most significantly the stellar mass and SFR (see Section \ref{sec:synthesizer}). An example of the SED of one of our sources and the corresponding zPDF which shows several peaks, together with the best-fitting models for each peak, are shown in Fig.~\ref{fig:zpdf} (see Figure~\ref{fig:SED} for another example). In this case, the most-probable redshift is found to be $z=1.4$.

In Fig.~\ref{fig:z} we show the distribution of the most-probable photometric redshifts for our sample, including the uncertainties based on the full consideration of the zPDF for each galaxy. We compare our results with the CANDELS-ALL sample, as well as with a subsample extracted from this catalog and limited in magnitude (the CANDELS-FAINT sample, $F160W>26$~mag). The latter is constructed as a better comparison sample for our faint galaxies. The redshifts of these comparison samples are taken from \citet{Barro}. They were obtained using a modified version of the \texttt{EAZY} code \citep{EAZY} adapted to take into account the spatial variation in the effective wavelength of the SHARDS filters depending on the galaxy position in the SHARDS mosaics. 

It can be seen that distinctively to the $H$-band selected catalogs, peaking at lower redshifts, our selection finds galaxies mostly between redshifts 1 and 4, with 1st and 3rd quartiles located at $z=1.5$ and $z=3.3$ respectively. The CANDELS-ALL sample 1st and 3rd quartiles are located at $z=0.8$ and 2.2, whereas for the CANDELS-FAINT sample, these numbers are $z=1.0$ and $z=2.6$.

In the following sections, we will focus on the redshift interval $1<z<3$ (highlighted in Figure \ref{fig:z}), where our sample is representative, allowing us to obtain meaningful statistical results. This redshift range contains 52\% of our objects (17,554). The B19 sample includes 21,216 objects in that redshift range.

We also compare our results with \texttt{EAZY}. This code uses a much more limited number of templates compared to \texttt{PZETA}. Still, it allows us to check the global performance of \texttt{PZETA}. Slightly more than 80\% of the sample located at $1<z<3$ according to \texttt{PZETA} also lies in that interval according to \texttt{EAZY}. In addition, both codes indicate that the cumulative probability for the $1<z<3$ interval is $>$60\% for 80\% of the sample.

Even though we could extend our analysis to higher redshifts ($z\sim4$), we see that those high-z sources count with fewer detections, SEDs are noisier and the photo-z uncertainty is important. However, it is worth noticing that, when looking at the SPS fitting of these galaxies, we mainly find best-fitting models that put the Balmer break bluewards of the $H$-band. Taking this into consideration, the literature estimations of the MS at $z>3$ might be affected by significant selection effects, more prominent than at lower redshifts. Indeed, the $H$-band is shifted to 320~nm at $z=4$ (400~nm at $z=3$), bluewards of the Balmer Break. This means that the $H$-band selection in the CANDELS-ALL catalog should be affected by a relatively strong discontinuity due to the presence of this spectral feature, i.e., it is more probable to miss galaxies when the $F160W$ photometric data point lies bluewards of the Balmer break, and the catalog completeness in terms of mass should be smaller.

\section{Properties of the stellar populations in the SHARDS/CANDELS faint galaxy sample}

\label{sec:synthesizer}

After estimating the photometric redshift for each galaxy, we fit the stellar populations by comparing the SEDs with the predictions of \citet{Bruzual} for an SFH described with a delayed $\tau$-model and a \citet{Chabrier} IMF, adopting a \citet{Calzetti} dust attenuation law. We use the \texttt{Synthesizer} code (\citealt{PZETA}, \citealt{SYNTHESIZER}), which takes into account the nebular continuum and emission lines. It is worth noticing that the slope from the dust attenuation curve depends on mass, and secondarily on SFR, as pointed out by \citet{Salim}. Slopes tend to be steeper in low-mass galaxies and this effect is enhanced when galaxies are located away from the MS in both directions. In \citet{SHARDSFF1} they study the Lyman Alpha emission of a galaxy at $z=5.75$ with a most probable stellar mass of $\sim10^{6.5}$M$_\odot$ and SFR$\sim$1.0 M$_\sun$yr$^{-1}$. To model the effect of the extinction they propose a model with no extinction and a Calzetti law with total-to-selective attenuation ratio, R$_V$, of 4.05 and color excess, E(B-V), ranging between 0 and 0.25~mag. They find that both approaches are consistent with the observations within their uncertainties, thus highlighting the degeneracy that affects the model parameters. Additionally, in \citet{Guo_low_mass}, based on 164 galaxies at $0.4<z<1$ reaching down to 10$^{8.5}$M$_\odot$, they calculate the extinction correction factor using 4 different attenuation curves (Milky Way, Large Magellanic Cloud, Small Magellanic Cloud, and Calzetti) and find that the average attenuation curves are not significantly different from their original results, ignoring dust. Constraining the attenuation law with our data is out of the scope of this research. Assuming a Calzetti law is reasonable given that most of our galaxies are starburst systems, but it is important not to forget that the modeling of dust in low-mass systems is still a matter of study.

\texttt{Synthesizer} provides the following stellar population parameters: the timescale $\tau$, the age $t_0$, the metallicity $Z$, the attenuation A(V), the stellar mass, and the mass-weighted age of the galaxies. The first 4 are free parameters in the fits, the last 2 are derived from the former. The timescale $\tau$ parameter, which is related to the duration of the burst, is allowed to vary between 100~Myr and 1~Gyr. The age t$_0$, which is defined as the time that passes between $t=0$, the time when the SFH starts, and the moment at which the galaxy is observed, is set between 1~Myr and 14~Gyr, limited by the age of the Universe corresponding to the redshift of each galaxy. The metallicity, $Z$, is allowed to show discrete values of $0.02~Z_\odot$, $0.2~Z_\odot$, $0.4~Z_\odot$, and $Z_\odot$. The attenuation is limited to $\mathrm{A(V)}\leq2$~mag. Flux upper limits are also taken into account in the fits, similarly to what is done for the photometric redshift determination with \texttt{PZETA}. The SPS models are compared with the observed photometric data using a maximum likelihood estimator that takes into account the uncertainties in each data point. Examples of the SED fitting performed by \texttt{Synthesizer} are shown in Fig.~\ref{fig:SED} and in Fig.~\ref{fig:zpdf}. 

We use the best-fitting model to estimate the emission at different wavelengths, deriving, the $U$, $V$, and $J$ rest-frame colors, and the UV-continuum slope. For a given galaxy, an estimation of the UV-continuum slope ($\upbeta$) can be computed by linearly fitting the best-fitting model between 1268-2580~\text{\r{A}}\ rest-frame. See more details in \citet{Barro}. 

In this work, we use SFRs calculated from rest-frame UV data. As our sample is composed of faint blue star-forming galaxies, with only a minor fraction detected in the mid- or far-IR, we derive SFRs considering the luminosity at rest-frame wavelength 280~nm, using the \citet{Kennicutt} relation for a \citet{Chabrier} IMF, and accounting for attenuation based on UV slopes. Using the results presented in Figures 25 and 26 from \citet{Barro}, we can estimate the attenuation-corrected UV-based SFRs.

In Table \ref{tab:subsample_physical} we show the latter physical parameters for the subsample shown in Fig.~\ref{fig:stamps_transmission}.

In order to estimate uncertainties taking into account the degeneracies inherent to any SPS fitting, we use a Monte Carlo method, varying the photometry according to a Gaussian with a standard deviation compatible with the photometric errors. We also account for the uncertainty in the determination of the photometric redshift, varying its value according to the zPDF and repeating the SPS analysis. 

\begin{figure*}
  \centering
  \includegraphics[width=18cm, height=15cm]{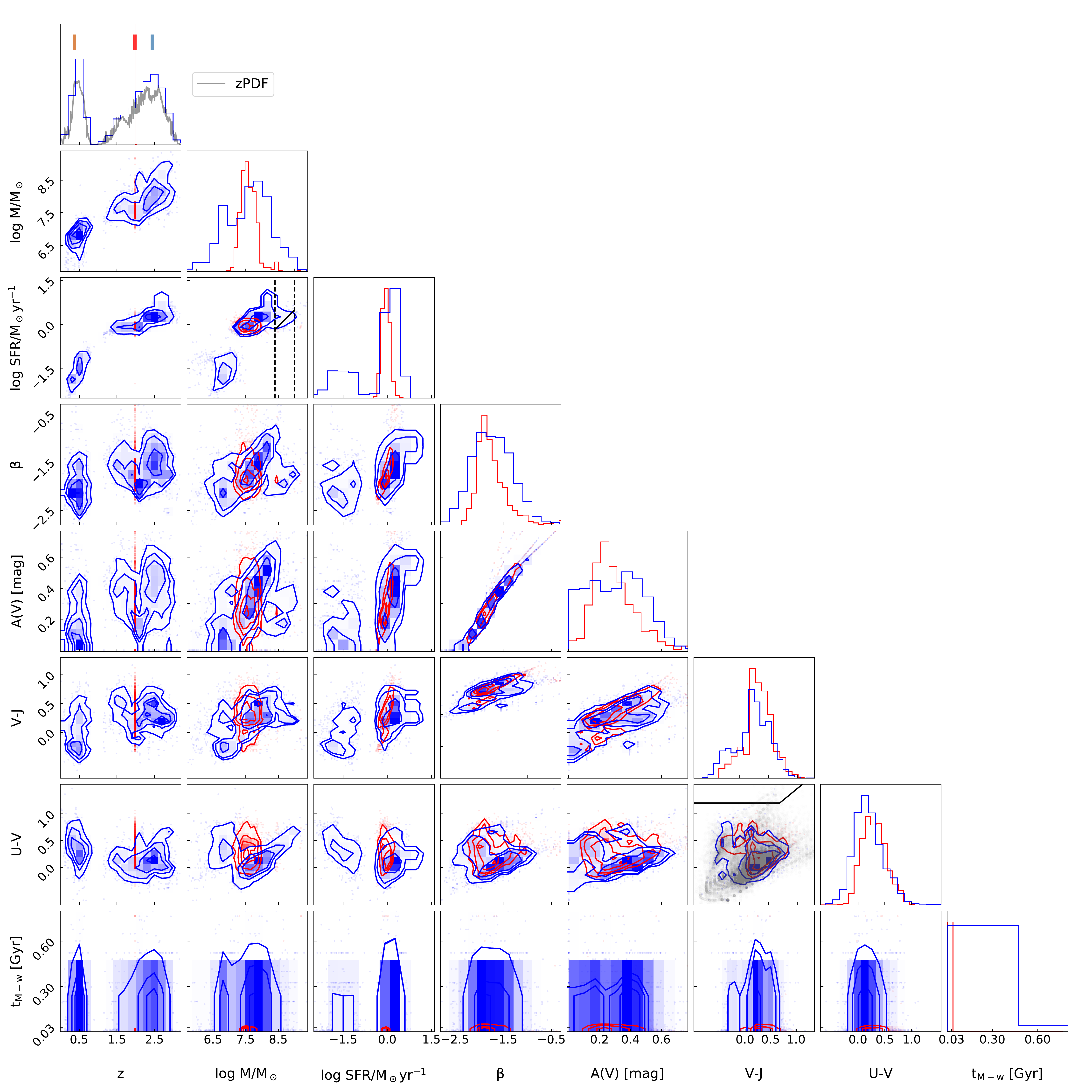}%
  \makebox[0pt][r]{
    \raisebox{29.2em}{%
      \includegraphics[width=8.7cm, height=5.7cm]{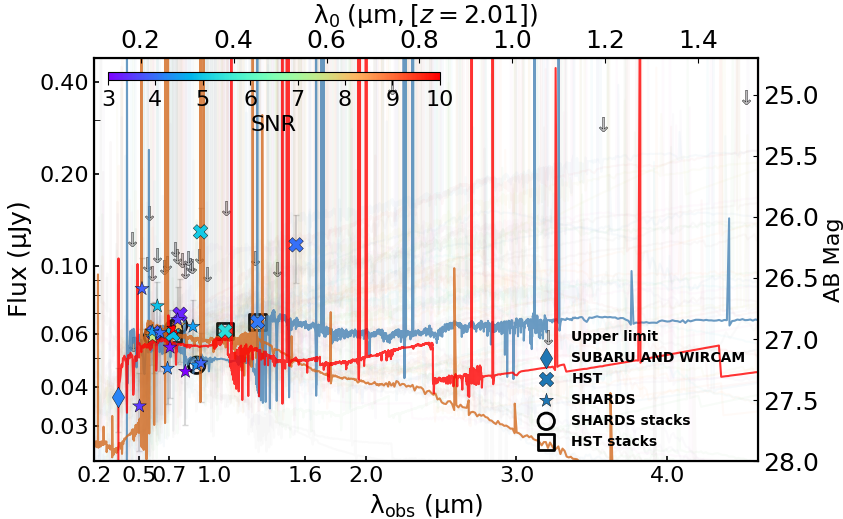}%
    }
  }%
  \makebox[0pt][r]{
    \raisebox{19em}{
      \includegraphics[width=5.8cm, height=2.5cm]{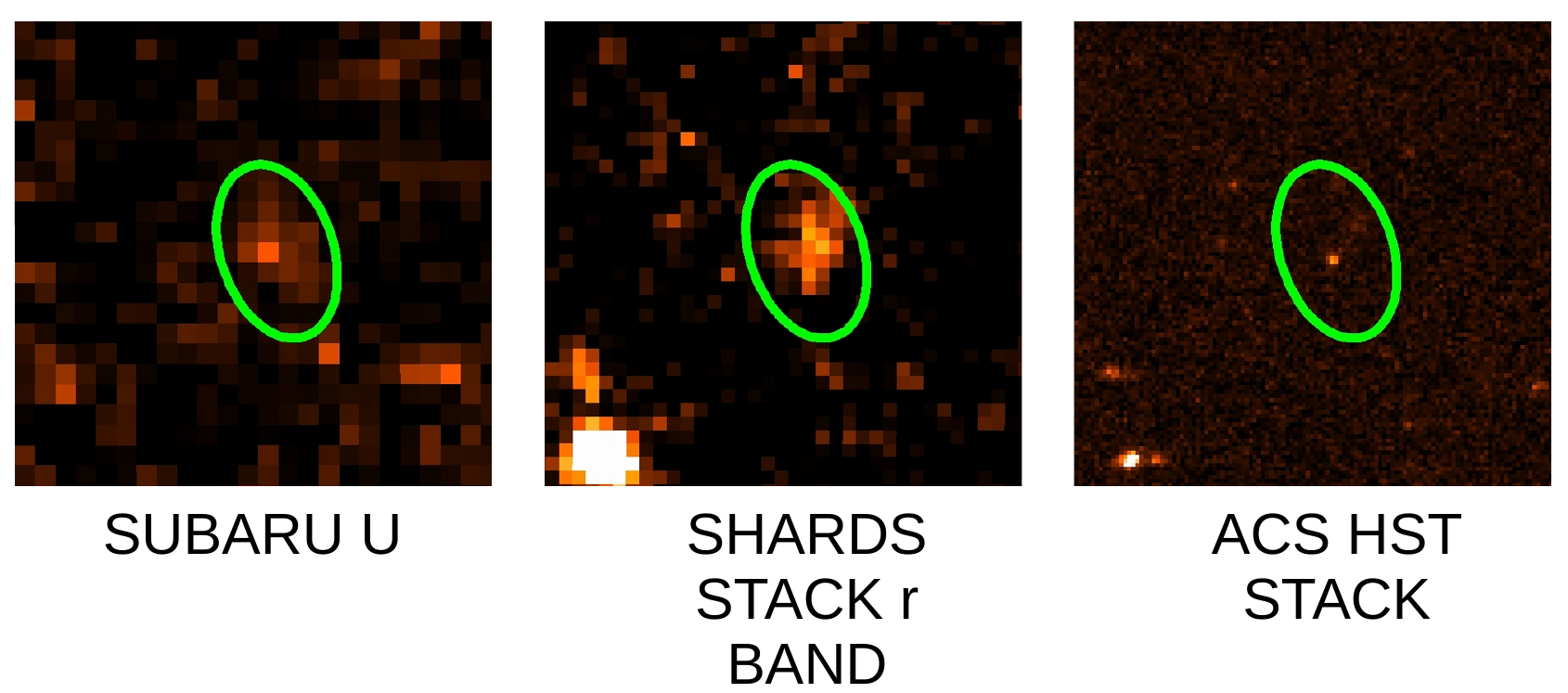}
    }}\hspace*{0em}
  \caption{Left panel: Corner plot  showing the results for the stellar population synthesis modeling Monte Carlo method applied to the galaxy shards$\_$faint$\_$p1$\_$043405. In blue, we show the results obtained by varying the photometric redshifts according to the zPDF and the photometry according to the photometric errors. In red, we depict results when the photometric redshift is fixed (to the most-probable value) and only the photometry is varied. The plots on the diagonal show the 1-D histogram for each parameter obtained by marginalizing over the other parameters. The off-diagonal plots show 2-D projections of the probability distributions for each pair of parameters. Contours are shown at 0.5, 1, 1.5, and 2$\sigma$. The panels show, from top to bottom, and left to right: photometric redshift, stellar mass M$_\star$, SFR, UV-slope $\upbeta$, attenuation A(V), rest-frame colors $V-J$ and $U-V$, and mass-weighted age t$_{\mathrm{M-w}}$. The A(V) extinction is the one derived from the UV-continuum slope. The first subplot in the first column, showing the histogram of the redshifts from all the iterations, includes the zPDF of the galaxy obtained by our photo-z code, overplotted in grey. The vertical segments, in brown, red, and blue, mark the redshifts of the different models that are represented in the figure on the right, showing the SED of the galaxy. The second subplot of the second column, showing log SFR vs. log M$_\star$, includes the Main Sequence (see Section \ref{sec:SFR}) in black. The vertical lines highlight the 50\% completeness level when considering the set of our galaxies and CANDELS-ALL (see Section \ref{sec:Stellar_mass}), and the CANDELS-ALL lower limit, respectively. The second subplot of the sixth column is a $UVJ$ diagram, where we include the quiescent galaxy wedge and the distribution of our sample of SHARDS/CANDELS faint objects in grey. Right panels: on the top, SED of the galaxy (see Fig.~\ref{fig:SED} for codes). Overplotted in brown, red, and blue, we show different stellar population models according to the Monte Carlo Method, whose redshift is signaled in the first subplot of the corner plot. The most-probable photo-z for this galaxy is $z=2.01$, its model is represented in red. Bottom: Cutouts of this object in the Subaru $U$-band, the SHARDS stack $r$-band, and the ACS HST stack.}
    \label{fig:Monte Carlo}
\end{figure*}

An example of the derivation of stellar population properties, including the error propagation and the scatter we obtain for each of those properties here considered can be found in Fig.~\ref{fig:Monte Carlo}. In this example, we show the error propagation obtained after varying the photometry and the photometric redshift, and only varying the photometry of the object. This galaxy is assigned a most-probable redshift $z=2.0$, but there are two more peaks in the zPDF, located at $z=0.4$ and $z=2.5$ respectively. The emission of this source in the $U$ band is crucial to determine the redshift, as noticed in the SED and the cutouts of the galaxy. 
The low-$z$ solution corresponds to a blue very low-mass galaxy with a low star formation rate. The other two solutions at higher redshifts correspond to low-mass galaxies with higher SFRs whose stellar mass is constrained within 0.4~dex. In terms of the log SFR, this translates to 0.2~dex. The low-$z$ solution covers 0.3~mag in the $U-V$ color and 0.4~mag in the $V-J$ color, whereas these numbers are 0.2~mag for both colors for the high-$z$ solutions. All iterations locate this galaxy in the star-forming region of the $UVJ$ diagram and point to low attenuations. The mass-weighted ages provide relatively young ages for all low- and high-$z$ solutions.

Based on this statistical analysis of the uncertainties and degeneracies, we conclude that stellar mass estimates of our whole sample are constrained within 0.32~dex. This difference translates to $\sim0.14$~dex for the SFR, 0.27~mag for the A(V) attenuation, 0.18 for the UV-continuum slope, 0.20~mag and 0.25~mag for $U-V$ and $V-J$, and 0.15 for the photometric redshift. Even though the stellar mass is traditionally better constrained than other parameters such as the UV SFR, our sources are blue and the SNR is in general higher in the blue part of the SEDs than in the red, more related to the stellar mass.

\subsection{Stellar masses of the SHARDS/CANDELS faint galaxy sample}
\label{sec:Stellar_mass}

The main goal of this work is to probe the behavior of the SFR vs. stellar mass relationship beyond the limits of precedent studies, mainly based on galaxies with M$_\star>10^9$M$_\odot$ (see Table \ref{tab:slopes} in Appendix~\ref{sec:app3}). It is thus necessary to define the mass range where our sample is complete. Given that the calculation of the MS will be performed with the set of our galaxies and CANDELS-ALL, the mass completeness has to take into account the whole sample (CANDELS-ALL + our sample). This calculation is performed following two approaches. The first one aims to relate the completeness in magnitude of our sample with the mass completeness. The completeness in magnitude is based on the ACS HST stack, since the major part of our sample is detected in HST. We slice the magnitude interval into 0.2 mag bins and introduce point-like sources at each bin in the stack. Afterward, we study the fraction of these objects that we can recover. As a result, we see that at 27.7~mag we reach the 80\% completeness level, coinciding with the peak of the histogram, and the 50\% at 28~mag. We try to translate this magnitude completeness into stellar mass for each redshift bin, but we see that these two quantities do not well correlate. Below the mass peak, where the completeness is expected to drop dramatically, we see a smooth decrease, which points to an overestimation of the mass completeness through this method. 

As a second approach, we relate the mass distribution of our sample + CANDELS-ALL with different stellar mass functions (SMF) (e.g. \citealt{SYNTHESIZER}, \citealt{Santini_SMF}, \citealt{Muzzin}, \citealt{Grazian}). The ratio between the SMF and our histogram allows us to define the stellar mass completeness. According to this method, the 50\% mass completeness is reached at 10$^{8.0}$M$_\odot$ at $z=1$ and $10^{8.5}$M$_\odot$ at $z=3$. According to the trajectories followed by the SMFs, our first method starts losing galaxies below $10^8$M$_\odot$, predicting 3 times fewer sources at $10^{7.5}$M$_\odot$ and $\sim$20 times fewer sources at $10^{6.5}$M$_\odot$. There is a population of faint galaxies, with ACS HST magnitudes below 28-29~mag, probably disk sources, with stellar masses between $10^6-10^8$ M$_\odot$, that we are not being able to select with our detection and makes the ACS HST magnitude a bad tracer for the stellar mass.

\begin{figure}
    \centering
    \includegraphics[width=8.5cm, height=7.3cm]{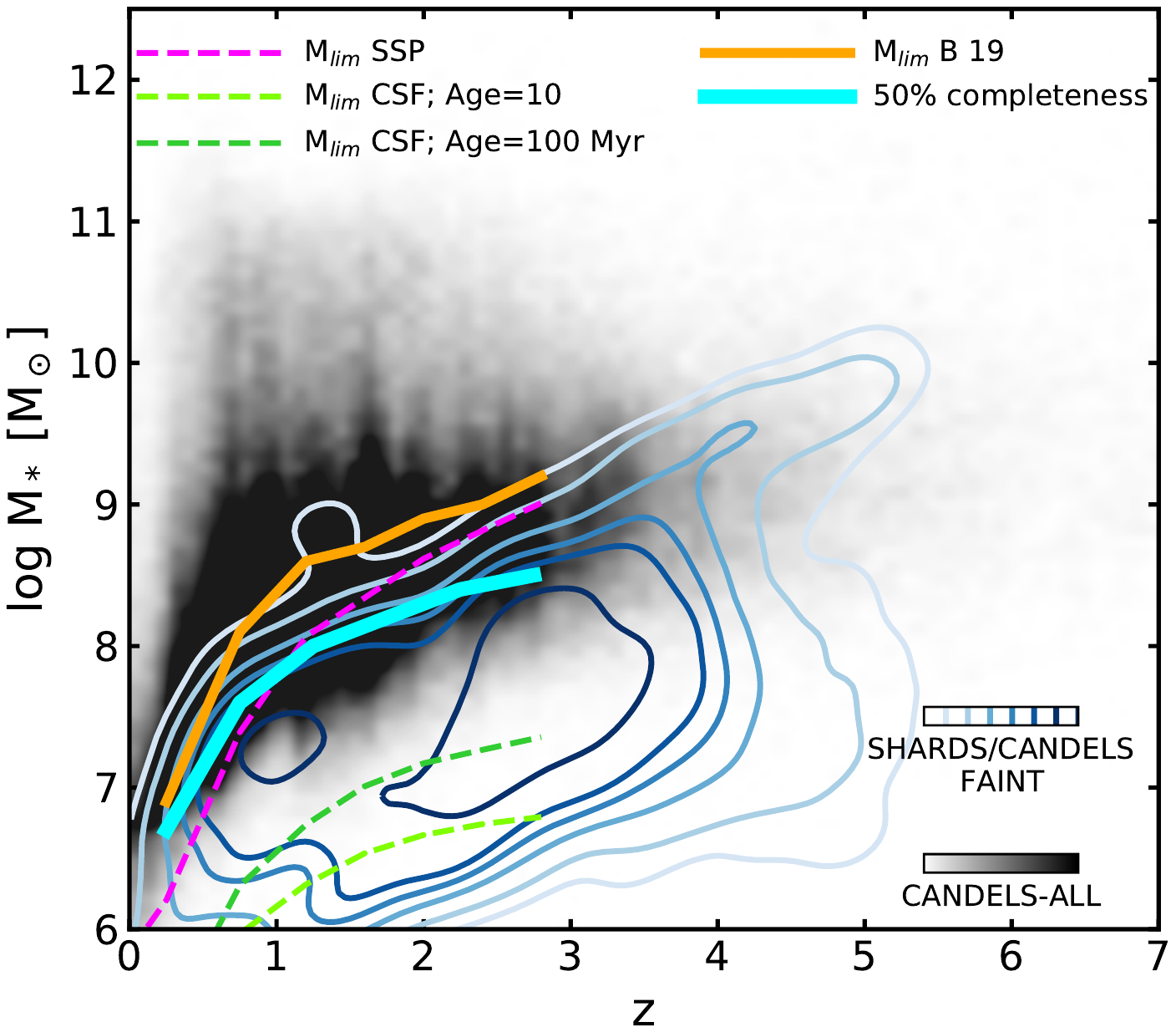}
    \includegraphics[width=8.5cm, height=7cm]{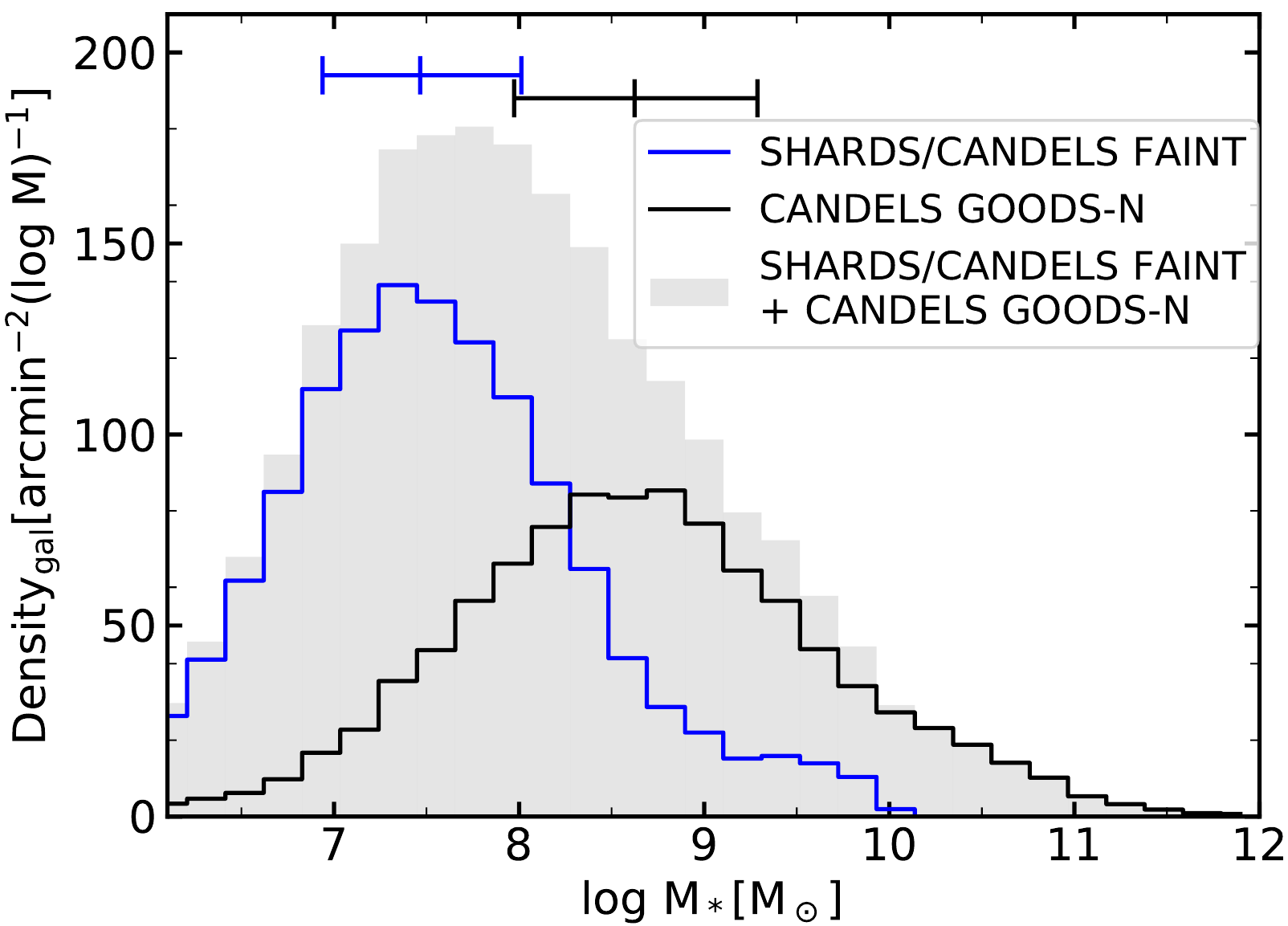}
    \caption{Top panel: stellar masses versus redshift of our sample shown as blue color contours compared to the CANDELS-ALL sample, in black. The density contours represent the 30th, 60th, 70th, 80th, 90th, and 100th percentiles. The solid cyan line shows the mass-representative limit, set at the 50\% mass completeness level. The mass limit derived using an SSP is shown in dashed magenta, as well as two CSF with ages of 10 and 100~Myr, respectively (dashed green lines). The mass completeness of B19 is represented as a solid orange line. Bottom panel: Mass histograms of our sample, not restricted in redshift, in blue, and B19, in black, obtained from the \texttt{Synthesizer} SPS fitting. Quartiles are shown as horizontal segments. The sum of the B19 sample and our galaxies is plotted in solid grey.}
    \label{fig:mass}
\end{figure}

In the top panel of Fig.~\ref{fig:mass}, we show the stellar mass of our sample versus the photometric redshift, compared to the CANDELS-ALL sample. We depict the new mass-representative limit, set at the 50\% completeness level. The B19 mass completeness limit is also represented. Including our sample makes these limits decrease around 0.6~dex in the redshift interval $1<z<3$.

Additionally, we derive the mass completeness limit for 2 different types of galaxies: one described by an instantaneous maximally-old star formation burst, and another characterized by a constant SFH, both unattenuated. For the former, we follow the procedure described in \citet{SYNTHESIZER}, which is based on assuming a passively evolving single stellar population (SSP) burst formed at z=$\infty$. The second calculation is based on a starburst with a constant star formation (CSF) and ages 10~Myr and 100~Myr. These are computed considering the typical fluxes of our sources in the optical.

\begin{figure*}
    \centering
    \includegraphics[width=18.5cm, height=11.8cm]{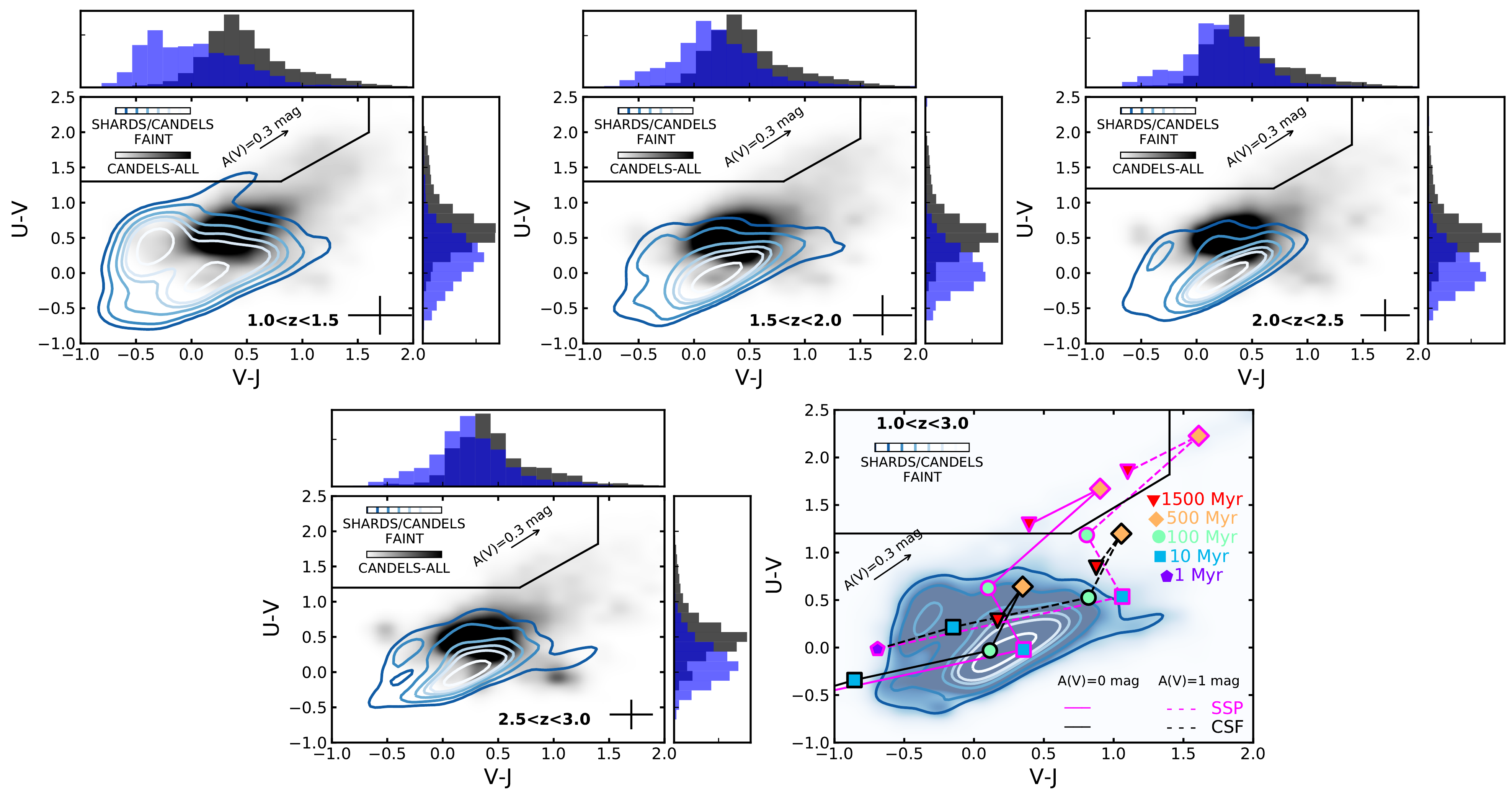}
    \caption{Rest-frame $U$-$V$ vs. $V$-$J$ color for our sample (in blue contours) and the CANDELS-ALL sample, in black. The density contours roughly represent the 20th, 40th, 50th, 60th, 80th, and 90th percentiles. The quiescent-galaxy wedge is taken from \citet{uvj}. Each panel represents a redshift interval, except for the last one, where our total sample in all bins is represented. For each bin, we also offer a median of the errors in the bottom right corner, as well as histograms for each color on the top and right of each panel. For the histograms, as well as for the final panel with the total samples, the same color code is followed: in blue we represent our sample whereas CANDELS-ALL is represented in black. The final panel depicts the tracks followed by two models, with and without dust attenuation, an SSP model in magenta and a CSF model in black, with ages from 1~Myr to 1.5~Gyr. The attenuated models (A(V)=1~mag) are shown as dashed lines and the models without attenuation are plotted as solid lines. The different markers are color-coded according to age. The arrow shows the typical attenuation for our sample, A(V)=0.30~mag.}
    \label{fig:UVJ}
\end{figure*}

CSF models are a good representation of the typical properties of the majority of the galaxies in our sample (see subsection \ref{sect:UVJ}). Our galaxies mainly occupy the low-mass regime between 10$^6$-10$^9$ M$_\odot$, which remained underpopulated by previous catalogs. Around the mass-representative limit, SSP models better reproduce the trend of our sources.

In the bottom panel of Fig.~\ref{fig:mass}, histograms of the stellar mass for our sample and B19 are shown. The median masses and quartiles of the samples are log M=8.5$_{8.0}^{9.1}\;$M$_\odot$ for B19 and log M=7.5$_{6.9}^{8.1}\;$M$_\odot$ for ours.

\begin{deluxetable*}{c|c|c|c|c|c|c|c|c|c|c|c|c}
\setlength{\tabcolsep}{1.2pt} 
\label{tab:subsample_physical}
\tabletypesize{\scriptsize}
\caption{Physical properties of the galaxies shown in Fig.\ref{fig:cutouts}. The A$_\upbeta$(V) extinction comes from the UV-continuum slope whereas A(V) is derived from the SED full spectral fitting.}
\tablehead{\colhead{Id.} & \colhead{z}&\colhead{log M[M$_\odot$]} & \colhead{SFR [M$_\odot yr^{-1}$]} & \colhead{$U-V$ [mag]}& \colhead{$V-J$ [mag]}& \colhead{$\upbeta$} &\colhead{A$_\upbeta$(V) [mag]}&\colhead{A(V) [mag]}&\colhead{t$_0$ [Gyr]}& \colhead{$\uptau$ [Myr]}& \colhead{t$_{M-w}$ [Gyr]}&\colhead{Z/Z$_\odot$}}
\startdata
shards$\_$faint$\_$p1$\_$112589&1.0&7.6$_{7.5}^{7.7}$&0.31$_{0.27}^{0.35}$&0.44$_{0.37}^{0.51}$&0.65$_{0.45}^{0.85}$&$-$1.4$_{-1.3}^{-1.5}$&0.47$_{0.00}^{1.17}$&0.90$_{0.77}^{1.03}$&0.18$_{0.14}^{0.22}$&100$_{74}^{130}$&0.080$_{0.059}^{0.101}$&1.0$_{0.4}^{1.0}$\\
hst$\_$faint$\_$p1$\_$28017&1.4&7.6$_{7.5}^{7.6}$&0.078$_{0.068}^{0.088}$&0.45$_{0.51}^{0.38}$&$-0.39_{-0.46}^{-0.31}$&$-$2.3$_{-2.3}^{-2.2}$&0.00$_{0.00}^{0.28}$&0.00$_{0.00}^{0.03}$&0.51$_{0.45}^{0.56}$&130$_{110}^{140}$&0.30$_{0.24}^{0.35}$&0.020$_{0.020}^{0.200}$\\
hst$\_$faint$\_$p1$\_$29334&1.6&7.8$_{7.7}^{7.8}$&0.084$_{0.074}^{0.094}$&0.66$_{0.55}^{0.77}$&$-0.32_{-0.55}^{-0.09}$&$-$1.9$_{-2.3}^{-1.5}$&0.22$_{0.00}^{0.50}$&0.00$_{0.00}^{0.03}$&0.72$_{0.64}^{0.80}$&130$_{110}^{140}$&0.48$_{0.41}^{0.55}$&0.020$_{0.020}^{0.200}$\\
\enddata
\end{deluxetable*}

\subsection{Rest-frame colors}
\label{sect:UVJ}

In  this subsection, we discuss the properties of our galaxies in terms of the rest-frame colors. We analyze the locus occupied by our sample of faint low-mass galaxies at $1<z<3$ in the $UVJ$ diagram, comparing them with the CANDELS-ALL sample. For this purpose, we divide the sample into 4 redshift bins, which roughly contain the same number of galaxies. The ranges are wide enough to remove any effect linked to redshift uncertainties up to $\Delta$(z)/(1+z)=0.25, which is $>8$ times the typical uncertainty of photo-z's for the CANDELS fields, and $>50$ times worse than the redshifts for the GOODS-N region obtained with SHARDS data \citep[see][]{Barro}. We acknowledge that our galaxies are much fainter than those samples and worse-quality photometric redshifts are expected for them, but no fair testing can be performed for them to date, probably until the James Webb Space Telescope (JWST) spectroscopic campaigns arrive.

In Fig.~\ref{fig:UVJ}, we compare the position of our galaxies with the CANDELS-ALL sample in this color-color diagram. It allows discriminating between star-forming, quiescent, and dusty star-forming galaxies. Whereas the CANDELS-ALL sample occupies the three regions of the diagram (especially the star-forming region), with galaxies in the quiescent quadrant in all redshift bins, our sample only shows some of these quiescent galaxies at 1$<z<$1.5 ($\sim$0.8\% of the whole sample in this interval). The median fraction of quiescent galaxies per redshift bin in the CANDELS-ALL sample is 2\%, with the maximum fraction of quiescent galaxies at 1$<z<$1.5 (4\%). Our sample overlaps with the star-forming galaxies of the CANDELS-ALL sample but is offsetted to bluer colors for both axes, $U-V$ and $V-J$, in all redshift bins, as illustrated by the histograms. This is also observed in Fig.~\ref{fig:COLOR_COLOR}, where instead we use the apparent magnitudes of the sample. The distributions of colors of the comparison dataset and our sample follow the direction of the extinction vector. 

The CANDELS-ALL sample presents a median $U-V\sim0.55_{0.38}^{0.76}$~mag whereas our sample always peaks below that value, differently according to the redshift bin, ranging from 0.22$_{-0.05}^{0.46}$~mag at $1<z<1.5$ to 0.11$_{-0.07}^{0.31}$~mag at $2.5<z<2$.
In terms of the $V-J$ color, our galaxies are also bluer than those from the CANDELS-ALL sample, but the difference between the distributions is smaller compared to what is found for the $U-V$ color, except for the first redshift bin. Most of the IRAC measurements are upper limits, so the $V-J$ colors should also be regarded as upper limits. IRAC detections account for 7\% of the sample, mainly at 3.6 and 4.5 $\upmu$m, with less than 1\% of galaxies being detected in the 4 IRAC bands. The SED fits count, however, with measured fluxes in the $V$-band spectral range. All this translates to poorly constrained $V-J$ colors with an increasingly lower probability as we move to redder values. The CANDELS-ALL sample shows a median $V-J$ color and quartiles of $\sim$0.42$_{0.24}^{0.71}$~mag, whereas our sample displays a median color of 0.23$_{0.01}^{0.46}$~mag for $z>2$ and $-0.014_{-0.324}^{0.318}$~mag at $1.0<z<1.5$.

To better understand the nature of these blue star-forming galaxies, the last panel (lower-right) of Fig.~\ref{fig:UVJ} includes some stellar population tracks. In particular, we show an SSP and a CSF model for different ages, with no attenuation and an attenuation of 1~mag. 
We see that a CSF model, and an SSP with an age younger than 500~Myr, can reproduce the trend followed by our galaxies. 
The attenuated models can also reproduce the trend when considering ages under 100~Myr for the CSF, and 10~Myr for the SSP. 

There is, however, one region of the diagram that cannot be reproduced by these models by changing the attenuation, located at $V-J>1$~mag and $U-V<0.5$~mag. This region is mainly occupied by galaxies with fewer detections (detected in less than 10 SHARDS and/or HST individual bands) and higher uncertainties in the rest-frame colors. It is mainly populated by the galaxies with the highest attenuation in our sample, with a median A(V)$\sim0.63$~mag. They are young ($\sim$1-3~Myr) galaxies, with a typical timescale, $\uptau$, of 130~Myr, and low metallicity. Therefore, they correspond to constant star formation galaxies which present strong emission lines and also a very bright nebular continuum in the optical and NIR, which is the main reason explaining the distinct colors.

\begin{figure*}
    \centering
    \includegraphics[width=18cm, height=4.2cm]{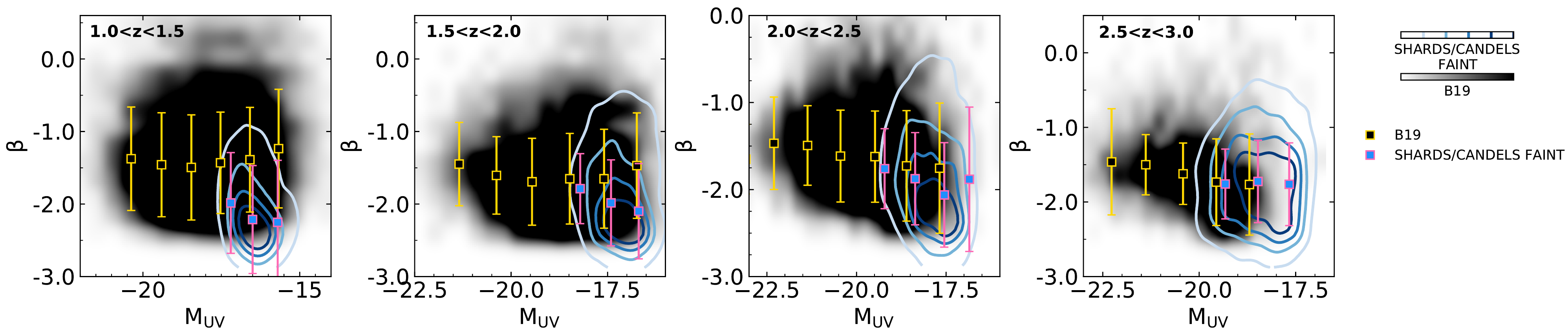}
    \caption{UV-slope $\upbeta$ vs. UV absolute magnitude M$_{\mathrm{UV}}$ for the B19 sample, in black, and our sample, shown as blue contours. The density contours roughly represent the 30th, 50th, 70th, and 90th percentiles. Running medians for the two samples in 1~mag bins are plotted with squares. The scatter associated with the running medians is also shown.}
    \label{fig:beta}
\end{figure*}

From these $UVJ$ diagrams, we conclude that our sample is composed of relatively blue star-forming (young) galaxies with low attenuations. In section \ref{sect:stellar_prop}, we will present a more detailed discussion of the stellar population ages of our sample.

\subsection{UV slope}

We estimate the UV spectral slope $\upbeta$ and UV absolute magnitude M$_{\mathrm{UV}}$ from the SEDs and the best-fitting model for all galaxies in our sample and those in B19. M$_{\mathrm{UV}}$ is obtained from the mean flux density in a spectral window of 100~nm around a rest-frame wavelength of 150~nm. In Fig.~\ref{fig:beta}, we show the $\upbeta$-M$_{\mathrm{UV}}$ relation for B19 and our sample in the same 4 redshift bins used in Fig.~\ref{fig:UVJ}. The running medians of both samples (B19 in black and our sample in blue) show a nearly flat distribution of the UV-continuum slope with respect to the M$_{\mathrm{UV}}$, especially for the B19 sample. For both samples and at all redshifts, the UV-continuum slope does not change much with luminosity, with median values ranging from $-1.5$ to $-1.7$ for the B19 sample, and bluer values ranging from $-1.8$ to $-2.0$ for our sample. Comparing both datasets, the median $\upbeta$ values of our sample tend to be bluer than the ones from B19, with a median difference of 0.33. 

In terms of the M$_{\mathrm{UV}}$, our sample shows values compatible with the faintest B19 galaxies and spans to even fainter values beyond $z>2.5$.

Deriving the attenuation from the UV-continuum slope \citep{Meurer}, using the calibration dependent on SFRs and stellar mass found in B19, we see that the B19 sample shows a median attenuation of A(V)=0.58$_{0.25}^{0.96}$~mag (0.66$_{0.32}^{1.10}$~mag at $1.0<z<1.5$ and 0.46$_{0.25}^{0.71}$~mag at $2.5<z<3$), whereas our sample shows an attenuation of 0.25$_{0.00}^{0.79}$~mag, 0.22$_{0.00}^{0.66}$~mag and 0.25$_{0.00}^{0.64}$~mag in the first redshift bins, and a median of 0.35$_{0.01}^{0.71}$~mag in the highest redshift one. The median A(V) extinction of the total sample is 0.28$_{0.00}^{0.68}$~mag.

\begin{figure*}
    \centering
    \includegraphics[width=17.5cm, height=11cm]{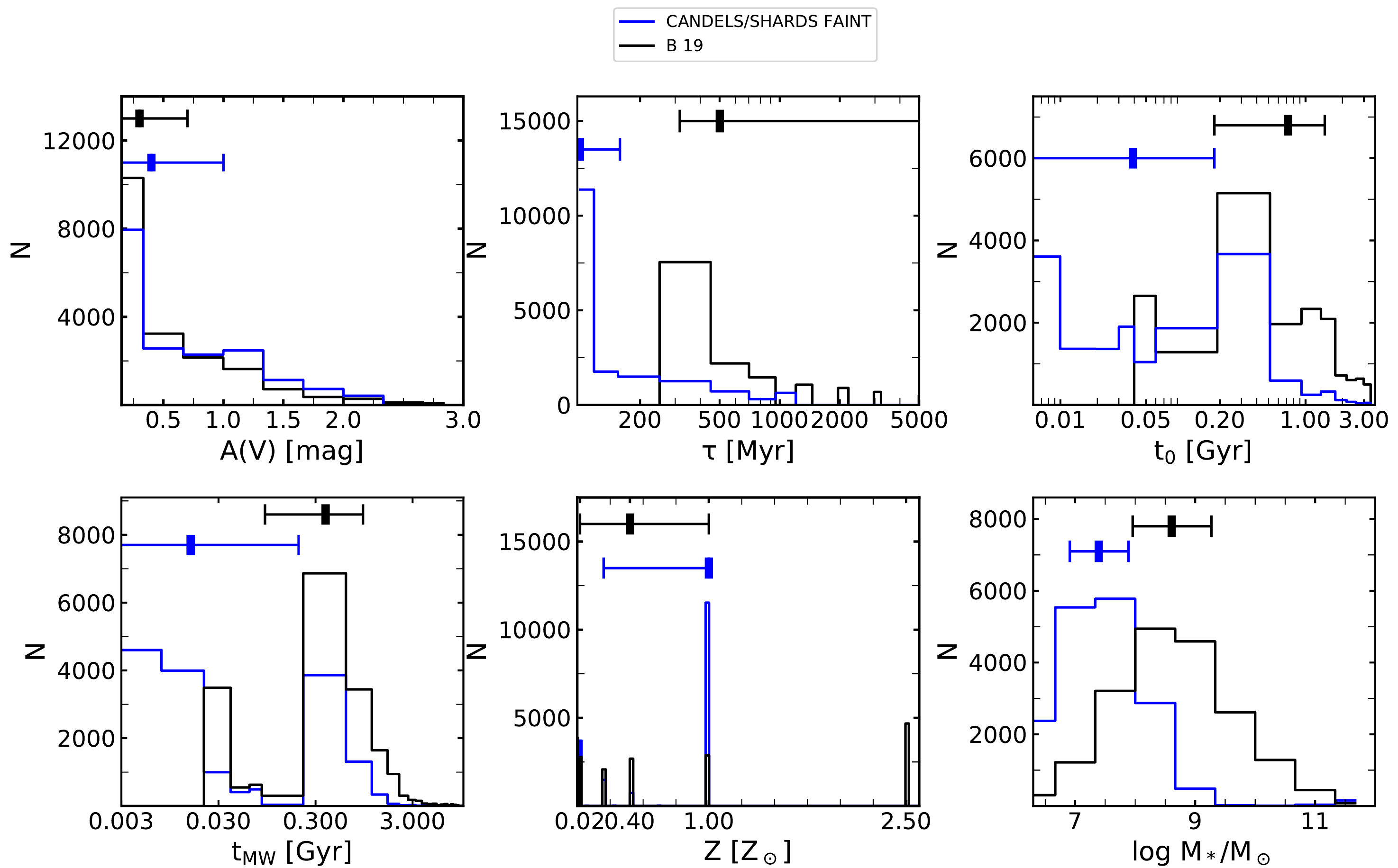}
    \caption{Histograms showing the distribution of the stellar population properties derived by \texttt{Synthesizer} for our sample, in blue, and the catalog in B19, in black. Quartiles are represented as horizontal segments, where the median is shown with a thicker vertical line. Starting at the first panel and rightwards we show the attenuation (A(V)), the timescale ($\uptau$), the age (t$_0$), the mass-weighted age (t$_{\mathrm{MW}}$), the metallicity (Z), and the stellar mass.}
    \label{fig:synthesizer}
\end{figure*}

\begin{deluxetable*}{ccccccccccccc}
\setlength{\tabcolsep}{0.1pt} 
\label{tab:summary}
\tabletypesize{\scriptsize}
\caption{Median and quartiles of the main stellar population properties of our sample (SHARDS/CANDELS FAINT), compared to CANDELS-ALL or B19 (CANDELS bright in the table), for each redshift bin and for the whole sample in the whole redshift interval here considered (last row). The A$_\upbeta$(V) extinction comes from the UV-continuum slope whereas A(V) is derived from the SED full spectral fitting.}
\tablehead{\colhead{z} & \colhead{Sample}&\colhead{log M$_\star$/M$_\odot$}&\colhead{SFR [M$_\odot$ yr$^{-1}$]} &\colhead{$U-V$}&\colhead{$V-J$}& \colhead{$\upbeta$} & \colhead{A$_\upbeta$(V) [mag]}& \colhead{A(V) [mag]}&\colhead{t$_0$ [Gyr]} & \colhead{$\uptau$ [Myr]} & \colhead{t$_{\mathrm{M-w}}$ [Gyr]} & \colhead{$Z/Z_\odot$}}
\startdata 
&SHARDS/CANDELS FAINT&7.2$_{6.7}^{7.7}$&0.13$_{0.07}^{0.29}$&0.22$_{-0.05}^{0.46}$&$-0.014_{-0.324}^{0.318}$&$-$1.9$_{-2.3}^{-1.3}$&0.25$_{0.00}^{0.79}$&0.10$_{0.00}^{0.60}$&0.18$_{0.04}^{0.51}$&100$_{100}^{130}$&0.014$_{0.002}^{0.201}$&0.40$_{0.02}^{1.00}$\\
$1.0<z<1.5$&&&&&&&\\
&CANDELS BRIGHT&8.4$_{8.0}^{9.0}$&0.66$_{0.28}^{2.34}$&0.61$_{0.43}^{0.82}$&0.44$_{0.26}^{0.72}$&$-1.5_{-1.9}^{-1.0}$&0.66$_{0.32}^{1.10}$&0.30$_{0.00}^{0.80}$&0.72$_{0.18}^{1.43}$&500$_{320}^{5000}$&0.41$_{0.09}^{1.02}$&0.40$_{0.02}^{1.00}$\\
&&&&&&&&\\
\hline
&SHARDS/CANDELS FAINT&7.1$_{6.7}^{7.6}$&0.22$_{0.12}^{0.47}$&0.060$_{-0.137}^{0.289}$&0.17$_{-0.04}^{0.43}$&$-2.0_{-2.3}^{-1.4}$&0.22$_{0.00}^{0.66}$&0.40$_{0.00}^{1.00}$&0.036$_{0.008}^{0.090}$&100$_{100}^{160}$&0.016$_{0.002}^{0.201}$&1.00$_{0.40}^{1.00}$\\
$1.5<z<2.0$&&&&&&&\\
&CANDELS BRIGHT&8.6$_{8.2}^{9.1}$&1.11$_{0.49}^{4.10}$&0.54$_{0.37}^{0.73}$&0.43$_{0.25}^{0.70}$&$-1.6_{-2.0}^{-1.2}$&0.48$_{0.17}^{0.81}$&0.30$_{0.00}^{0.80}$&0.72$_{0.18}^{1.43}$&790$_{320}^{7900}$&0.36$_{0.07}^{0.88}$&0.40$_{0.02}^{2.50}$ \\
&&&&&&&&\\
\hline
&SHARDS/CANDELS FAINT&7.3$_{6.9}^{7.8}$&0.33$_{0.19}^{0.73}$&0.038$_{-0.159}^{0.270}$&0.22$_{0.03}^{0.45}$&$-1.9_{-2.3}^{-1.4}$&0.25$_{0.00}^{0.64}$&0.50$_{0.10}^{0.90}$&0.036$_{0.010}^{0.053}$&100$_{100}^{130}$&0.016$_{0.002}^{0.201}$&1.00$_{0.40}^{1.00}$\\
$2.0<z<2.5$&&&&&&&\\
&CANDELS BRIGHT&8.8$_{8.4}^{9.3}$&2.14$_{0.89}^{8.07}$&0.53$_{0.37}^{0.73}$&0.40$_{0.22}^{0.69}$&$-1.6_{-2.0}^{-1.3}$&0.48$_{0.21}^{0.77}$&0.30$_{0.00}^{0.70}$&0.72$_{0.18}^{1.43}$&500$_{320}^{5000}$&0.36$_{0.07}^{1.17}$&0.40$_{0.02}^{2.50}$ \\
&&&&&&&&\\
\hline
&SHARDS/CANDELS FAINT&7.6$_{7.2}^{8.1}$&0.59$_{0.31}^{1.16}$&0.11$_{-0.07}^{0.31}$&0.23$_{0.04}^{0.44}$&$-1.8_{-2.2}^{-1.3}$&0.35$_{0.01}^{0.71}$&0.80$_{0.20}^{1.20}$&0.016$_{0.004}^{0.128}$&100$_{100}^{160}$&0.014$_{0.002}^{0.201}$&1.00$_{0.20}^{1.00}$\\
$2.5<z<3.0$&&&&&&&\\
&CANDELS BRIGHT&8.9$_{8.5}^{9.3}$&4.40$_{1.72}^{14.44}$&0.49$_{0.33}^{0.69}$&0.42$_{0.24}^{0.71}$&$-1.7_{-1.9}^{-1.3}$&0.46$_{0.25}^{0.71}$&0.30$_{0.00}^{0.70}$&0.72$_{0.18}^{1.43}$&500$_{320}^{5000}$ &0.36$_{0.07}^{1.17}$&0.40$_{0.02}^{2.50}$\\
&&&&&&&&\\
\hline
&SHARDS/CANDELS FAINT&7.4$_{6.9}^{7.9}$&0.31$_{0.14}^{0.67}$&0.10$_{-0.11}^{0.33}$&0.17$_{-0.06}^{0.41}$&$-1.9_{-2.3}^{-1.4}$&0.28$_{0.00}^{0.68}$&0.40$_{0.00}^{1.00}$&0.039$_{0.006}^{0.180}$&100$_{100}^{160}$&0.014$_{0.002}^{0.180}$&1.00$_{0.02}^{1.00}$\\
&&&&&&&&\\
&CANDELS BRIGHT&8.6$_{8.0}^{9.2}$&1.34$_{0.50}^{5.35}$&0.55$_{0.38}^{0.76}$&0.42$_{0.24}^{0.71}$&$-1.6_{-2.0}^{-1.2}$&0.58$_{0.25}^{0.96}$&0.30$_{0.00}^{0.70}$&0.72$_{0.18}^{1.43}$&500$_{320}^{5000}$ &0.38$_{0.09}^{0.93}$&0.40$_{0.02}^{1.00}$\\
\enddata
\end{deluxetable*}

\subsection{Ages, metallicities, and attenuations from full SED modeling}
\label{sect:stellar_prop}

Figure \ref{fig:synthesizer} shows the distribution of the different stellar population properties derived with \texttt{Synthesizer}, compared to those for the B19 sample. In \citet{Barro}, \texttt{Synthesizer} was used restricting the A(V) attenuation to the interval 0-4~mag, $Z$ was allowed to show discrete values of 0.005$Z_\odot$, 0.02$Z_\odot$, 0.2$Z_\odot$, 0.4$Z_\odot$, $Z_\odot$, and 2.5$Z_\odot$, $\uptau$ was restricted to 300~Myr-10~Gyr and the t$_0$ to 0.04~Gyr-13~Gyr.

Our galaxies are younger than those in the B19 sample, with median age and quartiles t$_0$=0.039$_{0.006}^{0.180}$~Gyr, compared to 0.72$_{0.18}^{1.43}$~Gyr for the B19 sample, and shorter star formation histories, with a median timescale and quartiles $\uptau$=100$_{100}^{160}$~Myr, compared to the 500$_{320}^{5000}$~Myr in the case of the B19 galaxies. In terms of the mass-weighted age, our sample shows a median t$_{\mathrm{M-w}}$ and quartiles of 0.014$_{0.002}^{0.180}$~Gyr, whereas this value increases to 0.38$_{0.09}^{0.93}$~Gyr for B19 sources. The metallicity is not well constrained by broad-band SEDs, so we cannot infer any difference between the samples.

In terms of the attenuation, the results are consistent with what is found using the UV-continuum slope. We remark that \texttt{Synthesizer} computes the extinction using the whole SED and not only the emission from the UV, so the 2 estimations of the attenuation are independent to some extent. Our galaxies are more concentrated in the lowest A(V) bin than the B19 objects. The median attenuation of our sample (0.40$_{0.00}^{1.00}$~mag) is, however, slightly higher according to \texttt{Synthesizer} than that of B19 (0.30$_{0.00}^{0.70}$~mag). 

After constraining our sample to the interval $1<z<3$, the median mass and quartiles of our galaxies are found at 7.4$_{6.9}^{7.9}$M$_\odot$. In the case of the B19 sample, these numbers are 8.6$_{8.0}^{9.2}$M$_\odot$.

Summarizing the information in Fig.~\ref{fig:synthesizer}, the typical galaxy in our sample is a low-mass star-forming galaxy with roughly constant star formation and mass-weighted age around 15~Myr, presenting an attenuation around A(V)=0.30~mag.

In Table \ref{tab:summary} we summarize all the information about stellar populations, described in Section~\ref{sec:synthesizer}, with the median and quartiles of the main properties considered in this work, compared with CANDELS-ALL/B19.

\section{The SFR-\texorpdfstring{M$_{\star}\:${r}} relation at low masses}
\label{sec:SFR}

\begin{figure*}
    \centering
    \includegraphics[width=18.5cm, height=10.4cm]{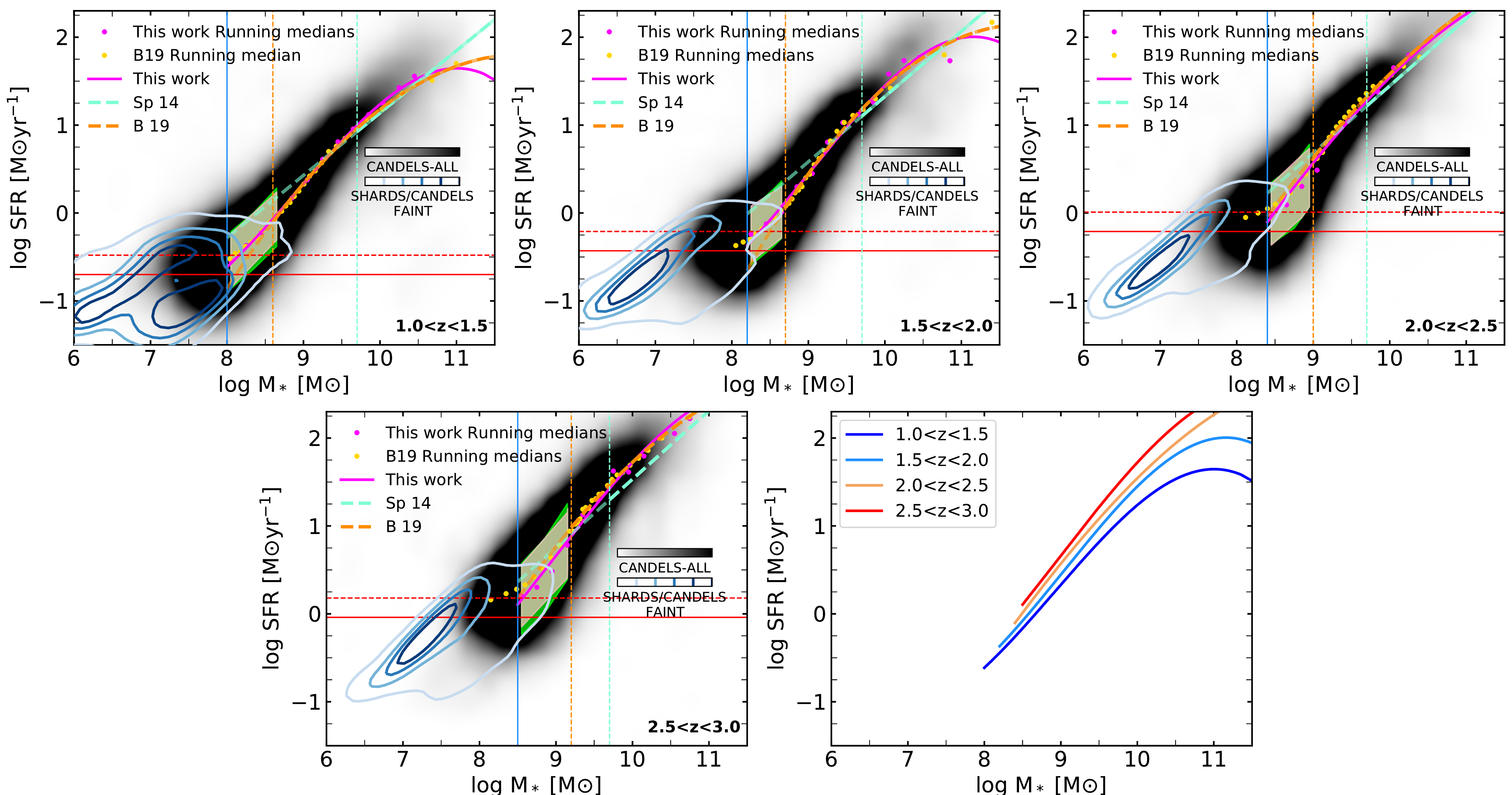}
    \caption{Main sequences, log SFR vs. log M$_{\star}$, for CANDELS-ALL (in black) and our sample (blue contours) in four redshift bins. The density contours roughly represent the 20th, 40th, 60th, and 90th percentiles. Vertical lines show the mass completeness limits for the CANDELS-ALL sample according to B19 and \citet{Speagle} (dashed orange and cyan lines, respectively), and the mass-representative limit that we have calculated including our sample (thick blue lines). The red horizontal lines show the SFR completeness limits: the dashed line corresponds to the limit calculated with A(V)=0.30~mag and the solid line corresponds to the limit calculated with no attenuation. The values of the running medians are shown as magenta points. They are fitted using a third-order polynomial. The running medians considered in \citet{Barro} are also shown as yellow points. The MS from B19 and \citet{Speagle} are represented in orange and cyan respectively, that turn lighter below their mass limits. The pink-shaded region around the fits denotes the statistical scatter around the running medians. The green lime-shaded region depicts the scatter/uncertainty linked to the typical observational errors, derived using a Monte Carlo method. The lower-right panel shows all our fits to the MS for the different redshift bins. }
    \label{fig:SFR}
\end{figure*}

Figure \ref{fig:SFR} shows the SFR-M$_{\star}$ relation, the so-called main sequence (MS), including  galaxies from previous catalogs and galaxies selected in this work, that sample the low-mass end.

\begin{deluxetable*}{c|cccc}
\label{tab:running_medians}
\centering                               
\caption{Running medians and scatter of the main sequence at low stellar masses presented in this work.}
\tablehead{\colhead{}&\colhead{1.0$<z<$1.5}&\colhead{1.5$<z<$2.0}&\colhead{2.0$<z<$2.5}&\colhead{2.5$<z<$3.0}}
\startdata
log M$_{\star}$[M$_\odot$] &  & & log SFR[M$_\odot$yr$^{-1}$]& \\    
\hline
    8.0 & $-$0.54 $\pm$ 0.31 &      &  &  \\
    8.2 & $-$0.37 $\pm$ 0.31 & $-$0.24 $\pm$ 0.32     &   & \\
    8.4 & $-$0.22 $\pm$ 0.30 & $-$0.12 $\pm$ 0.35     & $-$0.02 $\pm$ 0.35 & \\
    8.6 & $-$0.03 $\pm$ 0.27 & 0.01 $\pm$ 0.29      & 0.09 $\pm$ 0.35 &0.20 $\pm$ 0.35  \\
    8.8 & 0.17 $\pm$ 0.30 & 0.30 $\pm$ 0.33     & 0.30 $\pm$ 0.36 &0.30 $\pm$ 0.34  \\
    9.0 & 0.38 $\pm$ 0.31& 0.45 $\pm$ 0.32    & 0.49 $\pm$ 0.32 & 0.48 $\pm$ 0.33 \\
    9.2 & 0.61 $\pm$ 0.33 & 0.81 $\pm$ 0.31     & 0.88 $\pm$ 0.34 &0.78 $\pm$ 0.37  \\
    9.4 & 0.81 $\pm$ 0.32 & 1.03 $\pm$ 0.34      & 1.05 $\pm$ 0.33 & 1.13 $\pm$ 0.36\\
    9.6 & 0.96 $\pm$ 0.35&  1.12 $\pm$ 0.38    & 1.26 $\pm$ 0.34 &1.29 $\pm$ 0.36 \\
    9.8 & 1.12 $\pm$ 0.34 &  1.26 $\pm$ 0.41    & 1.37 $\pm$ 0.39& 1.63 $\pm$ 0.40 \\
    10.0 & 1.24 $\pm$ 0.31 &  1.58 $\pm$ 0.31    & 1.65 $\pm$ 0.36 & 1.61 $\pm$ 0.42 \\
    10.2 &  &     & 1.80 $\pm$ 0.29 & 1.79 $\pm$ 0.37\\
    10.4 & & & 1.90 $\pm$ 0.35 & 2.00 $\pm$ 0.39\\
\enddata
\end{deluxetable*}

We calculate the median value of the SFR (in log scale) and the corresponding standard deviation in bins of log~M/M$_\odot=0.2$ for the different redshift intervals defined in the previous sections. These values are listed in Table~\ref{tab:running_medians}. To combine our SHARDS/CANDELS faint sample, which only covers one field, with the CANDELS-ALL sample, we weigh the calculations with the relative area covered by our survey in GOODS-N with respect to the 5 CANDELS fields. Quantitatively, our GOODS-N work is restricted to 13\% of the total CANDELS area. 

We fit the running median values above the stellar mass limits for our SHARDS/CANDELS faint sample using a third-degree polynomial. This order is chosen to get a coefficient of determination, $R^2$, of 0.99 in all redshift bins. The coefficients defining these polynomials are shown in Table~\ref{tab:ms}. In Fig.~\ref{fig:SFR}, we offer two estimations of the MS scatter: one derived from the standard deviation in the calculation of the running medians and another using the Monte Carlo method based on the typical errors of the photometry and the photometric redshifts (see section \ref{sec:synthesizer}), which accounts for uncertainties jointly with possible real scatter. 

At the massive end of the MS (above the dashed orange vertical lines shown in the plots), the CANDELS-ALL sample dominates the number density of galaxies. Starting at that limit and for smaller masses, our sample is more representative of the galaxy population and provides the means to estimate the behavior of the MS in a regime that has not been studied before with such a large and well-characterized sample at this redshift range.

Fig.~\ref{fig:SFR} shows the mass-representative limit, as defined in Section \ref{sec:Stellar_mass}, of our sample + CANDELS-ALL, as well as the limit in $\log\;\mathrm{SFR}$. The $\log\mathrm{M}_\star$ completeness level of the B19 and \citet{Speagle} (Sp14) MS are also included. The $\log\;\mathrm{SFR}$ completeness limits are computed using the SHARDS and HST stacks, used for the detection of our sample (except for the CANDELS-ALL sample, for which we only count with the information from individual bands). The images are centered at the rest-frame UV. The limiting magnitudes for the stack(s) probing the spectral range around 280~nm rest-frame, which are used in our SFR calculations, are translated into a UV luminosity, and then to an SFR. We consider the case for no attenuation and $\mathrm{A(V)}=0.30$~mag, both being shown in Fig.~\ref{fig:SFR} as horizontal lines.

\begin{table*}
\caption{Main sequence fitting function coefficients. The second and fourth columns contain our mass-representative limits and our limits in SFR. The third column shows the stellar mass limits of the CANDELS-ALL sample. For each redshift bin, column 4 contains a pair of numbers, where the first one is the completeness limit in SFR considering no attenuation and the second corresponds to the limit calculated using A(V)=0.30~mag. Columns 5-8 contain the coefficients of the 3$^\mathrm{rd}$ order polynomial, in increasing order, for the main sequence.  }
\label{tab:ms}  
\centering       
\begin{tabular}{c c c c c c c c}          
\hline\hline                       
z & {log M$_{\star, \textrm{ lim ours}}$[M$_\odot$]} & {log M$_{\star, \textrm{ lim B19}}$[M$_\odot$]} & log SFR$_{\textrm{lim}}$[M$_\odot$yr$^{-1}$] & x$_0$ & x$_1$ & x$_2$ & x$_3$ \\    
\hline                                 
  $1.0<z<1.5 $& 8.0 & 8.6 & $-$0.70/$-$0.48 & 45.946 & $-$17.376 &2.061 & $-$0.077 \\     
   $ 1.5<z<2.0$ & 8.2 & 8.7 &$-$0.43/$-$0.21& 49.215 &$-$18.336 & 2.149 & $-$0.079  \\
  $  2.0<z<2.5 $& 8.4 & 9.0 & $-$0.21/0.01  & $-$7.666 &$-$0.209 &0.232 &$-$0.012\\
   $ 2.5<z<3.0$ & 8.5 & 9.2 &$-$0.04/0.18    & 28.629 &$-$11.657&1.430 &$-$0.053   \\
 
\hline                                             
\end{tabular}
\end{table*}

We will distinguish two regimes in the following discussion: an intermediate-mass regime probed by the CANDELS-ALL sample (M$_{ \mathrm{lim,\;B19}}<$M$_\star<$M$_{\mathrm{turnover}}$), and a low-mass regime (M$_{\mathrm{\;lim,\;ours}}<$M$_\star<$M$_{\mathrm{lim,\;B19}}$), just below the B19 limits and above the mass-representative limit, where our sample of faint galaxies dominates the number counts. There is a third regime, the high-mass one, above the turnover points defined by \citet{Whitaker_15}, where the slope flattens. This last regime, however, is out of the scope of this research.

The bottom-right panel in Fig.~\ref{fig:SFR} shows the best-fitted polynomials for all the redshift bins. At each redshift interval, we see a continuous transition between the mass regime defined by our sources and that defined by the CANDELS-ALL sample. 

We find a median scatter of 0.32~dex, with the scatter slightly increasing with redshift, ranging from 0.27~dex at $z=1$ to 0.36~dex at $z=3$.
These values are compatible with previous studies finding values of 0.20-0.30~dex \citep{Speagle}. The value of the statistical uncertainty is compatible with the uncertainty derived from our Monte Carlo method, although slightly lower at the last redshift bin, where the Monte Carlo method provides an uncertainty of 0.43~dex. This is expected due to the increase of the photometric uncertainties as we move to fainter galaxies. 

\section{Discussion}
\label{sec:discussion}

In this section, we compare our results with previous works, as well as with the values derived using the Ilustris TNG numerical simulations (\citealt{TNG}, \citealt{TNG_2}). TNG50 evolves 2 $\times$ 21,603 dark matter particles and gas cells in a volume 50 comoving Mpc across. The mass resolution for the baryonic element is $8\times10^4$~M$_\odot$ and the  average cell size lies between $\sim$100-140 pc in the star-forming regions of galaxies. We note that the TNG50 MS presents an offset with respect to the MS points derived from the observations at the redshifts considered in this paper. The existence of this offset was already pointed out by \citet{Donnari}. 

Our values of the slopes and their uncertainties, together with the mass range used for their calculation, can be found in Table \ref{tab:slopes}. In Appendix \ref{sec:app3}, we include the values extracted from the TNG50 simulation, as well as the slopes found by previous works and the stellar mass down to which they are representative.

\begin{figure*}
    \centering
    \includegraphics[width=8.2cm, height=7cm]{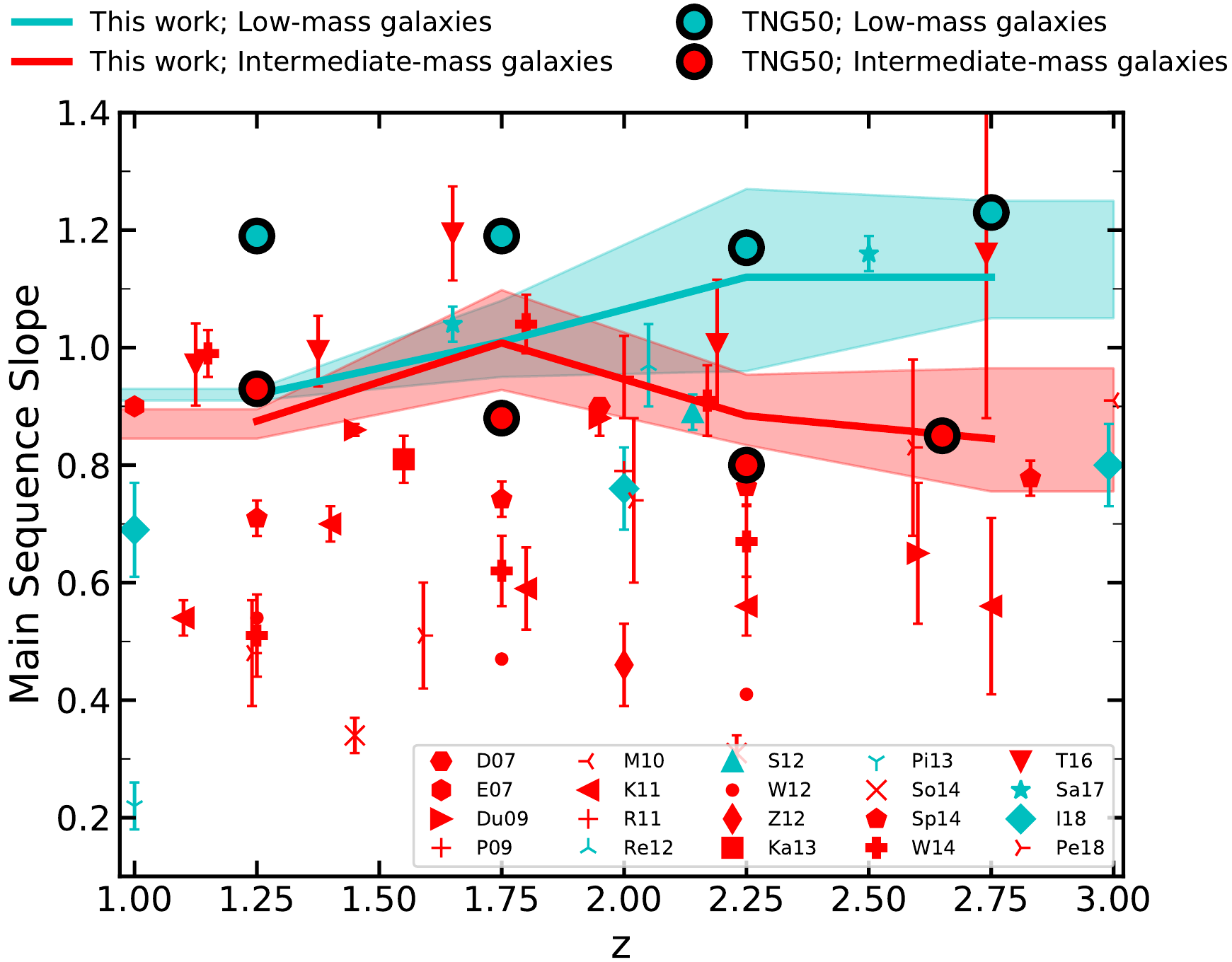}
    \includegraphics[width=8.4cm, height=7.4cm]{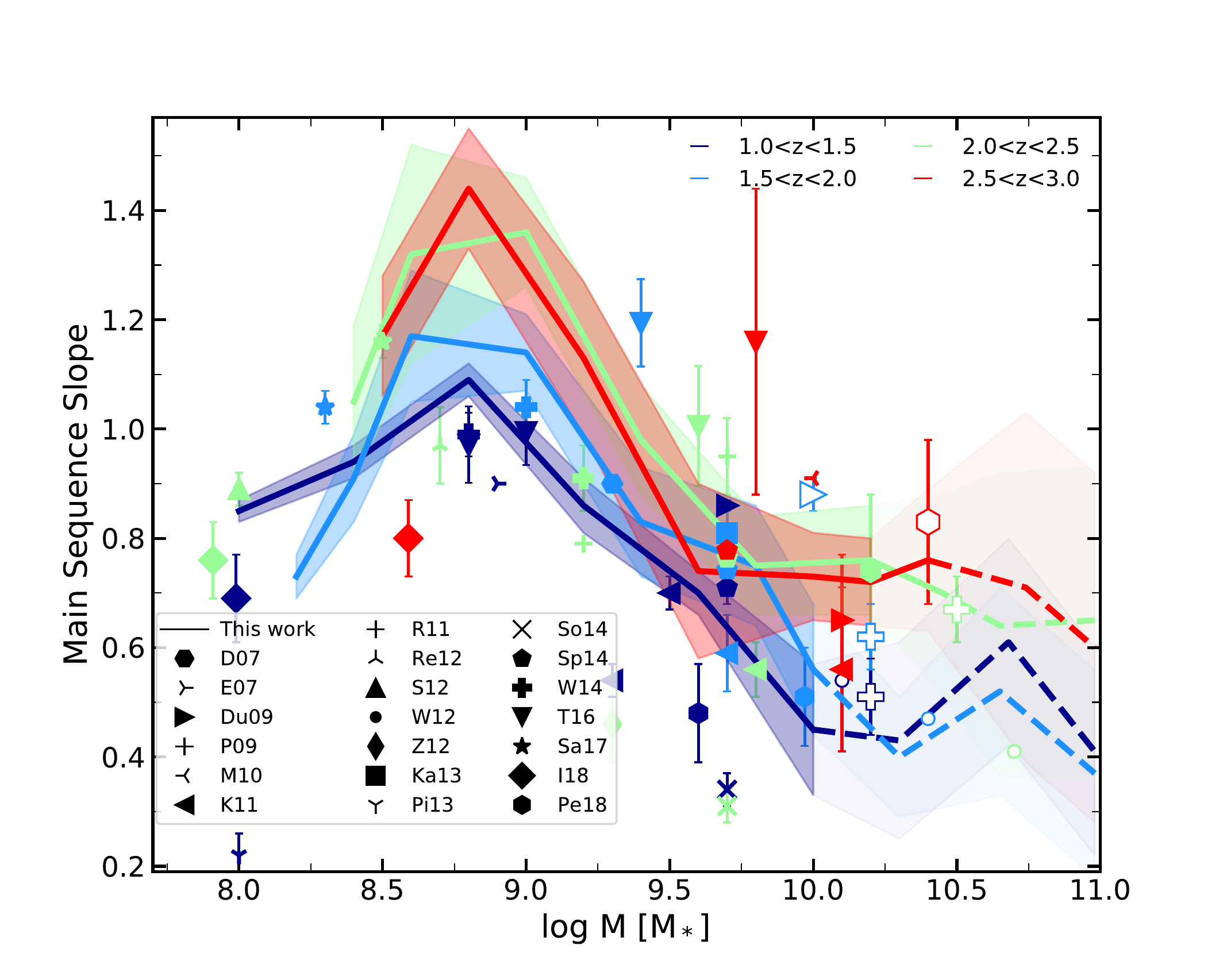}
    \caption{Left: main sequence slopes at different redshifts from precedent studies (markers) together with our calculations, derived by combining our sample of low-mass galaxies and the CANDELS-ALL  sample (thick lines). We also depict the slopes obtained from the Illustris TNG50 simulation (circles). We offer two estimations for the slope (including average value and uncertainties) based on our data and the TNG-50 simulation, in cyan and red. In cyan, the slopes are calculated in the mass range dominated by our low-mass galaxies (M$_{\mathrm{lim,\;ours}}<$M$_\star<$M$_{\mathrm{lim,\;B19}}$). In red, we show the slopes in the mass range dominated by intermediate-mass galaxies (M$_{\mathrm{lim,\;B19}}<$M$_\star<$M$_{\mathrm{turnover}}$), mainly CANDELS-ALL objects. The markers from precedent literature are color-coded according to their compatibility in mass with both regimes. Right: main sequence slopes versus their stellar mass lower bound. Colors highlight the different redshift bins. We derive the local main sequence slope at different mass bins for our sample of low-mass galaxies and CANDELS-ALL above our mass-representative limits (thick lines), together with the corresponding uncertainty. The dashed line corresponds to what it is called in the text the high-mass regime, derived using CANDELS-ALL data only, and only shown here for completeness.
    In both images, we include the estimation of the main sequence slope from \citet{Daddi} (D07), \citet{Elbaz_MS} (E07), \citet{Dunne} (Du09), \citet{Pannella} (P09), \citet{Magdis} (M10), \citet{Karim} (K11), \citet{Rodig} (R11), \citet{Reddy_MS} (Re12), \citet{Sawicki} (S12), \citet{Whitaker_12} (W12), \citet{Zahid} (Z12), \citet{Kashino} (Ka13), \citet{Pirzkal} (Pi13), \citet{Sobral} (So14), \citet{Speagle} (Sp14), \citet{Whitaker_15} (W14), \citet{Tomczak} (T16), \citet{Santini} (Sa17), \citet{Iyer} (I18) and \citet{Pearson} (Pe18). The values of the slope and mass limits here shown from \citet{Iyer} are the total ones, that take into account both, direct fits and trajectories. The uncertainties of each value of the slope are also plotted. Values corresponding to the high-mass regime are represented as empty markers. All the slopes, together with their uncertainties and the mass ranges where they are usable are listed in Table \ref{tab:slopes} and Appendix \ref{sec:app3}.}
    \label{fig:ms_slopes}
\end{figure*}

Figure~\ref{fig:ms_slopes} shows the evolution with redshift and mass of the MS slope. We separate the results for low- and intermediate-mass galaxies, as defined previously in Section~\ref{sec:SFR}.
The left panel from Fig.~\ref{fig:ms_slopes} shows the general local slopes (i.e., derived using all the running medians in the corresponding mass interval) at both mass regimes and their evolution with redshift. We first see that both regimes, defined by the set of our galaxies and CANDELS-ALL, are compatible with a single value of the slope below $z<2$, which would range between 0.9-1.0. At $z>2$, the slope derived in the low-mass regime becomes slightly steeper than that derived in the intermediate-mass regime. The low-mass regime between $2<z<3$ can be described with a slope of 1.12, whereas this value is 0.8-0.9 at the intermediate-mass regime.

When looking at the values from the precedent literature at higher masses, represented in red, there is apparently a huge scatter, with slopes ranging $\sim$0.3 to $\sim$1.2, but this is due to the fact that these researches are not tracing exactly the same mass range, that is why we include the panel on the right in Fig.~\ref{fig:ms_slopes}. However, we can already see in this plot that just a few works include galaxies with stellar masses compatible with the ones belonging to our sample. Ours is the only study that we have found in the literature that covers stellar masses as low as $10^{8-9}\,\mathrm{M}_\odot$ at $1<z<3$ via direct fits to galaxies with such a large sample. \citet{Reddy}, \citet{Pirzkal}, and \citet{Sawicki} based their calculations on 1,162, 302, and 91 galaxies, respectively. 

The results for the intermediate-mass regimes nicely match the TNG50 simulation. However, the TNG50 simulation predicts a steeper value of the slope at the low-mass range at $z<2$. At $z>2$ TNG50 matches our results again. Whereas according to the simulation, the slope at the low-mass regime evolves slightly with redshift, we find that it tends to increase. As stated by \citet{MUSE}, the slopes predicted by the models are a result of the growth rate of dark matter halos. According to models, supernova feedback or a decreased star formation efficiency would not affect the slope of the star formation MS, but they are crucial in low-mass galaxies.
Moreover, it is worth noticing that the resolution elements and the number of stellar particles are very close to the limits in the low-mass range. As pointed out by \citet{TNG_2}, the stellar-mass minimum enforces at least 140 stellar particles per galaxy. Stellar masses of 1-2·$10^7$M$_\odot$ correspond to a median of only 300 stellar particles and about 20,000 total resolution elements among dark matter, gas, and stars. In contrast, galaxies at 10$^{10}$M$_\odot$ are resolved with $10^5$ stellar particles and more than 1.5 million resolution elements in total.

On the right panel of Fig.~\ref{fig:ms_slopes} we show the MS slope calculated in discrete stellar mass bins, thus showing its evolution with mass. The slopes that describe the high-mass regime are obtained using the CANDELS-ALL sample only and are shown here for completeness. We include the different values from the literature, where the mass here represents their lower mass bound. 

First of all, looking at the evolution of the slopes as derived in this work, we notice a steepening for all redshift bins, with the slope peaking in the proximity of $10^{8.8}$M$_\odot$ at all redshifts (pointing to a possible second turnover mass, as that described by the literature for higher stellar masses). This steepening is smoother at $1<z<1.5$ and becomes stronger as we move higher in redshift. When looking at the average value of the slope, as in the left panel of Fig.~\ref{fig:ms_slopes}, it is not possible to see this effect, especially for the first redshift bins.
Above, the MS slope declines, reaching down to 0.4 at $z=1$ (0.7 at $z=3$) before the turnover mass. In this plot, we can clearly see that the lower values of the slope from the literature are tracing higher stellar masses. In general, precedent values are compatible with our findings, although at $z>2.5$ we lack points to compare. 

When extending the MS to even lower masses, (see Fig.~\ref{fig:SFR}) most of our galaxies lie in the region above the MS, especially at $z>1.5$ (on average $\sim$0.68~dex above the MS), i.e., they qualify as starburst galaxies. This is compatible with the short ages found in our stellar population synthesis study, which reveals typical t$_{0}\sim$40~Myr. 

Within the uncertainties, the slope of the MS at low masses is consistent with what is found at intermediate masses. In order to confirm or dismiss a possible change in the slope of the MS at the low-mass end, as could be hinted by the right panel from Fig.~\ref{fig:ms_slopes}, we require deeper observations. 

While it is usually established that the SFHs of massive galaxies vary slowly with time, low-mass galaxies are believed to have a more stochastic star formation history, with strong variations on timescales of $\sim10$~Myr. Feedback from supernovae can heat and expel gas from regions that can be as large as those where cold gas is present, leading to a temporary quenching of the star formation. The SFHs of these galaxies are thus characterized by being bursty. As discussed in \citet{bursty}, this burstiness can lead to strong selection effects when studying the statistical properties of these galaxies, especially at high redshifts, where galaxies in the burst phase, showing the strongest emission lines, will be preferentially selected.
This selection effect is also detected through simulations, as shown by \citet{Leja}, that studied the MS using stellar populations properties inferred by the \texttt{Prospector-$\upalpha$} code. At their low-mass limit ($\sim 10^9$M$_\odot$), they detect small upturns in the MS at some redshifts, which they translate to a combination of up-scattering of bright galaxies with high sSFRs below the limit and an incomplete census of low-sSFR galaxies above the limit. Additionally, in \citet{Stinson} they affirm that the shallow potential wells of these galaxies contribute to gas loss through supernova feedback and/or stripping.

On the other hand, there are already studies suggesting that there could be a population of dusty dwarf galaxies at high redshifts.
In \citet{Alvarez-Marquez}, they study the UV-to-FIR emission of  Lyman-break galaxies at $z\sim3$ in the COSMOS field that cannot be detected individually in the FIR. They find higher IR-excess than expected for galaxies with log M$_\star$/M$_\odot<10$. They suggest that the low-L$_\mathrm{UV}$ of these galaxies would only trace dust-free stars in galaxies that might be otherwise dusty. \citet{Bogdanoska}, using data compiled from the literature, study the UV dust attenuation as a function of stellar mass, finding a large scatter in the relationship of these two parameters for lower stellar masses with a non-zero average offset throughout most of the cosmic times. This offset can have different origins and they also mention a possible large dust content in low-mass galaxies, although they say this is unlikely. Taking advantage of the first 4 pointings of NIRCam, as part of the CEERS survey, \citet{Bisigello} find a sample of 145 $F200W$-dropouts with $\sim$82\% of the sample being dusty dwarf galaxies at $z<2$, with median A(V)=4.9~mag and median M$_\star$=10$^{7.3}$M$_\odot$. However, they say that this population of extremely dusty dwarfs has a minor contribution to the overall galaxy population at similar stellar masses and redshifts.

\movetableright=-1in
\startlongtable
\begin{deluxetable}{c|cc}
\tabletypesize{\normalsize}
\setlength{\tabcolsep}{4pt} 
\raggedleft
\tablecaption{Main sequence slopes from this work. The second and third columns show the redshifts and stellar mass limits where the slope from column 4 is applicable, i.e, where the samples used in the calculations are complete.}
\tablehead{\colhead{z}&\colhead{log M$_{\star}$/M$_\odot$}&\colhead{MS slope}}
\startdata
        &8.0$<$log M$<$8.6 &0.92$_{-0.01}^{+0.01}$ \\
    \textbf{ 1.00$<z<$1.5} & & \\
       &8.6$<$log M$<$10.2 & 0.88$^{+0.02}_{-0.03}$\\
\hline
       &8.2$<$log M$<$8.7 &1.01$^{+0.07}_{-0.06}$ \\
     \textbf{1.50$<z<$2.0} & & \\
       &8.7$<$log M$<$10.2 &1.01$^{+0.09}_{-0.08}$ \\
\hline
       &8.4$<$log M$<$9.0 &1.12$^{+0.13}_{-0.16}$ \\
    \textbf{2.0$<z<$2.5 }& & \\
       &9.0$<$log M$<$10.5 &0.88$^{+0.07}_{-0.05}$ \\
\hline
       &8.5$<$log M$<$9.2 &1.12$^{+0.13}_{-0.07}$ \\
   \textbf{2.5$<z<$3.0 }& & \\
       &9.2$<$log M$<$10.5 &0.85$^{+0.12}_{-0.09}$ \\
\enddata
\label{tab:slopes}
\end{deluxetable}

The combined action of stellar feedback and the shallow potential wells of these galaxies, as well as/or a bias toward optically bright sources might translate into a change in the MS slope. There are three possible scenarios: (a) the MS slope remains unchanged, (b) it gets steeper, or (c) it flattens.
We are seeing case (a) and only hints of case (b). Looking at the distribution of all of our galaxies, we could also talk about flattening if our mass-completeness limits allowed us to go beyond $10^8$M$_\odot$. A more complete census of galaxies in the low-mass regime, boosted by the advent of new telescopes and new spectroscopic campaigns, will bring light to how these and other physical mechanisms shape the main sequence of star-forming galaxies.

\section{Summary and conclusions}

\label{sect:conclusions}
Taking advantage of the spatial resolution of the HST images, together with the ability of SHARDS observations to detect emission line objects (thanks to its depth and spectral resolution), we obtain a sample of 34,061 faint objects in GOODS-N, virtually all of them not contained in previously published catalogs. Our sample is selected in the optical using stacked SHARDS and HST images centered at $\sim$700~nm.

The magnitudes of these objects are comparable to the limits of CANDELS in the HUDF, peaking at 28.3~mag in the optical and 27.8~mag in the NIR. This is 1.5~mag fainter than the sum of all the sources found by CANDELS in the five fields in the optical and NIR, respectively. Our sample surpasses the number of sources of the HUDF in the NIR beyond 28.3~mag. Our objects are mainly found in the redshift interval 1$<z<$3, with the redshift distribution peaking at $z\sim2.8$. 

According to the measurements of the UV-continuum slope, $UVJ$ rest-frame colors, and the ages derived through stellar population synthesis fitting, our sources are blue ($<U-V>\sim$0.10~mag; $<\upbeta> \sim -1.9$), young star-forming  ($<$t$_0>\sim$40~Myr) galaxies, showing very little obscuration ($<$A(V)$>\sim$0.30~mag). 

Their low stellar masses, with a median value of 10$^{7.5}$M$_\odot$, allow exploring the main sequence in the low-mass regime, extending to 10$^{8.0}$M$_\odot$, where we reach the 50\% mass-representative limit at $z=1$, and  10$^{8.5}$M$_\odot$ at $z=3$ ($\sim0.6$ order of magnitude deeper than previous estimations). The main sequence slope calculated in the mass regime dominated by our galaxies is compatible with the values found by previously published catalogs based on H-band selections. At $z<2$ we find an average value for the slope of 0.97, whereas this value is 1.12 at $z>2$.

Below our mass-representative limits, the sample (60\% of the objects, with stellar masses between 10$^{7.0}<$M$_{\star}<$M$_{\mathrm{lim,\;ours}}$) is mainly made up of starburst galaxies, especially at $z>1.5$, lying on average $\sim$0.68~dex above the main sequence.

\begin{acknowledgements}
RMM, PGPG, MA, and LC acknowledge support  from  Spanish  Ministerio  de  Ciencia e  Innovaci\'on MCIN/AEI/10.13039/501100011033 through grant PGC2018-093499-B-I00 and MDM-2017-0737 Unidad de Excelencia “Maria de Maeztu”-Centro de Astrobiología (INTA-CSIC), as well as by “ERDF A way of making Europe”. RMM acknowledges the support from the Instituto Nacional de T\'ecnica Aeroespacial SHARDS$^{JWST}$ project through the PRE-SHARDSJWST/2020 PhD fellowship, and by “ESF Investing in your future". AGA acknowledges the support of the Universidad Complutense de Madrid through the predoctoral grant CT17/17-CT18/17. MA and LC acknowledge financial support from Comunidad de Madrid under Atracci\'on de Talento grants 2020-T2/TIC-19971 and 2018-T2/TIC- 11612 respectively. This work has made use of the Rainbow Cosmological Surveys Database, which is operated by the Centro de Astrobiología (CAB/INTA), partnered with the University of California Observatories at Santa Cruz (UCO/Lick, UCSC).
\end{acknowledgements}

\software{astropy \citep{astropy}, corner \citep{corner}, EAZY \citep{EAZY}, matplotlib \citep{Hunter}, NumPy \citep{numpy}, photutils \citep{photutils}, PZETA (\citealt{PZETA}, Rainbow pipeline (\citealt{PZETA}, \citealt{SYNTHESIZER}, \citealt{Rainbow}), SciPy \citep{scipy}, SExtractor \citep{SExtractor}, Synthesizer (\citealt{PZETA}, \citealt{SYNTHESIZER}), TFIT \citep{TFIT}}

\appendix
\section{Photometric apertures}
\label{sec:app1}

For the SHARDS-faint catalog, we use the elliptical apertures provided by \texttt{SExtractor} (we use \citealt{Kron1980} photometric apertures to measure auto-consistent fluxes in all bands), setting the center to the flux-weighted centroid and imposing a minimum aperture radius of 0.7\arcsec\, (in both ground- and space-based images). This guarantees an aperture correction of less than 8\% for point-like sources, accounting for the seeing in all SHARDS and ground-based bands (see Table \ref{tab:filters}). For IRAC, 1.5\arcsec\, aperture radii are used for both sources from the HST- and SHARDS-faint catalogs.

We notice that, due to the different resolutions of GTC and HST, a fraction of the objects (4\%) that would appear as a single source in the SHARDS images are indeed different independent sources in the HST images. In order to define the apertures of the objects from the HST-faint catalog, we first need to identify which sources are isolated in the field and which sources can suffer from contamination of other close neighbors. We identify blended sources as those with a projected distance in the sky smaller than $r=0.55$\arcsec\, (which is the maximum radius that we will consider for the HST isolated sources) to other HST sources. These are then grouped into a single aperture of $r=0.7$\arcsec, setting the center to a weighted mean based on the \texttt{photutils} flux in $r=0.2$\arcsec\, apertures. This $r=0.7$\arcsec\, aperture is used to measure fluxes in all bands, ground- and space-based. Trying to disentangle the source into individual apertures would lead to the contamination of the photometry, especially in the SHARDS images. Given that this type of source only accounts for a minor fraction of the entire sample, the uncertainty in the nature of these galaxies has a negligible influence on the results presented in this work.

For the not blended objects, $r=0.55$\arcsec\, apertures are selected to recover the integrated flux when measuring in the ground-based images. This radius translates to an aperture correction of less than 15\% for all ground-based images. However, these are sometimes too large to account for the HST flux, as noticed in Fig. \ref{fig:cutouts}, where all the green HST apertures have a radius of 0.55\arcsec. The angular resolutions of the ground- and space-based images are quite different, with a PSF FWHM around 1\arcsec\, in SHARDS and 0.2\arcsec\, in HST. The size of the aperture must depend somehow on how spread the emission is in the HST images. For each object, three possible aperture radii are considered: 0.2\arcsec, 0.4\arcsec, and 0.55\arcsec. The final optimal radius ($r_\mathrm{OPTIMAL}$) is chosen in the following way:

$$ \textrm{If} \: \textrm{m}_{r_i} - 3*\textrm{$\Delta m$}_{r_i} > \textrm{m}_{r_{i+1}} \:  \: \& \: \: \textrm{SNR}_{r_{i+1}}>4, \:\: \textrm{then} \: \textrm{i}=\textrm{i}+1 $$
$$ \textrm{else} \: \textrm{r}=\textrm{r}_i $$
$$\textrm{With}\:\textrm{ r}=0.2\arcsec,0.4\arcsec,0.55\arcsec$$

\noindent m and $\Delta$m refer to the magnitude in the ACS HST stack and its uncertainty.

We additionally revise these optimal aperture radii according to the distance to the closest neighbor.
If this distance is smaller than 1.1\arcsec, the optimal radii of 0.55\arcsec\, are automatically reduced to 0.4\arcsec, and if this distance is smaller than 0.8\arcsec, then the optimal radii are set to 0.2\arcsec, regardless of the flux. 

To summarize, the aperture radii for the SHARDS-faint objects are  taken directly from \texttt{SExtractor} (Kron photometric apertures), imposing a minimum value of 0.7~\arcsec. The HST-faint objects are divided into two classes: blended and not blended sources. The blended ones, which represent a minor fraction of the total sample, are measured within an aperture radius of 0.7\arcsec, which is used for both, ground- and space-based images. The not blended ones are assigned two values of the radii, one that is used to measure within the ground-based images, which is 0.55\arcsec, and another, which we call the $r_\mathrm{OPTIMAL}$, that is used to measure within the space-based images. 40\% of the total sample is measured within an aperture of $r_\mathrm{OPTIMAL}$=0.2\arcsec, 35\% of it is assigned $r_\mathrm{OPTIMAL}$=0.4\arcsec, and 4\% is measured within a $r_\mathrm{OPTIMAL}$=0.55\arcsec aperture.

We also check the effect on colors and fluxes of fixing an aperture radius for all the HST-faint not blended objects, instead of setting a $r_\mathrm{OPTIMAL}$ for each galaxy. Colors are necessary to obtain robust photometric redshifts, while colors together with the level of the SEDs are essential to obtain reliable values of the stellar masses. 

In terms of the level of the continuum, when homogenizing the aperture radii taking the most common value, which is $r=0.2$\arcsec\, we lose on average 0.4~mag of the HST flux while getting similar SNR ($\sim$1.2 times the SNR obtained using $r_\mathrm{OPTIMAL}$=0.4\arcsec\, or $r_\mathrm{OPTIMAL}$=0.55\arcsec). This would happen to 39\% of the sample, whose $r_\mathrm{OPTIMAL}$ is not 0.2\arcsec. In terms of the colors, when comparing colors computed based on ACS and WFC3 bands, we get an average offset of $-0.005$~mag and a scatter of 0.400~mag, which are not statistically significant given that the typical color uncertainties are $\sim$0.6~mag.

We also study the opposite situation, computing the effect on flux and colors of setting a larger aperture radius than the optimal one.
If we increase the radius from $r_\mathrm{OPTIMAL}$=0.2\arcsec\, to $r=0.4$\arcsec, colors are again robust and there is not a significant difference (average offset of 0.009~mag and scatter of 0.500~mag) while on average we get 0.3~mag more flux within the $r=0.4$\arcsec, aperture, getting 40\% smaller SNR values.
Alternatively, we can increase the radius from $r_\mathrm{OPTIMAL}$=0.2\arcsec\, to $r=0.55$\arcsec. There is not a significant difference in the colors (average offset of $-0.03$~mag and scatter of 0.60~mag). We obtain 0.4~mag more flux within the $r=0.55$\arcsec\, aperture while getting a factor of 2 smaller SNR values.
These two scenarios would affect 40\% of the sample.

\begin{figure}
    \centering
    \gridline{\fig{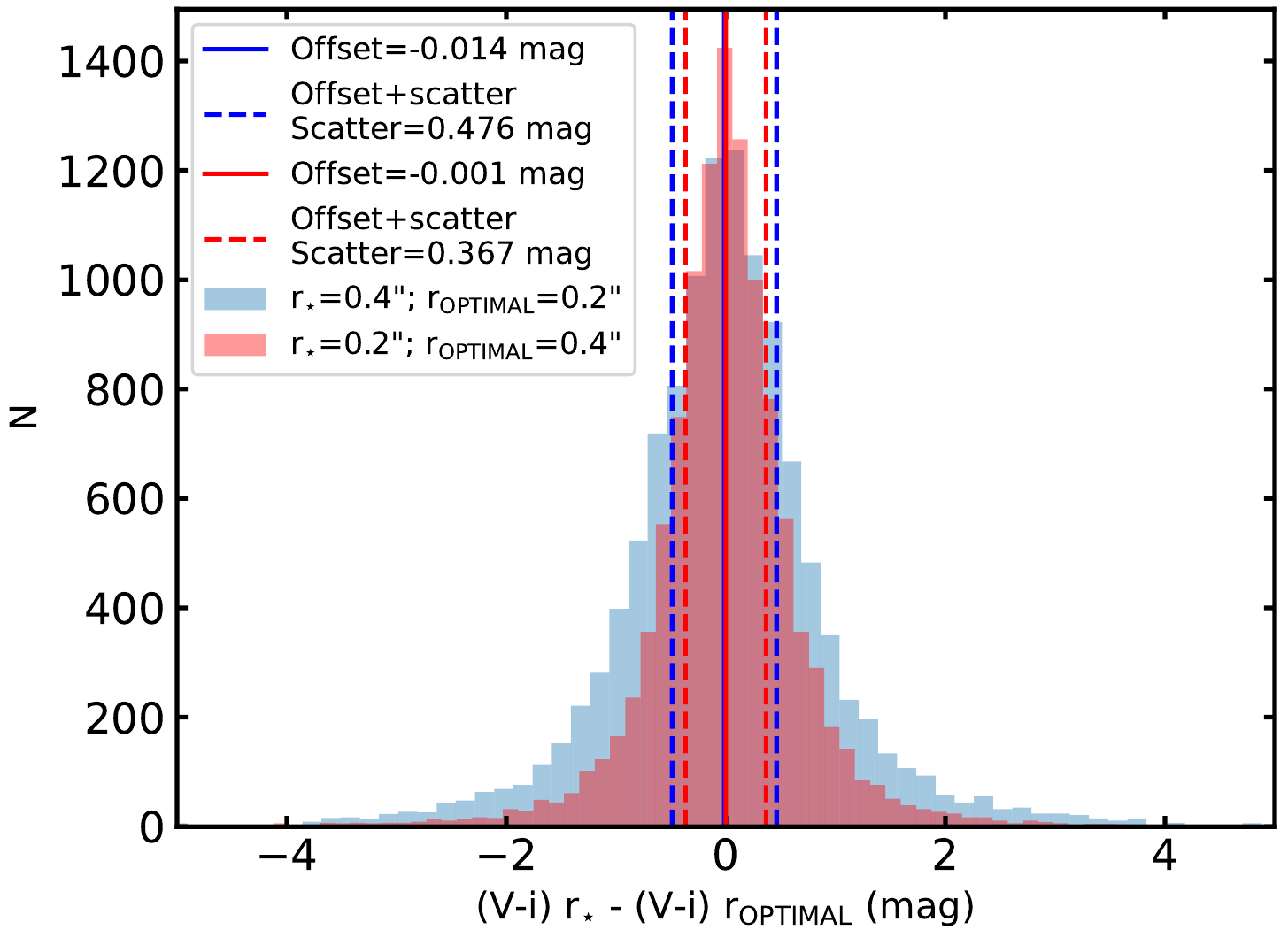}{0.5\textwidth}{}
          \fig{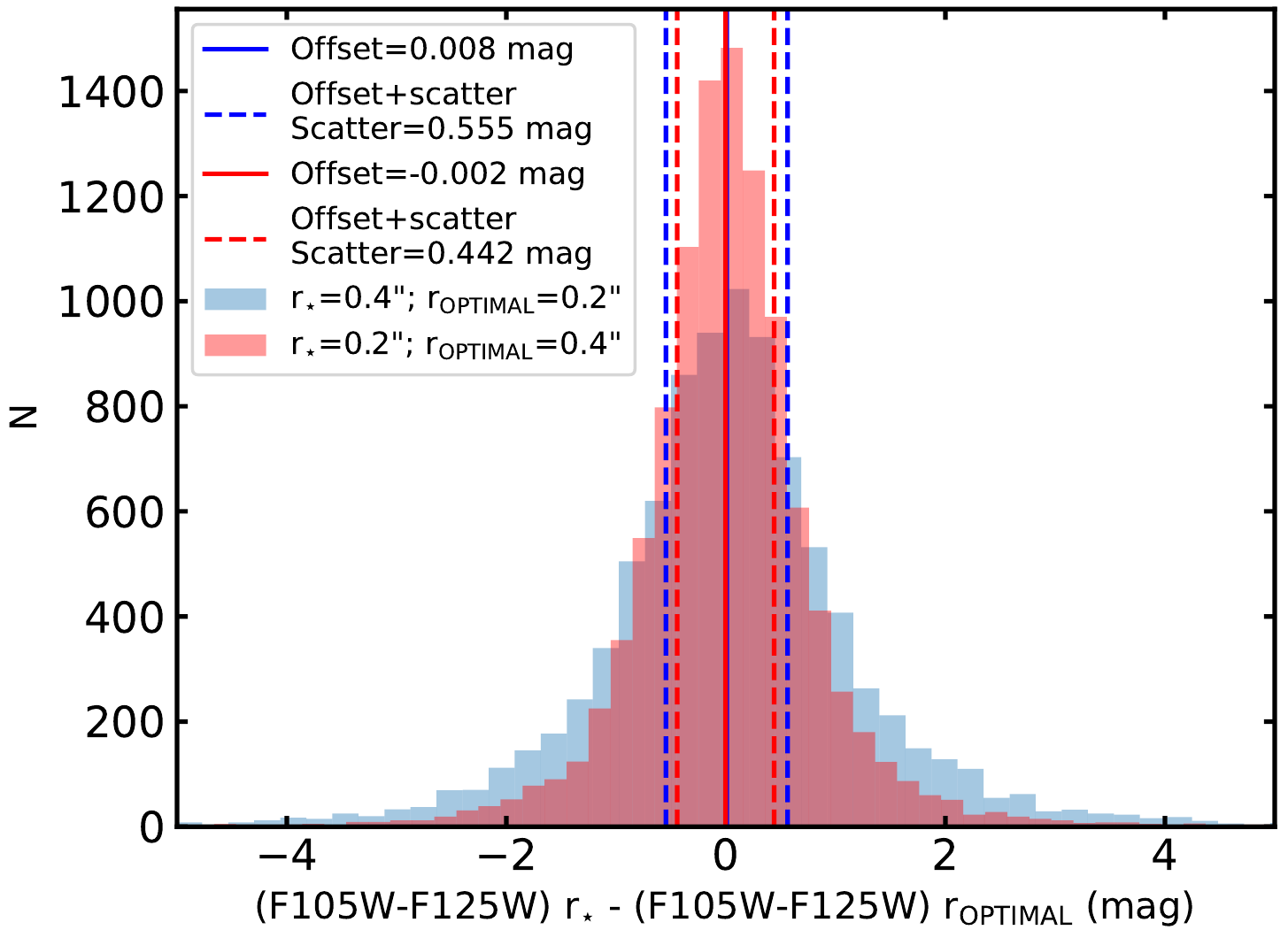}{0.5\textwidth}{}
           }
    \caption{Histograms showing the difference between colors. The red histogram shows the case where the aperture radius is larger than $r_\mathrm{OPTIMAL}$ whereas the blue one shows the case where the aperture radius is smaller than $r_\mathrm{OPTIMAL}$. We plot the weighted offset and scatter as vertical lines, following the same color code. The dashed lines show the scatter and the solid ones the offset. Left: comparing the $V-i$ color. Right: comparing the $F105W-F125W$ color.}
    \label{fig:colors}
\end{figure}

Lastly, we check the effect on colors and fluxes of changing $r_\mathrm{OPTIMAL}$=0.4\arcsec\, to $r=0.55$\arcsec. Colors are robust (average offset of 0.01~mag and scatter of 0.30~mag) and we get on average 0.2~mag more flux within the $r=0.55$\arcsec\, aperture while getting around 30\% smaller values of SNR. This situation would affect 35\% of the sample.

These tests show that our approach optimizes the SNR while keeping colors virtually unaffected and possible light losses under $\sim$0.3~mag (when we take into account the fraction of sources measured in different apertures). If the mass-to-light ratio of our stellar population fits does not change (given that the colors do not change), this could translate to mass systematic offsets of 0.1~dex. This value is significantly smaller than typical variations in mass (see Section \ref{sec:synthesizer} for more information).

In Figure \ref{fig:colors} we show the difference between two of the colors considered for two of the previously discussed cases.

\section{Point-like nature of our galaxies}
\label{sec:app2}

\begin{figure}
    \centering
         \centering
         \includegraphics[width=8.4cm, height=6.4cm]{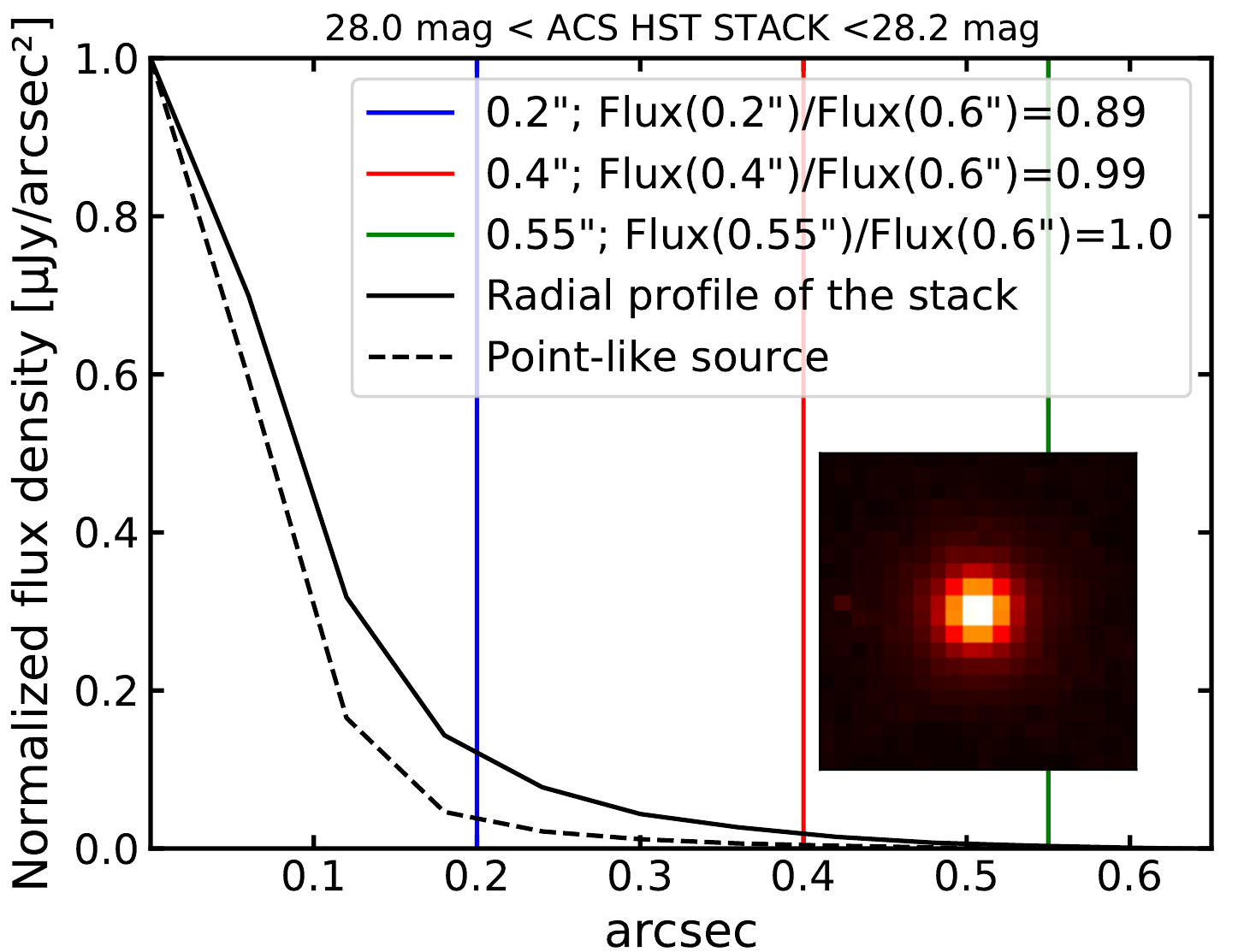}
         \includegraphics[width=8.4cm, height=6.4cm]{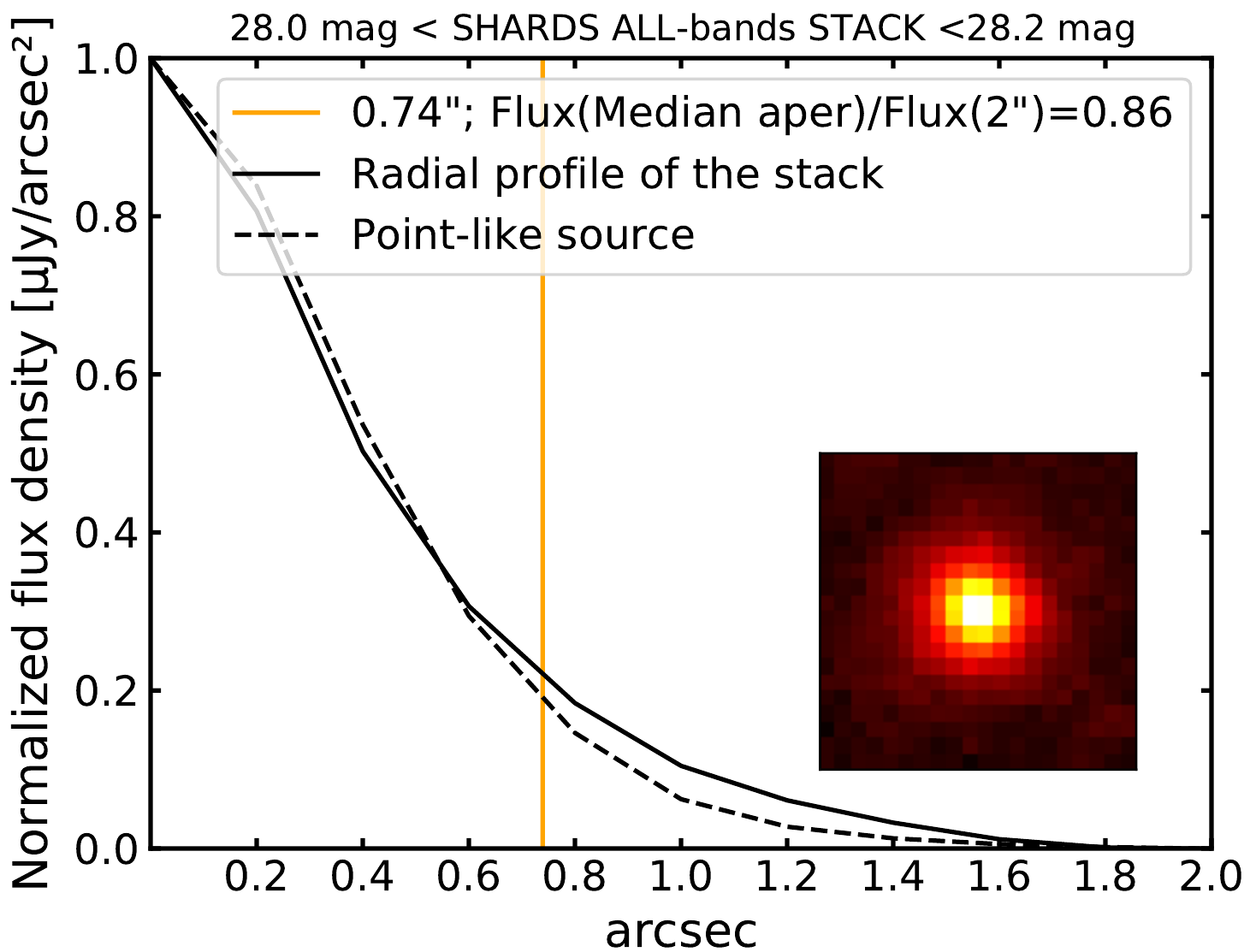}
    \caption{Intensity profile of the stacked galaxy at 28~mag$<$ACS HST STACK$<$28.2~mag (left) and intensity profile of the stacked galaxy at 28~mag$<$SHARDS ALL-bands stack$<$28.2~mag (right) (thick black). The profile of a point-like source is plotted as a dashed line. The colored vertical lines denote different values of the aperture radii. In the bottom panel, the vertical line corresponds to the median aperture radius. In the legend, we specify the fraction of the flux recovered at each radius with respect to the flux enclosed at a radius of 0.6~\arcsec (left) or 2~\arcsec (right). We include a cutout of the stacked galaxy that results from combining all the sources in the magnitude bin.}
    \label{fig:radial_profiles}
\end{figure}

We explore the morphology and shape of our galaxies by stacking sources. We stack sources in different magnitude bins as seen in the SHARDS ALL-bands and ACS HST stacks, our selection images. Let us note that all along the paper when referring to stacking, we talk about summing up different bands to get a stacked image. In this case, we are taking cutouts of galaxies in a certain image, centering them, and summing them up to obtain the image of the stacked galaxy. The radial profile of the stacked galaxies at $\sim$28~mag (the typical magnitude in the optical of our sources, see section \ref{sect:Photometric_properties} for more details), as well as a cutout in the corresponding stacked image, are shown in Fig.~\ref{fig:radial_profiles}. The median aperture radii for several magnitude intervals ranging from 26 to 29~mag in the SHARDS ALL-bands stack are 0.97\arcsec, 0.91\arcsec, 0.7\arcsec, and 0.55\arcsec. At those radii, we enclose 82-90\% of the total flux, with the total flux being defined as the flux enclosed within 2~arcsec, which is $\sim2$ times the SHARDS ALL-bands PSF FWHM. In the brightest magnitude bins, the sources are more extended because the sample is mainly made of objects that are detected within the SHARDS stack. At fainter magnitudes, the sources are not detected in the SHARDS ALL-bands stack, but only in the ACS HST stack, hence the median value of the aperture radii is the one imposed for the HST-faint sample when measuring within ground-based images. Looking at the radial profile we see that between 26-29~mag our stacked galaxies resemble that of a point-like source. The same is obtained when switching to the ACS HST stack. The median aperture turns smaller at fainter magnitudes and the resulting stacked galaxy can be approximated to a point-like source. The radii considered enclose $\sim$90\% of the total flux, with the total defined as the flux enclosed within 0.6~\arcsec, where the signal equalizes the level of the sky. Complementary to this last point, we calculate the isophotal radius for each HST-selected galaxy, defined as the radius where we reach 2 times the sky surface brightness. The median isophotal radius is 0.3~\arcsec.

\section{Main Sequence Slopes}
\label{sec:app3}

\startlongtable
\begin{deluxetable}{c|c|cc}
\setlength{\tabcolsep}{20pt} 
\tabletypesize{\normalsize}
\tablecaption{Main sequence slopes from the Illustris TNG50 simulation and previous literature values. The second and third columns show the redshifts and stellar mass limits where the slope from column 4 is applicable, i.e, where the samples used in the calculations are complete.}
\tablehead{\colhead{} &\colhead{z}&\colhead{log M$_{\star}$/M$_\odot$}&\colhead{MS slope}}
\startdata
\citet{Elbaz_MS} & 1.00 & $>8.9$ & 0.90 \\
\citet{Pirzkal} & 1.00 & $>8.0$ & 0.22$\pm$0.04\\
\citet{Iyer} & 1.00 & $>$7.99 (7.62$^{*}$) & 0.69$\pm$0.08\\
\citet{Karim} & 1.10 &$>9.3$ & 0.54$\pm$0.03\\
\citet{Pearson} & 1.24 & $>$9.6$^{**}$ & 0.48$\pm$0.09\\
\citet{Speagle} & 1.25 &$>9.7$& 0.71$\pm$0.03 \\
\citet{Whitaker_12} & 1.25 & $>$10.1 & 0.54\\
\citet{Karim} & 1.40 & $>9.5$ & 0.70$\pm$0.03 \\
\citet{Dunne} & 1.45 & $>9.7$ & 0.86$\pm$0.01 \\
\citet{Sobral} & 1.45 & $>9.7$ & 0.34$\pm$0.03 \\
&&$>8.8$& 0.99$\pm$0.04 \\
\citet{Whitaker_15}  & 1.00$<z<$1.50 & &\\
&&$>10.2$& 0.51$\pm$0.07\\
\citet{Tomczak} & 1.00$<z<$1.25 & $>8.8$& 0.97$\pm$0.07 \\
\citet{Tomczak} & 1.25$<z<$1.50 & $>9.0$& 0.99$\pm$0.06 \\ 
&       &8.0$<$log M$<$8.6 &1.19 \\
TNG50  &  1.00$<z<$1.50 & & \\
&       &8.6$<$log M$<$10.2 &0.93 \\
\hline
\citet{Kashino} & 1.55 & $>$9.7 & 0.81$\pm$0.04\\
\citet{Pearson} & 1.59 & $>$10$^{**}$ & 0.51$\pm$0.09\\
\citet{Speagle} & 1.75 &$>9.7$& 0.74$\pm$0.03 \\
\citet{Whitaker_12} & 1.75 & $>$10.4 & 0.47\\
\citet{Karim} & 1.80 &$>$9.7 & 0.59$\pm$0.07\\
\citet{Daddi} & 1.95 & $>$9.3 & 0.90 \\
\citet{Dunne} & 1.95 & $>$10 & 0.88$\pm$0.03\\
&&$>9.0$& 1.04$\pm$0.05\\
\citet{Whitaker_15}  & 1.50$<z<$2.0 & &  \\        
&&$>10.2$& 0.62$\pm$0.06\\
\citet{Tomczak} & 1.50$<z<$2.0 & $>9.4$& 1.19$\pm$0.08 \\
\citet{Santini} & 1.30$<z<$2.0 & $>8.3$ & 1.04$\pm$0.03\\
&       &8.2$<$log M$<$8.7 &1.18 \\
TNG50  &  1.50$<z<$2.0 & & \\
&       &8.7$<$log M$<$10.2 &0.88 \\
\hline
\citet{Pannella} & 2.00 & $>$9.7 & 0.95$\pm$0.07\\
\citet{Rodig} & 2.00 & $>$9.2 & 0.79\\
\citet{Zahid} & 2.00 & $>$9.3 & 0.46$\pm$0.07\\
\citet{Iyer} & 2.00 & $>$7.91 (7.74$^{*}$) & 0.76$\pm$0.07\\
\citet{Pearson} & 2.02 & $>$10.2$^{**}$& 0.74$\pm$0.14\\
\citet{Reddy} & 2.05 & $>$8.7 & 0.97$\pm$0.07\\
\citet{Sawicki} & 2.2 & $>8.0$ & 0.89$\pm$0.03\\
\citet{Sobral} & 2.23 & $>$9.7 & 0.31$\pm$0.03\\
\citet{Karim} & 2.25 & $>$9.8 & 0.56$\pm$0.05\\
\citet{Speagle} & 2.25 &$>9.7$& 0.76$\pm$0.03\\
\citet{Whitaker_12} & 2.25 & $>$10.7 & 0.41\\
&&$>9.2$& 0.91$\pm$0.06\\
\citet{Whitaker_15}  & 2.00$<z<$2.50 & &  \\        
&&$>10.5$& 0.67$\pm$0.06\\
\citet{Tomczak} & 2.00$<z<$2.50 & $>9.6$& 1.01$\pm$0.11 \\
&       &8.4$<$log M$<$9.0  &1.17 \\
TNG50  &   2.00$<z<$2.50 & & \\
&       &9.0$<$log M$<$10.5 &0.80 \\
\hline
\citet{Pearson} & 2.59 & $>$10.4$^{**}$ & 0.83$\pm$0.15\\
\citet{Dunne} & 2.60 & $>$10.1 & 0.65$\pm$0.12\\
\citet{Karim} & 2.75 & $>$10.1 & 0.56$\pm$0.15\\
\citet{Speagle} & 2.75 &$>9.7$& 0.78$\pm$0.03\\
\citet{Magdis} & 3.00 & $>$10 & 0.91\\
\citet{Iyer} & 3.00 & $>$8.59 (7.57$^{*}$) & 0.80$\pm$0.07\\
\citet{Tomczak} & 2.50$<z<$3.00 & $>9.8$& 1.16$\pm$0.28 \\
\citet{Santini} & 2.00$<z<$3.00 & $>8.5$ & 1.16$\pm$0.03\\
&       &8.5$<$log M$<$9.2  &1.23 \\
TNG50  &   2.50$<z<$3.00 & & \\
&       &9.2$<$log M$<$10.5 &0.85 \\
\enddata
\label{tab:slopes_app}
\tablecomments{$^{*}$\label{note4} The values of the slope here listed from \citet{Iyer} are the total ones, that take into account both, direct fits and trajectories. The value in brackets corresponds to the limit probed using trajectories.}
\tablecomments{$^{**}$\label{note5} Mass limit for the deep regions of COSMOS.}
\end{deluxetable}


\begin{thebibliography}{}
 \bibitem[Alavi et al.(2014)]{Alavi} Alavi, A., Siana, B., Richard, J., et al.\ 2014, \apj, 780, 143. doi:10.1088/0004-637X/780/2/143

 \bibitem[Alcalde Pampliega et al.(2019)]{Alcalde} Alcalde Pampliega, B., P{\'e}rez-Gonz{\'a}lez, P.~G., Barro, G., et al.\ 2019, \apj, 876, 135. doi:10.3847/1538-4357/ab14f2

 \bibitem[{\'A}lvarez-M{\'a}rquez et al.(2019)]{Alvarez-Marquez} {\'A}lvarez-M{\'a}rquez, J., Burgarella, D., Buat, V., et al.\ 2019, \aap, 630, A153. doi:10.1051/0004-6361/201935719
 
 \bibitem[Anderson et al.(2017)]{Anderson} Anderson, L., Governato, F., Karcher, M., et al.\ 2017, \mnras, 468, 4077. doi:10.1093/mnras/stx709
 
 \bibitem[Arrabal Haro et al.(2018)]{Arrabal} Arrabal Haro, P., Rodr{\'\i}guez Espinosa, J.~M., Mu{\~n}oz-Tu{\~n}{\'o}n, C., et al.\ 2018, \mnras, 478, 3740. doi:10.1093/mnras/sty1106

 \bibitem[Arrigoni Battaia et al.(2016)]{Arrigoni} Arrigoni Battaia, F., Hennawi, J.~F., Cantalupo, S., et al.\ 2016, \apj, 829, 3. doi:10.3847/0004-637X/829/1/3

 \bibitem[Ashby et al.(2013)]{IRAC1} Ashby, M.~L.~N., Willner, S.~P., Fazio, G.~G., et al.\ 2013, \apj, 769, 80. doi:10.1088/0004-637X/769/1/80
 
 \bibitem[Astropy Collaboration et al.(2022)]{astropy} Astropy Collaboration, Price-Whelan, A.~M., Lim, P.~L., et al.\ 2022, \apj, 935, 167. doi:10.3847/1538-4357/ac7c74

 \bibitem[Bacon et al.(2017)]{MUSE_2} Bacon, R., Conseil, S., Mary, D., et al.\ 2017, \aap, 608, A1. doi:10.1051/0004-6361/201730833
    
 \bibitem[Barro et al.(2011)]{Rainbow} Barro, G., P{\'e}rez-Gonz{\'a}lez, P.~G., Gallego, J., et al.\ 2011, \apjs, 193, 30. doi:10.1088/0067-0049/193/2/30

 \bibitem[Barro et al.(2019)]{Barro} Barro, G., P{\'e}rez-Gonz{\'a}lez, P.~G., Cava, A., et al.\ 2019, \apjs, 243, 22. doi:10.3847/1538-4365/ab23f2

 \bibitem[Bertin \& Arnouts(1996)]{SExtractor} Bertin, E. \& Arnouts, S.\ 1996, \aaps, 117, 393. doi:10.1051/aas:1996164

\bibitem[Bisigello et al.(2023)]{Bisigello} Bisigello, L., Gandolfi, G., Grazian, A., et al.\ 2023, arXiv:2302.12270. doi:10.48550/arXiv.2302.12270

\bibitem[Bogdanoska \& Burgarella(2020)]{Bogdanoska} Bogdanoska, J. \& Burgarella, D.\ 2020, \mnras, 496, 5341. doi:10.1093/mnras/staa1928

 \bibitem[Boogaard et al.(2018)]{MUSE} Boogaard, L.~A., Brinchmann, J., Bouch{\'e}, N., et al.\ 2018, \aap, 619, A27. doi:10.1051/0004-6361/201833136
 
 \bibitem[Bouwens et al.(2010)]{Bouwens_beta_2010} Bouwens, R.~J., Illingworth, G.~D., Oesch, P.~A., et al.\ 2010, \apjl, 708, L69. doi:10.1088/2041-8205/708/2/L69

 \bibitem[Bouwens et al.(2012)]{2012ApJ...754...83B} Bouwens, R.~J., Illingworth, G.~D., Oesch, P.~A., et al.\ 2012, \apj, 754, 83. doi:10.1088/0004-637X/754/2/83

 \bibitem[Bouwens et al.(2015)]{Bouwens} Bouwens, R.~J., Illingworth, G.~D., Oesch, P.~A., et al.\ 2015, \apj, 803, 34. doi:10.1088/0004-637X/803/1/34
 
 \bibitem[Bradley et al.(2019)]{photutils} Bradley, L., Sip{\H{o}}cz, B., Robitaille, T., et al.\ 2019, Zenodo

 \bibitem[Brammer et al.(2008)]{EAZY} Brammer, G.~B., van Dokkum, P.~G., \& Coppi, P.\ 2008, \apj, 686, 1503. doi:10.1086/591786
 
 \bibitem[Bruzual \& Charlot(2003)]{Bruzual} Bruzual, G. \& Charlot, S.\ 2003, \mnras, 344, 1000. doi:10.1046/j.1365-8711.2003.06897.x
  
  \bibitem[Calzetti et al.(2000)]{Calzetti} Calzetti, D., Armus, L., Bohlin, R.~C., et al.\ 2000, \apj, 533, 682. doi:10.1086/308692
  
  \bibitem[Capak et al.(2004)]{Capak} Capak, P., Cowie, L.~L., Hu, E.~M., et al.\ 2004, \aj, 127, 180. doi:10.1086/380611
  
  \bibitem[Caputi et al.(2021)]{Caputi} Caputi, K.~I., Caminha, G.~B., Fujimoto, S., et al.\ 2021, \apj, 908, 146. doi:10.3847/1538-4357/abd4d0
  
  \bibitem[Cava et al.(2015)]{Cava} Cava, A., P{\'e}rez-Gonz{\'a}lez, P.~G., Eliche-Moral, M.~C., et al.\ 2015, \apj, 812, 155. doi:10.1088/0004-637X/812/2/155

  \bibitem[Cepa(1998)]{Osiris} Cepa, J.\ 1998, \apss, 263, 369. doi:10.1023/A:1002144913887

  \bibitem[Cirasuolo et al.(2007)]{2007MNRAS.380..585C} Cirasuolo, M., McLure, R.~J., Dunlop, J.~S., et al.\ 2007, \mnras, 380, 585. doi:10.1111/j.1365-2966.2007.12038.x
  
  \bibitem[Chabrier(2003)]{Chabrier} Chabrier, G.\ 2003, \pasp, 115, 763. doi:10.1086/376392
  
  \bibitem[Chaikin et al.(2022)]{Chaikin} Chaikin, E., Schaye, J., Schaller, M., et al.\ 2022, arXiv:2203.07134
  
  \bibitem[Costantin et al.(2021)]{constantin_1} Costantin, L., P{\'e}rez-Gonz{\'a}lez, P.~G., M{\'e}ndez-Abreu, J., et al.\ 2021, \apj, 913, 125. doi:10.3847/1538-4357/abef72
  
  \bibitem[Costantin et al.(2022)]{constantin_2} 
  Costantin, L., P{\'e}rez-Gonz{\'a}lez, P.~G., M{\'e}ndez-Abreu, J., et al.\ 2022, arXiv:2202.02332
  
  \bibitem[Daddi et al.(2007)]{Daddi} Daddi, E., Dickinson, M., Morrison, G., et al.\ 2007, \apj, 670, 156. doi:10.1086/521818

  \bibitem[Davis et al.(2007)]{EGS} Davis, M., Guhathakurta, P., Konidaris, N.~P., et al.\ 2007, \apjl, 660, L1. doi:10.1086/517931
  
  \bibitem[DeGraf et al.(2017)]{DeGraf} DeGraf, C., Dekel, A., Gabor, J., et al.\ 2017, \mnras, 466, 1462. doi:10.1093/mnras/stw2777
  
  \bibitem[Dekel et al.(2021)]{Dekel} Dekel, A., Freundlich, J., Jiang, F., et al.\ 2021, \mnras, 508, 999. doi:10.1093/mnras/stab2416

  \bibitem[D{\'\i}az et al.(2021)]{Diaz} D{\'\i}az, C.~G., Ryan-Weber, E.~V., Karman, W., et al.\ 2021, \mnras, 502, 2645. doi:10.1093/mnras/staa3129

  \bibitem[Dickinson et al. (2003)]{IRAC2} Dickinson, M., Bergeron, J., Casertano, S., al. 2003,     
  Great Observatories Origins Deep Survey (GOODS) Validation Observations, Spitzer Proposal
  
  \bibitem[Dickinson(2008)]{GOODS} Dickinson, M.\ 2008, \aas
  
   \bibitem[Dom{\'\i}nguez et al.(2015)]{bursty} Dom{\'\i}nguez, A., Siana, B., Brooks, A.~M., et al.\ 2015, \mnras, 451, 839. doi:10.1093/mnras/stv1001
   
   \bibitem[Dom{\'\i}nguez S{\'a}nchez et al.(2016)]{Helena} Dom{\'\i}nguez S{\'a}nchez, H., P{\'e}rez-Gonz{\'a}lez, P.~G., Esquej, P., et al.\ 2016, \mnras, 457, 3743. doi:10.1093/mnras/stw201

   \bibitem[Donnari et al.(2019)]{Donnari} Donnari, M., Pillepich, A., Nelson, D., et al.\ 2019, \mnras, 485, 4817. doi:10.1093/mnras/stz712
   
   \bibitem[Dunne et al.(2009)]{Dunne} Dunne, L., Ivison, R.~J., Maddox, S., et al.\ 2009, \mnras, 394, 3. doi:10.1111/j.1365-2966.2008.13900.x

   \bibitem[Elbaz et al.(2007)]{Elbaz_MS} Elbaz, D., Daddi, E., Le Borgne, D., et al.\ 2007, \aap, 468, 33. doi:10.1051/0004-6361:20077525

   \bibitem[Elbaz et al.(2011)]{Elbaz} Elbaz, D., Dickinson, M., Hwang, H.~S., et al.\ 2011, \aap, 533, A119. doi:10.1051/0004-6361/201117239
   
   \bibitem[Finkelstein et al.(2015)]{Finkelstein} Finkelstein, S.~L., Ryan, R.~E., Papovich, C., et al.\ 2015, \apj, 810, 71. doi:10.1088/0004-637X/810/1/71
   
   \bibitem[Foreman-Mackey(2016)]{corner} Foreman-Mackey, D.\ 2016, The Journal of Open Source Software, 1, 24. doi:10.21105/joss.00024
   
   \bibitem[Galametz et al.(2013)]{uds_cat} Galametz, A., Grazian, A., Fontana, A., et al.\ 2013, \apjs, 206, 10. doi:10.1088/0067-0049/206/2/10

   \bibitem[Giavalisco et al.(2004)]{ACS} Giavalisco, M., Ferguson, H.~C., Koekemoer, A.~M., et al.\ 2004, \apjl, 600, L93. doi:10.1086/379232
   
   \bibitem[Grazian et al.(2015)]{Grazian} Grazian, A., Fontana, A., Santini, P., et al.\ 2015, \aap, 575, A96. doi:10.1051/0004-6361/201424750

   \bibitem[Griffiths et al.(2021)]{SHARDSFF2} Griffiths, A., Conselice, C.~J., Ferreira, L., et al.\ 2021, \mnras, 508, 3860. doi:10.1093/mnras/stab2566

   \bibitem[Grogin et al.(2011)]{CANDELS} Grogin, N.~A., Kocevski, D.~D., Faber, S.~M., et al.\ 2011, \apjs, 197, 35. doi:10.1088/0067-0049/197/2/35
   
   \bibitem[Guo et al.(2013)]{Guo} Guo, Y., Ferguson, H.~C., Giavalisco, M., et al.\ 2013, \apjs, 207, 24. doi:10.1088/0067-0049/207/2/24
   
   \bibitem[Guo et al.(2016)]{Guo_low_mass} Guo, Y., Rafelski, M., Faber, S.~M., et al.\ 2016, \apj, 833, 37. doi:10.3847/1538-4357/833/1/37
   
   \bibitem[Hern{\'a}n-Caballero et al.(2013)]{Hernan_absorption} Hern{\'a}n-Caballero, A., Alonso-Herrero, A., P{\'e}rez-Gonz{\'a}lez, P.~G., et al.\ 2013, \mnras, 434, 2136. doi:10.1093/mnras/stt1165
   
   \bibitem[Hern{\'a}n-Caballero et al.(2014)]{Hernan_absorption_2} Hern{\'a}n-Caballero, A., Alonso-Herrero, A., P{\'e}rez-Gonz{\'a}lez, P.~G., et al.\ 2014, \mnras, 443, 3538. doi:10.1093/mnras/stu1413

   \bibitem[Hern{\'a}n-Caballero et al.(2017)]{SHARDSFF1} Hern{\'a}n-Caballero, A., P{\'e}rez-Gonz{\'a}lez, P.~G., Diego, J.~M., et al.\ 2017, \apj, 849, 82. doi:10.3847/1538-4357/aa917f

   \bibitem[Hsu et al.(2019)]{WIRCAM} Hsu, L.-T., Lin, L., Dickinson, M., et al.\ 2019, \apj, 871, 233. doi:10.3847/1538-4357/aaf9a7
   
   \bibitem[Hunter(2007)]{Hunter} Hunter, J.~D.\ 2007, Computing in Science and Engineering, 9, 90. doi:10.1109/MCSE.2007.55

   \bibitem[Iyer et al.(2018)]{Iyer} Iyer, K., Gawiser, E., Dav{\'e}, R., et al.\ 2018, \apj, 866, 120. doi:10.3847/1538-4357/aae0fa
   
   \bibitem[Karim et al.(2011)]{Karim} Karim, A., Schinnerer, E., Mart{\'\i}nez-Sansigre, A., et al.\ 2011, \apj, 730, 61. doi:10.1088/0004-637X/730/2/61

   \bibitem[Karman et al.(2017)]{Karman} Karman, W., Caputi, K.~I., Caminha, G.~B., et al.\ 2017, \aap, 599, A28. doi:10.1051/0004-6361/201629055

   \bibitem[Kashino et al.(2013)]{Kashino} Kashino, D., Silverman, J.~D., Rodighiero, G., et al.\ 2013, \apjl, 777, L8. doi:10.1088/2041-8205/777/1/L8

   \bibitem[Kennicutt(1998)]{Kennicutt} Kennicutt, R.~C.\ 1998, \araa, 36, 189. doi:10.1146/annurev.astro.36.1.189
   
   \bibitem[Koekemoer et al.(2011)]{WFC3} Koekemoer, A.~M., Faber, S.~M., Ferguson, H.~C., et al.\ 2011, \apjs, 197, 36. doi:10.1088/0067-0049/197/2/36
   
   \bibitem[Koudmani et al.(2021)]{Koudmani} Koudmani, S., Henden, N.~A., \& Sijacki, D.\ 2021, \mnras, 503, 3568. doi:10.1093/mnras/stab677

   \bibitem[Kron(1980)]{Kron1980} Kron, R.~G.\ 1980, \apjs, 43, 305. doi:10.1086/190669

   \bibitem[Kurczynski et al.(2016)]{Kurczynski} Kurczynski, P., Gawiser, E., Acquaviva, V., et al.\ 2016, \apjl, 820, L1. doi:10.3847/2041-8205/820/1/L1

   \bibitem[Laidler et al.(2006)]{TFIT} Laidler, V.~G., Grogin, N., Clubb, K., et al.\ 2006, Astronomical Data Analysis Software and Systems XV, 351, 228
   
   \bibitem[Lawrence et al.(2007)]{UDS1} Lawrence, A., Warren, S.~J., Almaini, O., et al.\ 2007, \mnras, 379, 1599. doi:10.1111/j.1365-2966.2007.12040.x
   
   \bibitem[Lee et al.(2015)]{Lee} Lee, N., Sanders, D.~B., Casey, C.~M., et al.\ 2015, \apj, 801, 80. doi:10.1088/0004-637X/801/2/80
   
   \bibitem[Leja et al.(2021)]{Leja} Leja, J., Speagle, J.~S., Ting, Y.-S., et al.\ 2021, arXiv:2110.04314

   \bibitem[Lumbreras-Calle et al.(2019)]{Lumbreras} Lumbreras-Calle, A., Mu{\~n}oz-Tu{\~n}{\'o}n, C., M{\'e}ndez-Abreu, J., et al.\ 2019, \aap, 621, A52. doi:10.1051/0004-6361/201731670
   
   \bibitem[Magdis et al.(2010)]{Magdis} Magdis, G.~E., Rigopoulou, D., Huang, J.-S., et al.\ 2010, \mnras, 401, 1521. doi:10.1111/j.1365-2966.2009.15779.x

   \bibitem[Magnelli et al.(2013)]{Magnelli} Magnelli, B., Popesso, P., Berta, S., et al.\ 2013, \aap, 553, A132. doi:10.1051/0004-6361/201321371
   
   \bibitem[Maseda et al.(2018)]{Maseda} Maseda, M.~V., van der Wel, A., Rix, H.-W., et al.\ 2018, \apj, 854, 29. doi:10.3847/1538-4357/aaa76e

   \bibitem[McGaugh et al.(2017)]{Mcgaugh} McGaugh, S.~S., Schombert, J.~M., \& Lelli, F.\ 2017, \apj, 851, 22. doi:10.3847/1538-4357/aa9790

   \bibitem[Matthee et al.(2016)]{Matthee} Matthee, J., Sobral, D., Oteo, I., et al.\ 2016, \mnras, 458, 449. doi:10.1093/mnras/stw322

   \bibitem[Meurer et al.(1999)]{Meurer} Meurer, G.~R., Heckman, T.~M., \& Calzetti, D.\ 1999, \apj, 521, 64. doi:10.1086/307523
   
   \bibitem[Muzzin et al.(2013)]{Muzzin} Muzzin, A., Marchesini, D., Stefanon, M., et al.\ 2013, \apj, 777, 18. doi:10.1088/0004-637X/777/1/18

   \bibitem[Nayyeri et al.(2017)]{cosmos_cat} Nayyeri, H., Hemmati, S., Mobasher, B., et al.\ 2017, \apjs, 228, 7. doi:10.3847/1538-4365/228/1/7
   
   \bibitem[Nelson et al.(2019)]{TNG} Nelson, D., Springel, V., Pillepich, A., et al.\ 2019, Computational Astrophysics and Cosmology, 6, 2. doi:10.1186/s40668-019-0028-x

   \bibitem[Nilsson et al.(2009)]{Nilsson} Nilsson, K.~K., Tapken, C., M{\o}ller, P., et al.\ 2009, \aap, 498, 13. doi:10.1051/0004-6361/200810881

   \bibitem[Noeske et al.(2007)]{Noeske} Noeske, K.~G., Weiner, B.~J., Faber, S.~M., et al.\ 2007, \apjl, 660, L43. doi:10.1086/517926
   
   \bibitem[Noeske et al.(2007)]{Downsizing} Noeske, K.~G., Faber, S.~M., Weiner, B.~J., et al.\ 2007, \apjl, 660, L47. doi:10.1086/517927
   
   \bibitem[Oke \& Gunn(1983)]{okegunn1987} Oke, J.~B. \& Gunn, J.~E.\ 1983, \apj, 266, 713. doi:10.1086/160817
   
   \bibitem[Pannella et al.(2009)]{Pannella} Pannella, M., Carilli, C.~L., Daddi, E., et al.\ 2009, \apjl, 698, L116. doi:10.1088/0004-637X/698/2/L116
   
   \bibitem[Pearson et al.(2018)]{Pearson} Pearson, W.~J., Wang, L., Hurley, P.~D., et al.\ 2018, \aap, 615, A146. doi:10.1051/0004-6361/201832821

   \bibitem[P{\'e}rez-Gonz{\'a}lez et al.(2005)]{PZETA} P{\'e}rez-Gonz{\'a}lez, P.~G., Rieke, G.~H., Egami, E., et al.\ 2005, \apj, 630, 82. doi:10.1086/431894
   
   \bibitem[P{\'e}rez-Gonz{\'a}lez et al.(2008)]{SYNTHESIZER} P{\'e}rez-Gonz{\'a}lez, P.~G., Rieke, G.~H., Villar, V., et al.\ 2008, \apj, 675, 234. doi:10.1086/523690
   
   \bibitem[P{\'e}rez-Gonz{\'a}lez et al.(2013)]{SHARDS} P{\'e}rez-Gonz{\'a}lez, P.~G., Cava, A., Barro, G., et al.\ 2013, \apj, 762, 46. doi:10.1088/0004-637X/762/1/46

   \bibitem[Pillepich et al.(2019)]{TNG_2} Pillepich, A., Nelson, D., Springel, V., et al.\ 2019, \mnras, 490, 3196. doi:10.1093/mnras/stz2338
   
   \bibitem[Pirzkal et al.(2013)]{Pirzkal} Pirzkal, N., Rothberg, B., Ly, C., et al.\ 2013, \apj, 772, 48. doi:10.1088/0004-637X/772/1/48

   \bibitem[Popesso et al.(2019)]{Popesso} Popesso, P., Concas, A., Morselli, L., et al.\ 2019, \mnras, 483, 3213. doi:10.1093/mnras/sty3210

   \bibitem[Quinn et al.(1996)]{photoionization} Quinn, T., Katz, N., \& Efstathiou, G.\ 1996, \mnras, 278, L49. doi:10.1093/mnras/278.4.L49
   
   \bibitem[Reddy(2009)]{Reddy} Reddy, N.~A.\ 2009, Galaxy Evolution: Emerging Insights and Future Challenges, 419, 313
   
   \bibitem[Reddy et al.(2012)]{Reddy_MS} Reddy, N.~A., Pettini, M., Steidel, C.~C., et al.\ 2012, \apj, 754, 25. doi:10.1088/0004-637X/754/1/25

   \bibitem[Rinaldi et al.(2021)]{Rinaldi} Rinaldi, P., Caputi, K.~I., van Mierlo, S., et al.\ 2021, arXiv:2112.03935
   
   \bibitem[Rodighiero et al.(2011)]{Rodig} Rodighiero, G., Daddi, E., Baronchelli, I., et al.\ 2011, \apjl, 739, L40. doi:10.1088/2041-8205/739/2/L40
      
   \bibitem[Rodr{\'\i}guez-Mu{\~n}oz et al.(2022)]{Rodriguez-Munoz} Rodr{\'\i}guez-Mu{\~n}oz, L., Rodighiero, G., P{\'e}rez-Gonz{\'a}lez, P.~G., et al.\ 2022, \mnras, 510, 2061. doi:10.1093/mnras/stab3558
   
   \bibitem[Salim et al.(2018)]{Salim} Salim, S., Boquien, M., \& Lee, J.~C.\ 2018, \apj, 859, 11. doi:10.3847/1538-4357/aabf3c

   \bibitem[Salmon et al.(2015)]{Salmon} Salmon, B., Papovich, C., Finkelstein, S.~L., et al.\ 2015, \apj, 799, 183. doi:10.1088/0004-637X/799/2/183

   \bibitem[Santos et al.(2020)]{Santos} Santos, S., Sobral, D., Matthee, J., et al.\ 2020, \mnras, 493, 141. doi:10.1093/mnras/staa093
   
   \bibitem[Santini et al.(2012)]{Santini_SMF} Santini, P., Fontana, A., Grazian, A., et al.\ 2012, \aap, 538, A33. doi:10.1051/0004-6361/201117513

   \bibitem[Santini et al.(2017)]{Santini} Santini, P., Fontana, A., Castellano, M., et al.\ 2017, \apj, 847, 76. doi:10.3847/1538-4357/aa8874
   
   \bibitem[Sawicki(2012)]{Sawicki} Sawicki, M.\ 2012, \mnras, 421, 2187. doi:10.1111/j.1365-2966.2012.20452.x

   \bibitem[Scoville et al.(2007)]{COSMOS} Scoville, N., Aussel, H., Brusa, M., et al.\ 2007, \apjs, 172, 1. doi:10.1086/516585

   \bibitem[Schreiber et al.(2015)]{Schreiber} Schreiber, C., Pannella, M., Elbaz, D., et al.\ 2015, \aap, 575, A74. doi:10.1051/0004-6361/201425017

   \bibitem[Skelton et al.(2014)]{3DHST} Skelton, R.~E., Whitaker, K.~E., Momcheva, I.~G., et al.\ 2014, \apjs, 214, 24. doi:10.1088/0067-0049/214/2/24
   
   \bibitem[Sobral et al.(2013)]{Sobral} Sobral, D., Smail, I., Best, P.~N., et al.\ 2013, \mnras, 428, 1128. doi:10.1093/mnras/sts096

   \bibitem[Speagle et al.(2014)]{Speagle} Speagle, J.~S., Steinhardt, C.~L., Capak, P.~L., et al.\ 2014, \apjs, 214, 15. doi:10.1088/0067-0049/214/2/15
   
   \bibitem[Stark et al.(2014)]{Stark} Stark, D.~P., Richard, J., Siana, B., et al.\ 2014, \mnras, 445, 3200. doi:10.1093/mnras/stu1618
   
   \bibitem[Stefanon et al.(2017)]{egs_cat} Stefanon, M., Yan, H., Mobasher, B., et al.\ 2017, \apjs, 229, 32. doi:10.3847/1538-4365/aa66cb
   
   \bibitem[Stinson et al.(2007)]{Stinson} Stinson, G.~S., Dalcanton, J.~J., Quinn, T., et al.\ 2007, \apj, 667, 170. doi:10.1086/520504

   \bibitem[Sun et al.(2021)]{Sun} Sun, F., Egami, E., P{\'e}rez-Gonz{\'a}lez, P.~G., et al.\ 2021, \apj, 922, 114. doi:10.3847/1538-4357/ac2578

   \bibitem[Teyssier et al.(2013)]{Teyssier} Teyssier, R., Pontzen, A., Dubois, Y., et al.\ 2013, \mnras, 429, 3068. doi:10.1093/mnras/sts563

   \bibitem[Tomczak et al.(2016)]{Tomczak} Tomczak, A.~R., Quadri, R.~F., Tran, K.-V.~H., et al.\ 2016, \apj, 817, 118. doi:10.3847/0004-637X/817/2/118
   
   \bibitem[Torrey et al.(2020)]{Torrey} Torrey, P., Hopkins, P.~F., Faucher-Gigu{\`e}re, C.-A., et al.\ 2020, \mnras, 497, 5292. doi:10.1093/mnras/staa2222
   
   \bibitem[van der Walt et al.(2011)]{numpy} van der Walt, S., Colbert, S.~C., \& Varoquaux, G.\ 2011, Computing in Science and Engineering, 13, 22. doi:10.1109/MCSE.2011.37
   
   \bibitem[Virtanen et al.(2020)]{scipy} Virtanen, P., Gommers, R., Oliphant, T.~E., et al.\ 2020, Nature Methods, 17, 261. doi:10.1038/s41592-019-0686-2

   \bibitem[Weisz et al.(2012)]{Weisz} Weisz, D.~R., Johnson, B.~D., Johnson, L.~C., et al.\ 2012, \apj, 744, 44. doi:10.1088/0004-637X/744/1/44

   \bibitem[Whitaker et al.(2011)]{uvj} Whitaker, K.~E., Labb{\'e}, I., van Dokkum, P.~G., et al.\ 2011, \apj, 735, 86. doi:10.1088/0004-637X/735/2/86

   \bibitem[Whitaker et al.(2012)]{Whitaker_12} Whitaker, K.~E., van Dokkum, P.~G., Brammer, G., et al.\ 2012, \apjl, 754, L29. doi:10.1088/2041-8205/754/2/L29

   \bibitem[Whitaker et al.(2014)]{Whitaker_15} Whitaker, K.~E., Franx, M., Leja, J., et al.\ 2014, \apj, 795, 104. doi:10.1088/0004-637X/795/2/104

   \bibitem[Yamaguchi et al.(2019)]{Yamaguchi} Yamaguchi, Y., Kohno, K., Hatsukade, B., et al.\ 2019, \apj, 878, 73. doi:10.3847/1538-4357/ab0d22

   \bibitem[Zahid et al.(2012)]{Zahid} Zahid, H.~J., Dima, G.~I., Kewley, L.~J., et al.\ 2012, \apj, 757, 54. doi:10.1088/0004-637X/757/1/54
\end{thebibliography}
\end{document}